%% file: main.tex
\documentclass[longauth]{aa} 
\input{definitions}
\graphicspath{{figures/}}

\newcommand{\nmuse}{19}

\newcommand{\nalmagmcs}{88}

\newcommand{\TabGalaxySample}{%
  \begin{table*}
  \centering
    \caption{Properties of our galaxy sample}
    \label{tab:galaxy:sample}
    \centering %
    \begin{tabular}{lrrcrrr}
      \hline
      \hline
      Galaxy & Distance  & Incl.     & Morph. & Stellar Mass           & Linear Res.   & Linear Res. \\
             & $[$Mpc$]$ & $[$deg$]$ &        & $[\log_{10}{M_\sun}]$  & $[$ALMA,\,pc$]$ & $[$MUSE,\,pc$]$ \\
      \hline
      IC\,5332   &   9.0  &  27  &  SABc &   9.7  &   32  &   38  \\
      NGC\,0628  &   9.8  &   9  &  Sc   &  10.3  &   53  &   44  \\
      NGC\,1087  &  15.8  &  43  &  Sc   &   9.9  &  123  &   71  \\
      NGC\,1300  &  19.0  &  32  &  Sbc  &  10.6  &  112  &   82  \\
      NGC\,1365  &  19.6  &  55  &  Sb   &  11.0  &  131  &  109  \\
      NGC\,1385  &  17.2  &  44  &  Sc   &  10.0  &  106  &   56  \\
      NGC\,1433  &  18.6  &  29  &  SBa  &  10.9  &   98  &   82  \\
      NGC\,1512  &  18.8  &  42  &  Sa   &  10.7  &  115  &  114  \\
      NGC\,1566  &  17.7  &  30  &  SABb &  10.8  &   95  &   69  \\
      NGC\,1672  &  19.4  &  43  &  Sb   &  10.7  &  182  &   90  \\
      NGC\,2835  &  12.2  &  41  &  Sc   &  10.0  &   50  &   68  \\
      NGC\,3351  &  10.0  &  45  &  Sb   &  10.4  &   70  &   51  \\
      NGC\,3627  &  11.3  &  57  &  Sb   &  10.8  &   89  &   58  \\
      NGC\,4254  &  13.1  &  34  &  Sc   &  10.4  &  113  &   57  \\
      NGC\,4303  &  17.0  &  24  &  Sbc  &  10.5  &  152  &   64  \\
      NGC\,4321  &  15.2  &  38  &  SABb &  10.7  &  126  &   86  \\
      NGC\,4535  &  15.8  &  45  &  Sc   &  10.5  &  119  &   43  \\
      NGC\,5068  &   5.2  &  36  &  Sc   &   9.4  &   26  &   26  \\
      NGC\,7496  &  18.7  &  36  &  Sb   &  10.0  &  152  &   81  \\
      \hline                
    \end{tabular}\\
    Table references: \citep{Anand2021}, \citep{Emsellem2022}, \citep{Lang2020}, \citep{Leroy2021a}.
  \end{table*}}

\newcommand{\TabStatOverlapFullSample}{%
  \begin{table*}
    \caption{The fraction of GMCs and \HII{} regions that are identified as
      matched objects across our sample.}
    \label{tab:overlap:full:sample}
    \centering{} %
    \begin{tabular}{ccccc}
      \hline %
      \hline %
      & \multicolumn{2}{c}{GMCs} & \multicolumn{2}{c}{\HII{} regions}\\
     MOP &  By Number & By Mass  & By Number & By Luminosity  \\
      \%      & \%    & \%       & \%    & \%       \\

      \hline
10 & 38.8 & 41.4 & 19.5 & 48.0  \\
40 & 10.4 & 11.6 & 3.8 & 27.8  \\
70 & 2.8 & 2.6 & 1.0 & 10.9  \\
\hline
    \end{tabular}
    \tablefoot{The fractions are indicated by number relative to the total
      number of GMCs (10866) and \HII{} regions (29904), and their
      contribution to the combined molecular gas mass
      ($5.1 \times 10^{10}M_{\odot}$) and \Ha{} luminosity
      ($5.4 \times 10^{42}$erg~s$^{-1}$) of the galaxies in our
      sample. Results are reported for a MOP of 10, 40 and 70\% (see main text).}
  \end{table*}}

\newcommand{\TabStatGalRegion}{%
  \begin{table*}
    \caption{Number and luminosity of GMCs and \HII{} regions per galactic
      environment of the studied sample, for a MOP
      of 40\%.}
    \label{tab:environmental:dependence}
    \centering{} %
    \begin{tabular}{ccccccc}
      \hline %
      \hline %
      \multirow{2}{*}{Galactic Environment} & \multicolumn{2}{c}{GMC} & \multicolumn{2}{c}{\HII{} regions} & CO Luminosity & \Ha{} Luminosity\\
      \cline{2-7}
      & \#  & \%\tablefootmark{a} & \#  & \%\tablefootmark{a} & \%\tablefootmark{b} & \%\tablefootmark{b} \\
      \hline
      Nucleus \& Bars          &  113 &  10.1 &  115 &  10.2 & 17.6 &  19.7 \\
      Arms                     &  605 &  53.4 &  610 &  53.4 &  60.2 &  54.7 \\
      Inter-arms \& Outer Disk &  414 &  36.5 &  417 &  36.4 &   22.3 &  25.6 \\
      All environments         & 1121 & 100.0 & 1142 & 100.0 & 100.0 & 100.0 \\
      \hline
    \end{tabular}\\
    \tablefoot{%
      \tablefoottext{a}{Proportion of the matched GMC/\HII{} region pairs
        that are in a specific region with respect to the total number of
        pairs.} %
      \tablefoottext{b}{Proportion of the luminosity for the matched
        GMC/\HII{} region pairs that are in a specific region with respect
        to the total luminosity of the matched GMG/\HII{} regions.}
    }\\[\bigskipamount]
    \resizebox{\linewidth}{!}{%
      \begin{tabular}{ccccccccc}
        \hline %
        \hline %
        \multirow{2}{*}{Galactic Environment} & \multicolumn{3}{c}{GMC} & \multicolumn{3}{c}{\HII{} regions} & CO Luminosity & \Ha{} Luminosity\\
        \cline{2-9}
        & Matched & Total & Matched & Matched & Total & Matched &  Matched & Matched \\
        &  \# & \# & \%\tablefootmark{a} & \# & \# & \%\tablefootmark{a} & \%\tablefootmark{b} & \%\tablefootmark{b} \\
        \hline
        Nucleus \& Bars          & 113 & 2509 &  4.5 & 115 &  2710 & 4.2 &  5.2 & 12.8 \\
        Arms                     & 605 & 4405 & 13.6 & 610 & 12033 & 5.1 & 19.6 & 41.4 \\
        Inter-arms \& Outer Disk & 414 & 3951 & 10.3 & 417 & 15161 & 2.7 & 10.0 & 34.7 \\
        \hline
      \end{tabular}}\\ 
    \tablefoot{%
      \tablefoottext{a}{Proportion of the matched GMC/\HII{} region pairs
        that are in a specific environment with respect to the total number of
        GMCs or \HII{} regions in this environment.} %
      \tablefoottext{b}{Proportion of the luminosity for the matched
        GMC/\HII{} region pairs that are in a specific environment with respect
        to the total luminosity of the GMCs or \HII{} regions in this
        environment.}
    }
  \end{table*}}

\newcommand{\TabStatOverlapPerGalaxies}{%
  \begin{table*}
    \caption{Number and luminosity of GMCs and \HII{} regions per galaxy of
      the studied sample as a function of the MOP.}
    \label{tab:overlap:per:galaxies}
    \centering{} %
    \begin{tabular}{cccccccccccc}
      \hline %
      \hline %
      \multirow{3}{*}{Galaxy} & \multirow{3}{*}{\%} & \multicolumn{2}{c}{GMC} & \multicolumn{2}{c}{\HII{} regions} & \multicolumn{2}{c}{CO mass} & \multicolumn{2}{c}{H$\alpha$ luminosity} \\
      \cline{3-10}
                               &                     & Total & Matched         & Total & Matched                  & Total           & Matched & Total              & Matched             \\
                               &                     & \#    & \%              & \#    & \%                       &\unit{M_{\odot}} & \%      & \unit{erg\,s^{-1}} & \%                  \\
      \hline
NGC\,0628 & 10 & 809 & 56.7 & 2869 & 23.5 & 9.69\e{8} & 69.0 & 9.58\e{40} & 81.1 \\
& 40 & & 15.0 &  & 4.3 & 9.69\e{8} & 18.7 & 9.58\e{40} & 51.0 \\
& 70 & & 3.5 &  & 1.0 & 9.69\e{8} & 4.1 & 9.58\e{40} & 21.5 \\
NGC\,1087 & 10 & 308 & 60.4 & 1011 & 27.7 & 1.07\e{9} & 64.7 & 1.76\e{41} & 58.5 \\
& 40 & & 14.3 &  & 4.4 & 1.07\e{9} & 14.2 & 1.76\e{41} & 32.7 \\
& 70 & & 2.6 &  & 0.8 & 1.07\e{9} & 3.3 & 1.76\e{41} & 15.4 \\
NGC\,1300 & 10 & 434 & 37.6 & 1478 & 15.0 & 1.21\e{9} & 39.7 & 9.16\e{40} & 52.1 \\
& 40 & & 14.3 &  & 4.2 & 1.21\e{9} & 16.9 & 9.16\e{40} & 32.1 \\
& 70 & & 2.8 &  & 0.8 & 1.21\e{9} & 2.1 & 9.16\e{40} & 10.2 \\
NGC\,1365 & 10 & 1092 & 7.1 & 1455 & 6.0 & 1.62\e{10} & 10.5 & 1.38\e{42} & 7.6 \\
& 40 & & 3.7 &  & 2.7 & 1.62\e{10} & 4.5 & 1.38\e{42} & 5.6 \\
& 70 & & 1.4 &  & 1.0 & 1.62\e{10} & 1.5 & 1.38\e{42} & 3.2 \\
NGC\,1385 & 10 & 407 & 49.4 & 1029 & 26.1 & 1.16\e{9} & 65.3 & 2.97\e{41} & 66.4 \\
& 40 & & 20.1 &  & 8.0 & 1.16\e{9} & 28.5 & 2.97\e{41} & 50.2 \\
& 70 & & 5.4 &  & 2.1 & 1.16\e{9} & 9.1 & 2.97\e{41} & 27.1 \\
NGC\,1433 & 10 & 355 & 25.1 & 1736 & 6.0 & 1.03\e{9} & 20.2 & 6.70\e{40} & 26.4 \\
& 40 & & 12.4 &  & 2.5 & 1.03\e{9} & 10.5 & 6.70\e{40} & 21.6 \\
& 70 & & 3.9 &  & 0.8 & 1.03\e{9} & 2.4 & 6.70\e{40} & 11.5 \\
NGC\,1512 & 10 & 317 & 20.2 & 632 & 10.8 & 6.36\e{8} & 21.7 & 6.04\e{40} & 42.9 \\
& 40 & & 13.2 &  & 6.8 & 6.36\e{8} & 14.2 & 6.04\e{40} & 38.7 \\
& 70 & & 5.4 &  & 2.7 & 6.36\e{8} & 7.0 & 6.04\e{40} & 32.7 \\
NGC\,1566 & 10 & 1127 & 38.9 & 2404 & 24.1 & 5.30\e{9} & 48.9 & 4.20\e{41} & 63.5 \\
& 40 & & 12.8 &  & 6.1 & 5.30\e{9} & 20.2 & 4.20\e{41} & 49.0 \\
& 70 & & 2.7 &  & 1.2 & 5.30\e{9} & 4.1 & 4.20\e{41} & 22.0 \\
NGC\,1672 & 10 & 517 & 31.5 & 1581 & 13.9 & 6.14\e{9} & 43.4 & 6.77\e{41} & 52.3 \\
& 40 & & 7.9 &  & 2.6 & 6.14\e{9} & 14.9 & 6.77\e{41} & 33.5 \\
& 70 & & 1.4 &  & 0.4 & 6.14\e{9} & 3.5 & 6.77\e{41} & 9.3 \\
NGC\,2835 & 10 & 211 & 26.5 & 1121 & 5.4 & 1.05\e{8} & 44.5 & 5.78\e{40} & 32.0 \\
& 40 & & 22.3 &  & 4.3 & 1.05\e{8} & 37.6 & 5.78\e{40} & 28.8 \\
& 70 & & 17.1 &  & 3.3 & 1.05\e{8} & 26.9 & 5.78\e{40} & 26.5 \\
NGC\,3351 & 10 & 370 & 42.2 & 1284 & 14.8 & 5.64\e{8} & 34.3 & 6.35\e{40} & 18.4 \\
& 40 & & 8.9 &  & 2.6 & 5.64\e{8} & 11.2 & 6.35\e{40} & 13.4 \\
& 70 & & 1.9 &  & 0.5 & 5.64\e{8} & 1.1 & 6.35\e{40} & 2.0 \\
NGC\,3627 & 10 & 983 & 35.2 & 1635 & 27.6 & 5.56\e{9} & 41.1 & 3.77\e{41} & 59.2 \\
& 40 & & 7.0 &  & 4.3 & 5.56\e{9} & 10.0 & 3.77\e{41} & 34.2 \\
& 70 & & 2.4 &  & 1.5 & 5.56\e{9} & 2.7 & 3.77\e{41} & 15.0 \\
NGC\,4254 & 10 & 918 & 61.8 & 2960 & 29.4 & 5.86\e{9} & 76.2 & 4.44\e{41} & 78.0 \\
& 40 & & 10.1 &  & 3.2 & 5.86\e{9} & 13.0 & 4.44\e{41} & 26.8 \\
& 70 & & 0.9 &  & 0.3 & 5.86\e{9} & 1.5 & 4.44\e{41} & 5.0 \\
NGC\,4303 & 10 & 874 & 49.7 & 3067 & 21.6 & 5.98\e{9} & 63.6 & 6.22\e{41} & 73.2 \\
& 40 & & 10.1 &  & 2.9 & 5.98\e{9} & 14.1 & 6.22\e{41} & 35.1 \\
& 70 & & 1.7 &  & 0.5 & 5.98\e{9} & 1.7 & 6.22\e{41} & 8.6 \\
NGC\,4321 & 10 & 1214 & 40.4 & 1847 & 34.4 & 4.94\e{9} & 54.5 & 3.03\e{41} & 66.2 \\
& 40 & & 6.6 &  & 4.3 & 4.94\e{9} & 11.7 & 3.03\e{41} & 34.8 \\
& 70 & & 1.2 &  & 0.8 & 4.94\e{9} & 2.9 & 3.03\e{41} & 12.8 \\
NGC\,4535 & 10 & 638 & 32.9 & 1938 & 15.3 & 1.78\e{9} & 46.8 & 1.26\e{41} & 57.3 \\
& 40 & & 5.5 &  & 1.8 & 1.78\e{9} & 8.2 & 1.26\e{41} & 24.7 \\
& 70 & & 0.9 &  & 0.3 & 1.78\e{9} & 2.8 & 1.26\e{41} & 11.0 \\
NGC\,5068 & 10 & 292 & 39.4 & 1857 & 8.6 & 1.01\e{8} & 56.8 & 2.76\e{40} & 50.3 \\
& 40 & & 22.3 &  & 3.6 & 1.01\e{8} & 27.8 & 2.76\e{40} & 35.7 \\
& 70 & & 13.4 &  & 2.1 & 1.01\e{8} & 15.0 & 2.76\e{40} & 29.4 \\
      \hline
    \end{tabular}
  \end{table*}}

\newcommand{\TabCorrelations}{%
  \begin{table*}
    \caption{Correlation properties (\Rsquared{} coefficient and power law
      index) for various GMC properties as a function of the \Ha{}
      luminosity for each galaxy in the sample. The minimum overlap
      percentage used for the matching is 40\%. The galaxies are sorted by
      decreasing \Rsquared{} coefficient for each individual sub-table.}
    \label{tab:correlations}
    \centering %
    \begin{minipage}{0.245\linewidth}
    \centering %
      \begin{tabular}{lrr}
        \hline
        \hline
        \multicolumn{3}{c}{\thead{Molecular Mass \\ $ $}} \\
        \hline
        Galaxy & \Rsquared{} & slope \\
        \hline
        NGC\,4535 & 0.65 & 0.51 \\
        NGC\,1672 & 0.60 & 0.59 \\
        NGC\,4321 & 0.54 & 0.55 \\
        NGC\,3351 & 0.51 & 0.47 \\
        NGC\,1566 & 0.48 & 0.57 \\
        NGC\,0628 & 0.43 & 0.46 \\
        NGC\,3627 & 0.43 & 0.46 \\
        NGC\,4254 & 0.42 & 0.59 \\
        NGC\,1385 & 0.42 & 0.38 \\
        NGC\,4303 & 0.38 & 0.47 \\
        NGC\,1300 & 0.37 & 0.37 \\
        NGC\,1512 & 0.36 & 0.24 \\
        NGC\,1365 & 0.35 & 0.33 \\
        NGC\,2835 & 0.25 & 0.21 \\
        NGC\,1087 & 0.17 & 0.21 \\
        NGC\,5068 & 0.15 & 0.18 \\
        NGC\,1433 & 0.10 & 0.13 \\
        \hline                
      \end{tabular}
    \end{minipage}
    \begin{minipage}{0.245\linewidth}
      \centering %
      \begin{tabular}{lrr}
        \hline
        \hline
        \multicolumn{3}{c}{\thead{Surface density \\ $ $}} \\
        \hline
        Galaxy & \Rsquared{} & slope \\
        \hline
        NGC\,3627 & 0.37 & 0.34 \\
        NGC\,4535 & 0.35 & 0.34 \\
        NGC\,4321 & 0.25 & 0.30 \\
        NGC\,1566 & 0.23 & 0.32 \\
        NGC\,0628 & 0.15 & 0.20 \\
        NGC\,1672 & 0.13 & 0.21 \\
        NGC\,3351 & 0.11 & 0.22 \\
        NGC\,4254 & 0.11 & 0.24 \\
        NGC\,5068 & 0.10 & 0.18 \\
        NGC\,1512 & 0.09 & 0.15 \\
        NGC\,1300 & 0.06 & 0.14 \\
        NGC\,1385 & 0.05 & 0.13 \\
        NGC\,4303 & 0.05 & 0.17 \\
        NGC\,2835 & 0.04 & 0.11 \\
        NGC\,1087 & 0.04 & 0.12 \\
        NGC\,1365 & 0.02 & 0.08 \\
        NGC\,1433 & 0.00 & -0.03 \\
        \hline                
      \end{tabular}      
    \end{minipage}
    \begin{minipage}{0.245\linewidth}
      \centering %
      \begin{tabular}{lrr}
        \hline %
        \hline %
        \multicolumn{3}{c}{\thead{Peak Temperature \\ $ $}} \\
        \hline
        Galaxy  & \Rsquared{} & slope \\
        \hline
        NGC\,1672 & 0.71 & 0.43 \\
        NGC\,4535 & 0.69 & 0.30 \\
        NGC\,4321 & 0.57 & 0.34 \\
        NGC\,3627 & 0.45 & 0.26 \\
        NGC\,3351 & 0.44 & 0.28 \\
        NGC\,1566 & 0.41 & 0.31 \\
        NGC\,1385 & 0.38 & 0.24 \\
        NGC\,1365 & 0.34 & 0.16 \\
        NGC\,1512 & 0.34 & 0.15 \\
        NGC\,4303 & 0.33 & 0.29 \\
        NGC\,0628 & 0.32 & 0.19 \\
        NGC\,4254 & 0.32 & 0.38 \\
        NGC\,1300 & 0.31 & 0.16 \\
        NGC\,1087 & 0.16 & 0.14 \\
        NGC\,2835 & 0.15 & 0.08 \\
        NGC\,5068 & 0.06 & 0.07 \\
        NGC\,1433 & 0.06 & 0.05 \\
        \hline                
      \end{tabular}
    \end{minipage}
    \begin{minipage}{0.245\linewidth}
      \centering %
      \begin{tabular}{lrr}
        \hline
        \hline
        \multicolumn{3}{c}{ \thead{\modifref{Characteristic Turbulent} \\ \modifref{Linewidth}}} \\
        \hline
        Galaxy & \Rsquared{} & slope \\
        \hline
        \modifref{NGC}\,\modifref{3627} & \modifref{0.26} & \modifref{0.10} \\
        \modifref{NGC}\,\modifref{4321} & \modifref{0.23} & \modifref{0.15} \\
        \modifref{NGC}\,\modifref{1566} & \modifref{0.23} & \modifref{0.15} \\
        \modifref{NGC}\,\modifref{4254} & \modifref{0.17} & \modifref{0.15} \\
        \modifref{NGC}\,\modifref{1433} & \modifref{0.12} & \modifref{0.10} \\
        \modifref{NGC}\,\modifref{1300} & \modifref{0.12} & \modifref{0.10} \\
        \modifref{NGC}\,\modifref{1512} & \modifref{0.11} & \modifref{0.16} \\
        \modifref{NGC}\,\modifref{4535} & \modifref{0.10} & \modifref{0.09} \\
        \modifref{NGC}\,\modifref{1672} & \modifref{0.09} & \modifref{0.07} \\
        \modifref{NGC}\,\modifref{1365} & \modifref{0.08} & \modifref{0.07} \\
        \modifref{NGC}\,\modifref{0628} & \modifref{0.07} & \modifref{0.11} \\
        \modifref{NGC}\,\modifref{4303} & \modifref{0.05} & \modifref{0.07} \\
        \modifref{NGC}\,\modifref{2835} & \modifref{0.05} & \modifref{0.08} \\
        \modifref{NGC}\,\modifref{1385} & \modifref{0.04} & \modifref{0.08} \\
        \modifref{NGC}\,\modifref{5068} & \modifref{0.03} & \modifref{0.08} \\
        \modifref{NGC}\,\modifref{1087} & \modifref{0.01} & \modifref{0.04} \\
        \modifref{NGC}\,\modifref{3351} & \modifref{0.00} & \modifref{0.02} \\
        \hline                
      \end{tabular}      
    \end{minipage}
  \end{table*}}

\newcommand{\TabKSTest}{%
  \begin{table*} %
    \caption{Kolmogorov-Smirnov tests performed to assess whether the
      distributions are significantly different. \textbf{Top:} Comparison
      of the distributions of matched GMC/\HII{}-regions with their parent
      distributions.  \textbf{Bottom:} Comparison of the distributions of
      matched GMC/\HII{}-regions for different minimum overlap
      percentages.}
    \label{tab:KS:test}
    \centering %
    \resizebox{0.75\linewidth}{!}{\centering %
    \begin{tabular}{cccccccccc}
      \hline
      \hline
      \multicolumn{10}{c}{Parent distribution vs matched GMC/\HII{}-regions}\\
      \hline
      \parbox{1.5cm}{\centering Minimum \\ Overlap} & \multicolumn{9}{c}{GMC properties}\\
      Percentage & & Size & $L_\emr{CO}$ & $\sigma_{v}$  & $\Sigma_\emr{mol}$ & $\sigma_0$ & $\alpha_\emr{vir}$ &$T_\emr{peak}$  & $\tau_\emr{ff}$\\
      \hline
\hline
\textbf{10\%} & s value     & 1.00 & 1.00 & 0.73 & 1.00 & 0.98 & 0.78 & 0.92 & 0.99 \\ 
& p value  & $\ll$1e-3 & $\ll$1e-3 & $\ll$1e-3 & $\ll$1e-3 & $\ll$1e-3 & $\ll$1e-3 & $\ll$1e-3 & $\ll$1e-3 \\ 
& Different ? & yes & yes & yes & yes & yes & yes & yes & yes \\

\hline
\textbf{40\%} & s value     & 1.00 & 1.00 & 0.73 & 1.00 & 0.98 & 0.80 & 0.88 & 0.99 \\ 
& p value  & $\ll$1e-3 & $\ll$1e-3 & $\ll$1e-3 & $\ll$1e-3 & $\ll$1e-3 & $\ll$1e-3 & $\ll$1e-3 & $\ll$1e-3 \\ 
& Different ? & yes & yes & yes & yes & yes & yes & yes & yes \\

\hline
\textbf{70\%} & s value     & 1.00 & 1.00 & 0.73 & 1.00 & 0.98 & 0.83 & 0.86 & 0.99 \\ 
& p value  & $\ll$1e-3 & $\ll$1e-3 & $\ll$1e-3 & $\ll$1e-3 & $\ll$1e-3 & $\ll$1e-3 & $\ll$1e-3 & $\ll$1e-3 \\ 
& Different ? & yes & yes & yes & yes & yes & yes & yes & yes \\ 
      \hline
      \\
    \end{tabular}}
    \\
    \resizebox{0.75\linewidth}{!}{\centering %
    \begin{tabular}{cccccccccc}
      \hline
      \hline
      \multicolumn{10}{c}{Matched GMC/\HII{}-regions vs Matched GMC/\HII{}-regions}\\
      \hline
      \parbox{1.5cm}{\centering Minimum \\ Overlap} & \multicolumn{9}{c}{GMC properties}\\
      Percentage & & Size & $L_\emr{CO}$  &  $\sigma_{v}$& $\Sigma_\emr{mol}$ & $\sigma_0$ & $\alpha_\emr{vir}$ & $T_\emr{peak}$ & $\tau_\emr{ff}$\\
      \hline
\hline
\textbf{10\% vs 40\%} & s value     & 0.20 & 0.09 & 0.05 & 0.13 & 0.06 & 0.07 & 0.11 & 0.17 \\ 
& p value  & $\ll$1e-3 & $\ll$1e-3 &1.60e-02 & $\ll$1e-3 &5.20e-03 & $\ll$1e-3 & $\ll$1e-3 & $\ll$1e-3 \\ 
& Different ? & yes & yes & yes & yes & yes & yes & yes & yes \\

\hline
\textbf{10\% vs 70\%} & s value     & 0.35 & 0.22 & 0.13 & 0.24 & 0.10 & 0.11 & 0.20 & 0.32 \\ 
& p value  & $\ll$1e-3 & $\ll$1e-3 & $\ll$1e-3 & $\ll$1e-3 &3.94e-03 &2.05e-03 & $\ll$1e-3 & $\ll$1e-3 \\ 
& Different ? & yes & yes & yes & yes & yes & yes & yes & yes \\

\hline
\textbf{40\% vs 70\%} & s value     & 0.20 & 0.14 & 0.09 & 0.12 & 0.05 & 0.04 & 0.10 & 0.17 \\ 
& p value  & $\ll$1e-3 & $\ll$1e-3 &3.98e-02 &3.05e-03 &4.86e-01 &8.18e-01 &1.40e-02 & $\ll$1e-3 \\ 
& Different ? & yes & yes & yes & yes & yes & no & yes & yes \\ 
      \hline                
    \end{tabular}}
  \end{table*}}

\newcommand{\TabADTest}{%
  \begin{table*} %
    \caption{Anderson-Darling tests performed to assess whether the
      distributions are significantly different. \textbf{Top:} Comparison
      of the distributions of matched GMC/\HII{}-regions with their parent
      distributions.  \textbf{Bottom:} Comparison of the distributions of
      matched GMC/\HII{}-regions for different minimum overlap
      percentages.}
    \label{tab:AD:test}
    \centering %
    \resizebox{0.75\linewidth}{!}{\centering %
    \begin{tabular}{cccccccccc}
      \hline
      \hline
      \multicolumn{10}{c}{Parent distribution vs matched GMC/\HII{}-regions}\\
      \hline
      \parbox{1.5cm}{\centering Minimum \\ Overlap} & \multicolumn{9}{c}{GMC properties}\\
      Percentage & & Size & $L_\emr{CO}$ & $\sigma_{v}$  & $\Sigma_\emr{mol}$ & $\sigma_0$ & $\alpha_\emr{vir}$ &$T_\emr{peak}$  & $\tau_\emr{ff}$\\
\hline
\hline
\textbf{10\%} & s value     & $\gg$1e2 & $\gg$1e2 & $\gg$1e2 & $\gg$1e2 & $\gg$1e2 & $\gg$1e2 & $\gg$1e2 & $\gg$1e2 \\ 
& c value  & 1.96 & 1.96 & 1.96 & 1.96 & 1.96 & 1.96 & 1.96 & 1.96 \\ 
& p value  & $\ll$1e-3 & $\ll$1e-3 & $\ll$1e-3 & $\ll$1e-3 & $\ll$1e-3 & $\ll$1e-3 & $\ll$1e-3 & $\ll$1e-3 \\ 
& Different ? & yes & yes & no & yes & yes & yes & yes & yes \\

\hline
\textbf{40\%} & s value     & $\gg$1e2 & $\gg$1e2 & $\gg$1e2 & $\gg$1e2 & $\gg$1e2 & $\gg$1e2 & $\gg$1e2 & $\gg$1e2 \\ 
& c value  & 1.96 & 1.96 & 1.96 & 1.96 & 1.96 & 1.96 & 1.96 & 1.96 \\ 
& p value  & $\ll$1e-3 & $\ll$1e-3 & $\ll$1e-3 & $\ll$1e-3 & $\ll$1e-3 & $\ll$1e-3 & $\ll$1e-3 & $\ll$1e-3 \\ 
& Different ? & yes & yes & yes & yes & yes & yes & yes & yes \\

\hline
\textbf{70\%} & s value     & $\gg$1e2 & $\gg$1e2 & $\gg$1e2 & $\gg$1e2 & $\gg$1e2 & $\gg$1e2 & $\gg$1e2 & $\gg$1e2 \\ 
& c value  & 1.96 & 1.96 & 1.96 & 1.96 & 1.96 & 1.96 & 1.96 & 1.96 \\ 
& p value  & $\ll$1e-3 & $\ll$1e-3 & $\ll$1e-3 & $\ll$1e-3 & $\ll$1e-3 & $\ll$1e-3 & $\ll$1e-3 & $\ll$1e-3 \\ 
& Different ? & yes & yes & yes & yes & yes & yes & yes & yes \\ 
      \hline
      \\
    \end{tabular}}
    \\
    \resizebox{0.75\linewidth}{!}{\centering %
    \begin{tabular}{cccccccccc}
      \hline
      \hline
      \multicolumn{10}{c}{Matched GMC/\HII{}-regions vs Matched GMC/\HII{}-regions}\\
      \hline
      \parbox{1.5cm}{\centering Minimum \\ Overlap} & \multicolumn{9}{c}{GMC properties}\\
      Percentage & & Size & $L_\emr{CO}$  &  $\sigma_{v}$& $\Sigma_\emr{mol}$ & $\sigma_0$ & $\alpha_\emr{vir}$ & $T_\emr{peak}$ & $\tau_\emr{ff}$\\
      \hline
\hline
\textbf{10\% vs 40\%} & s value     & $\gg$1e2 & 21.72 & 6.79 & 63.95 & 7.13 & 14.61 & 35.91 & 97.71 \\ 
& c value  & 1.96 & 1.96 & 1.96 & 1.96 & 1.96 & 1.96 & 1.96 & 1.96 \\ 
& p value  & $\ll$1e-3 & $\ll$1e-3 & $\ll$1e-3 & $\ll$1e-3 & $\ll$1e-3 & $\ll$1e-3 & $\ll$1e-3 & $\ll$1e-3 \\ 
& Different ? & yes & yes & yes & yes & yes & yes & yes & yes \\

\hline
\textbf{10\% vs 70\%} & s value     & $\gg$1e2 & 48.69 & 13.61 & 55.43 & 8.62 & 8.24 & 37.88 & $\gg$1e2 \\ 
& c value  & 1.96 & 1.96 & 1.96 & 1.96 & 1.96 & 1.96 & 1.96 & 1.96 \\ 
& p value  & $\ll$1e-3 & $\ll$1e-3 & $\ll$1e-3 & $\ll$1e-3 & $\ll$1e-3 & $\ll$1e-3 & $\ll$1e-3 & $\ll$1e-3 \\ 
& Different ? & yes & yes & yes & yes & yes & yes & yes & yes \\

\hline
\textbf{40\% vs 70\%} & s value     & 30.20 & 12.92 & 2.85 & 7.35 & 0.74 & -0.50 & 6.15 & 20.06 \\ 
& c value  & 1.96 & 1.96 & 1.96 & 1.96 & 1.96 & 1.96 & 1.96 & 1.96 \\ 
& p value  & $\ll$1e-3 & $\ll$1e-3 &2.22e-02 & $\ll$1e-3 &1.63e-01 &2.50e-01 &1.38e-03 & $\ll$1e-3 \\ 
& Different ? & yes & yes & yes & yes & no & no & yes & yes \\ 
      \hline                
    \end{tabular}}
  \end{table*}}

\newcommand{\FigHistSizeGMCHII}{%
  \begin{figure}
    \centering %
    \includegraphics[width=\linewidth]{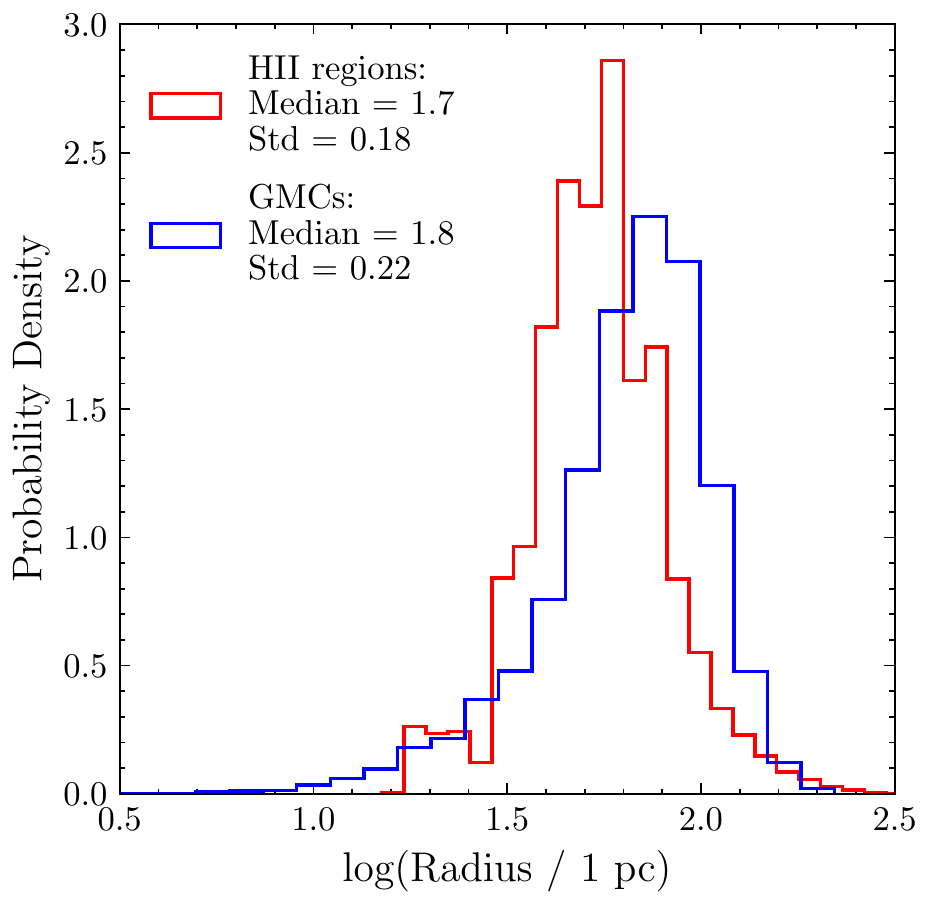}
    \caption{Probability Distribution Functions of the size of GMCs in blue
      and \HII{} regions in red. Means, medians, and standard deviations of
      the two distributions are listed on the left.}
    \label{fig:hist:sizeGMCHII}
  \end{figure}}

\newcommand{\FigMatchingExample}{%
  \begin{figure*}
    \centering %
    \includegraphics[width=0.95\linewidth]{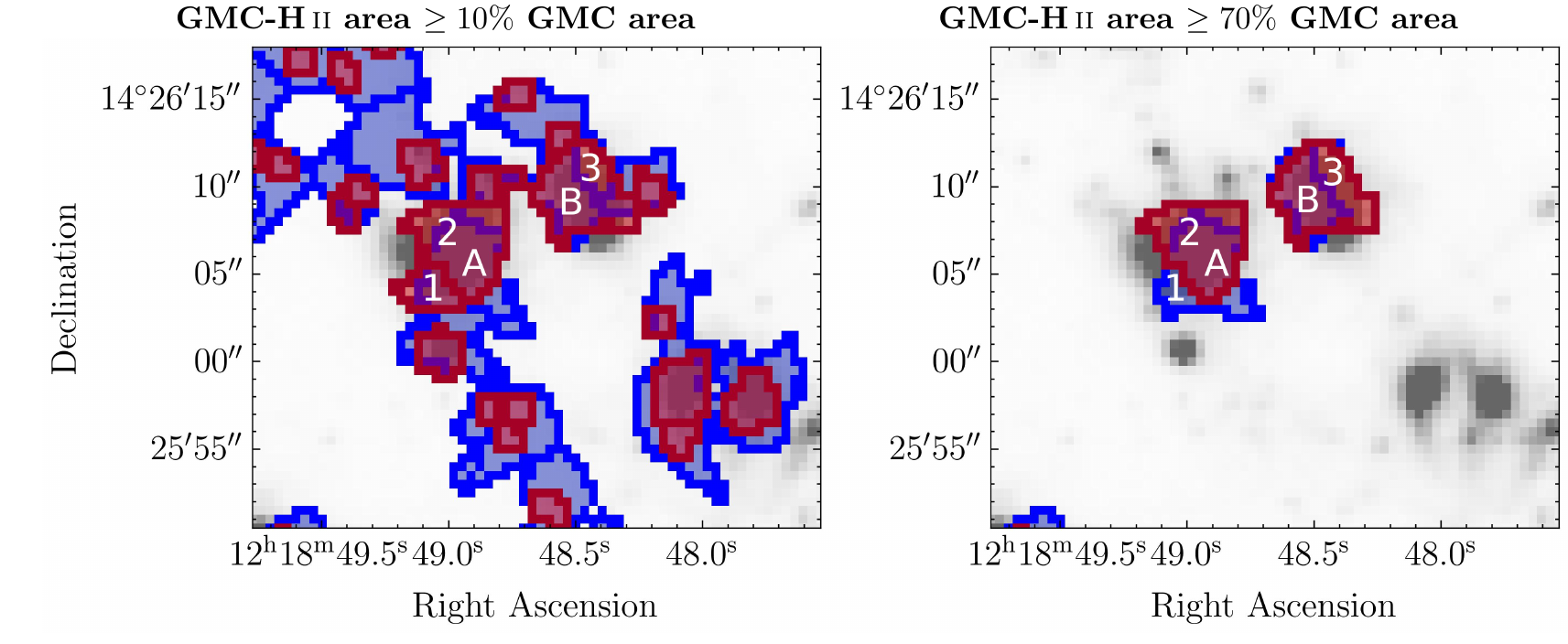}
    \caption{Example results of the algorithm used to match GMCs (blue) and
      \HII{} regions (red) for the same field-of-view. The background image
      is \Ha{} line emission from a spiral arm segment in NGC\,4254. The
      right and left panels show the matches obtained when the area of the
      GMC/\HII{} overlap region is at least 10\% and 70\%, respectively, of
      projected area of a GMC. Imposing a higher overlap percentage leads
      to fewer identifications of matched GMC/\HII{}-regions.}
    \label{fig:matching:example:zoom}
  \end{figure*}}

\newcommand{\FigHistVelOffandMOP}{%
  \begin{figure*}
    \centering %
    \includegraphics[height=0.47\linewidth,trim={0.0cm 0.1cm 0.1cm
      0.1cm},clip]{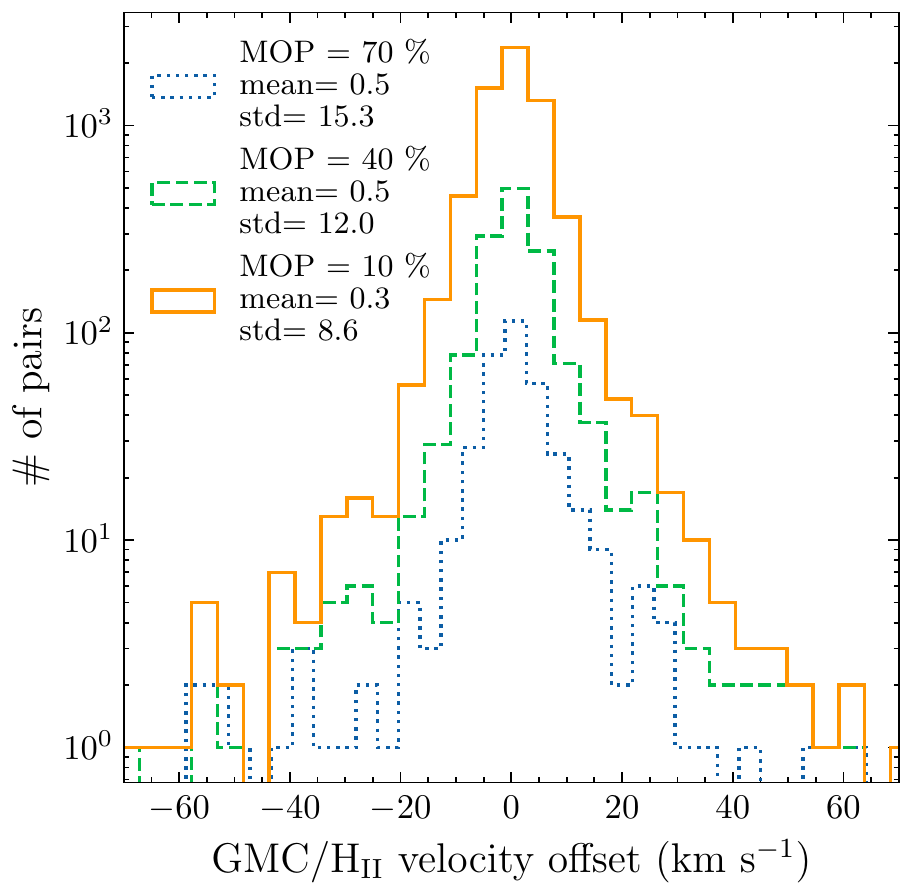}
    \hfill{} %
    \includegraphics[height=0.47\linewidth,clip]{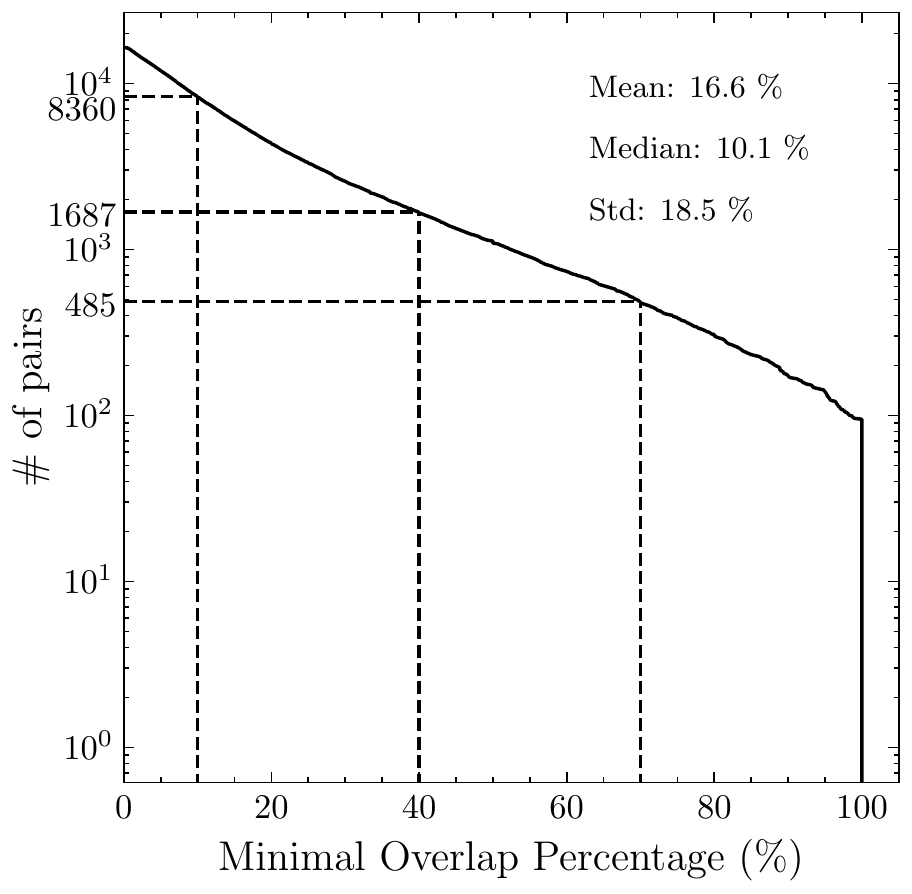}
    \caption{Histogram of the relative velocity (left) and number of pairs
      as a function of the MOP (right) between
      associated GMCs and \HII{} regions for all galaxies in the
      sample. One GMC/\HII{}-region pair is present in this histogram when
      at least one pixel is overlapping. The mean, median, and standard
      deviation of the overlap distribution are listed on the top right
      corner. The vertical dashed lines in the right panel represent the
      10\%, 40\% and 70\% MOP.}
    \label{fig:hist:overlap}
  \end{figure*}}

\newcommand{\FigHistGMCandHII}{%
  \begin{figure*}
    \centering %
     \includegraphics[width=0.84\linewidth]{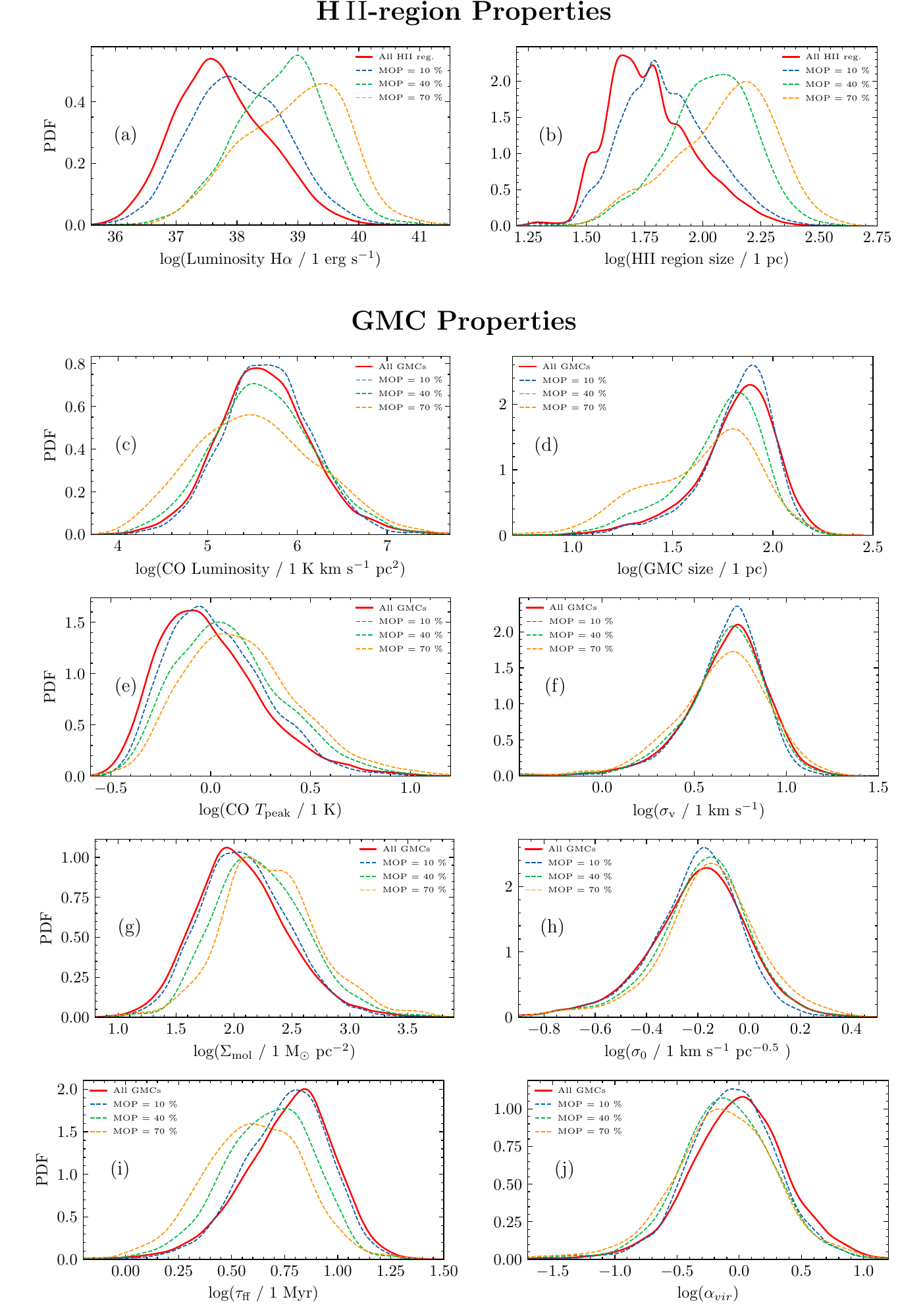}
     \caption{Probability density functions of key physical properties of
       \HII{} regions \modifreftwo{and of GMCs}. \textbf{Top, from left to
         right:} extinction-corrected \Ha\ luminosity and
       size. \textbf{Bottom, from top left to bottom right:} CO luminosity,
       size, peak temperature, velocity dispersion, surface density,
       characteristic turbulent linewidth, free-fall time and virial
       parameter. The distribution for the full GMC (\HII{} region)
       population is shown in red, while the matched GMCs/\HII{} regions
       adopting a MOP of 10, 40 and 70\% are shown in blue, green and gold
       respectively.}
     \label{fig:hist:overview}
\end{figure*}}

\newcommand{\FigCorrMatrix}{%
  \begin{figure*}
    \centering %
    \includegraphics[width=\linewidth]{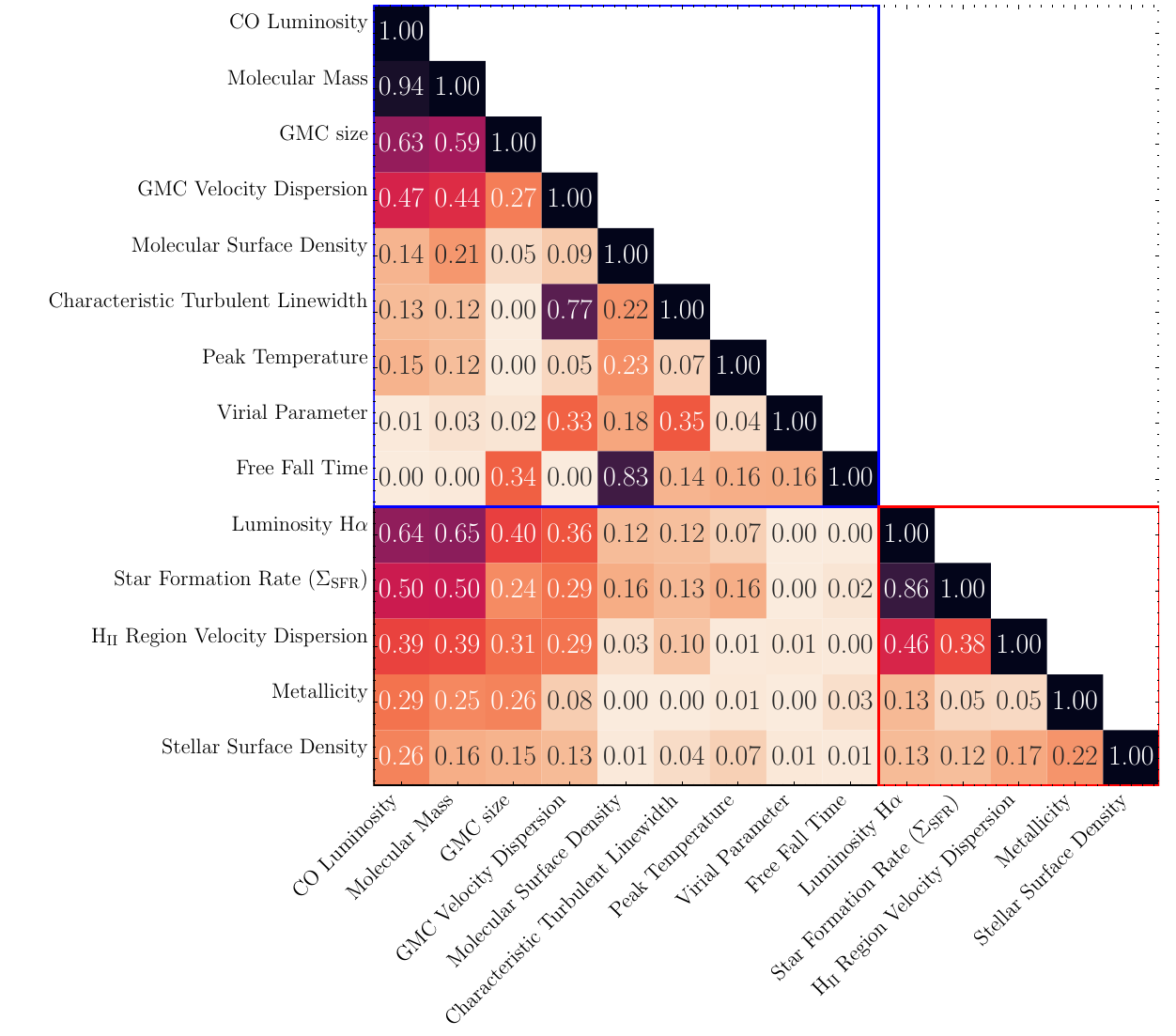}
    \caption{Correlation matrix of the different GMC and \HII{} region
      properties investigated. The \textbf{top left blue square} shows the
      correlations between GMC properties \modifreftwo{considering the entire sample}. The \textbf{bottom right red
        square} shows the correlations between the properties of \HII{}
      regions \modifreftwo{considering the entire sample}. The \textbf{bottom left black rectangle} shows the strengths
      of the correlations between matched GMC and \HII{} regions for a
      MOP of 70\%.}
    \label{fig:corr:matrix}
  \end{figure*}}
  
\newcommand{\FigCorrVsPercentage}{%
  \begin{figure*}
    \centering %
    \includegraphics[width=0.87\linewidth]{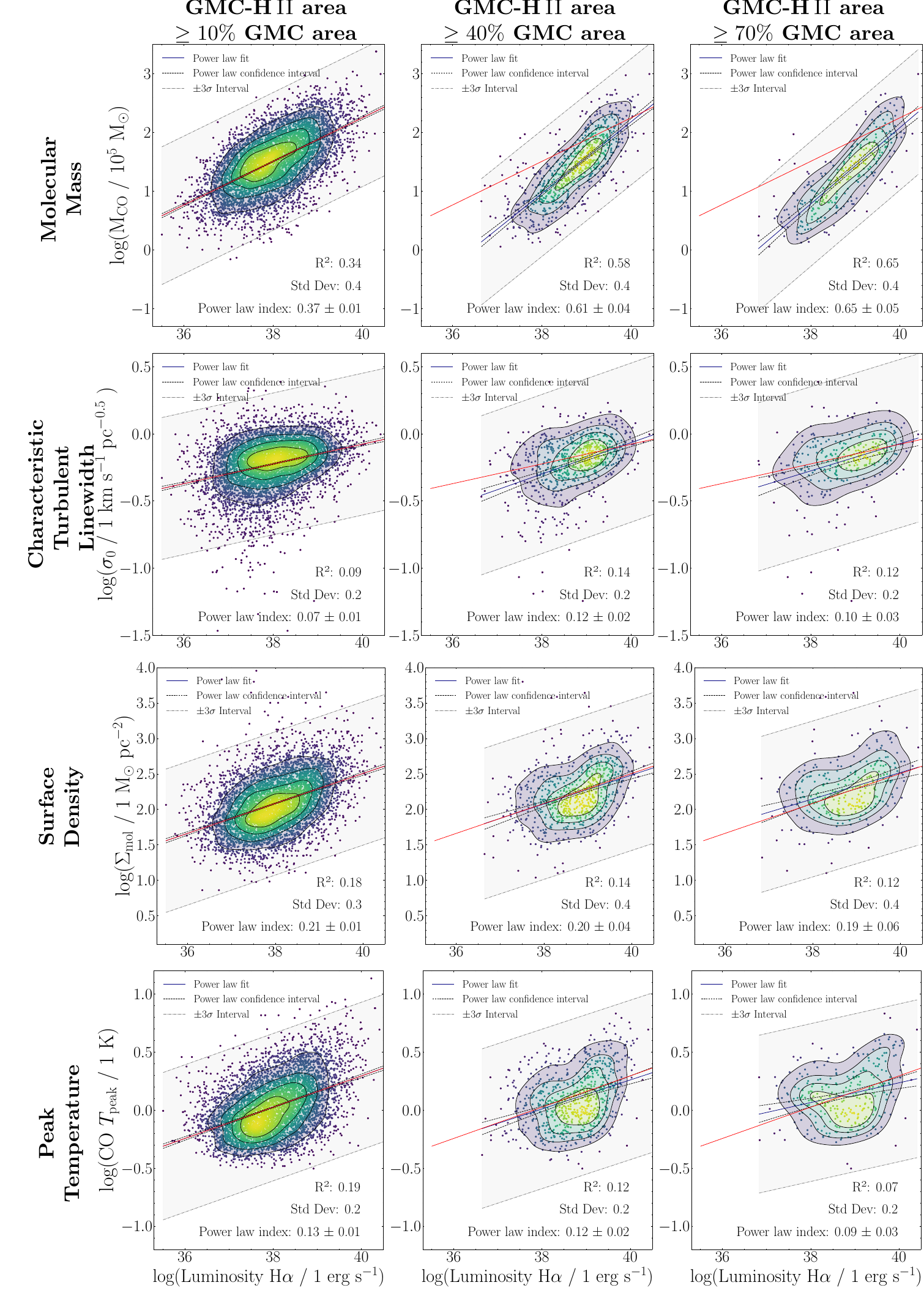}
    \caption{Scatter plots of various GMC properties as a function of the
      \Ha{} luminosity for three different MOP of
      the \HII/GMC pairs: 10\% for the \textbf{left column,} 40\% for the
      \textbf{middle} one, and 70\% for the \textbf{right} one. The GMC
      properties are \textbf{from top to bottom:} \modifreftwo{the molecular mass, the
        characteristic turbulent linewidth, the
      surface density and the peak temperature}. The black solid lines show a
      power law fit and its uncertainty for each scatter plot. The dashed
      lines show the $\pm 3\sigma$ levels from the fitted line.  The solid
      red line represents the linear regressions obtained in the leftmost
      column, for comparison. The fitted line properties are displayed on
      the bottom right corner of each panel.}
    \label{fig:corr:vs:percentage}
  \end{figure*}}

\newcommand{\FigCorrFitVsPercentage}{%
  \begin{figure*}
    \centering %
    \includegraphics[width=\linewidth]{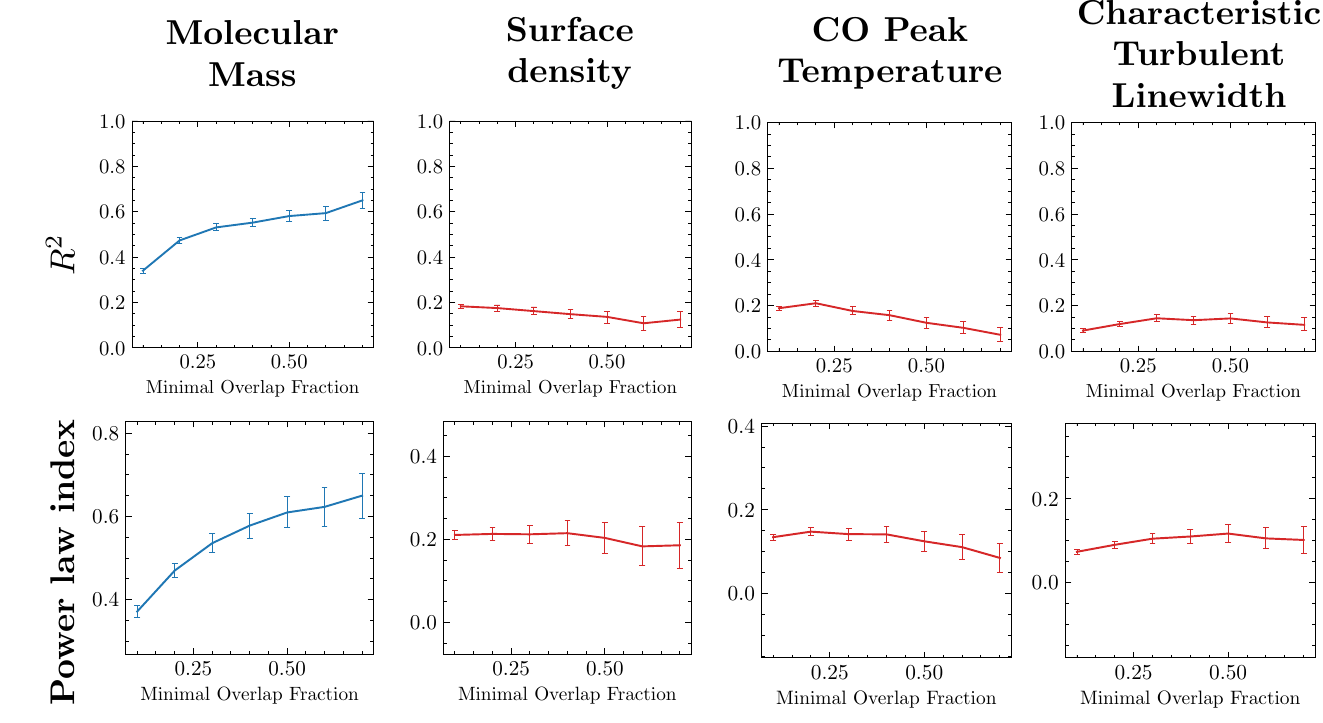}
    \caption{\Rsquared{} \textbf{(top)} and power law index
      \textbf{(bottom)} of the fits of various GMC properties (as a
      function of the \Ha{} luminosity of the matched \HII{} region)
      against the MOP. The fitted GMC properties are
      from left to right: the molecular mass, the peak temperature, the
      surface density, \modifref{and the characteristic turbulent
        linewidth}. Blue (resp. red) lines show an increase
      (resp. decrease) of the \Rsquared{} coefficient with the minimal
      overlap percentage. The uncertainty intervals are computed using a
      standard bootstrap method.}
    \label{fig:fits:vs:percentage}
  \end{figure*}}
  
\newcommand{\FigMatchedLocation}{%
  \begin{figure*}
    \centering %
    \includegraphics[width=\linewidth]{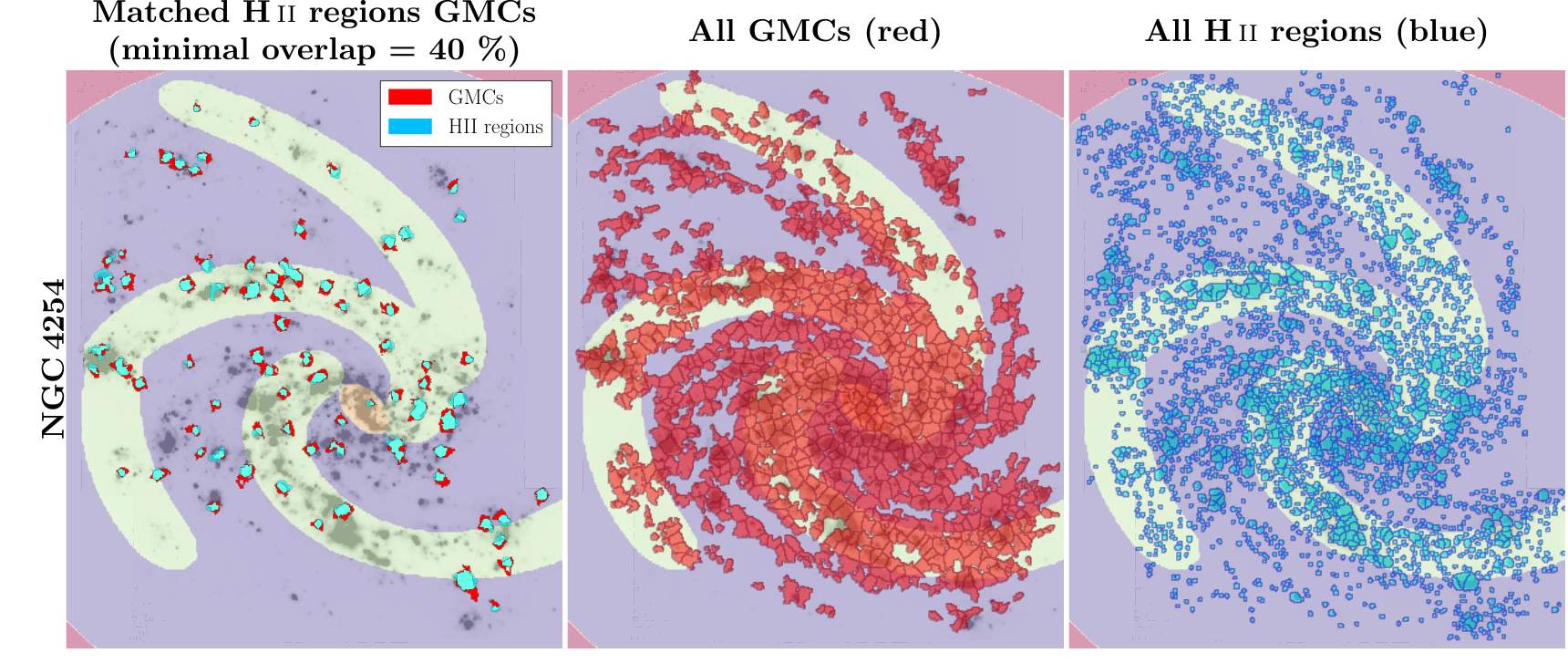}
    \caption{Comparison of the footprints of the matched GMCs/\HII{}
      regions (left panel) with the footprints of all the GMCs (center
      panel) and of all the \HII{} regions (right panel) for the NGC\,4254
      galaxy. The background image shows the \Ha{} line emission in grey
      and the environmental masks: the center in yellow, the spiral arms in
      white, and the inter-arms and disk violet. In this case, regions are
      matched when the overlap area represents at least 40\% of the GMC
      region area.}
    \label{fig:matching:example:full}
  \end{figure*}}

\newcommand{\FigBarplotEnvir}{%
  \begin{figure}
    \centering %
    \includegraphics[width=\linewidth]{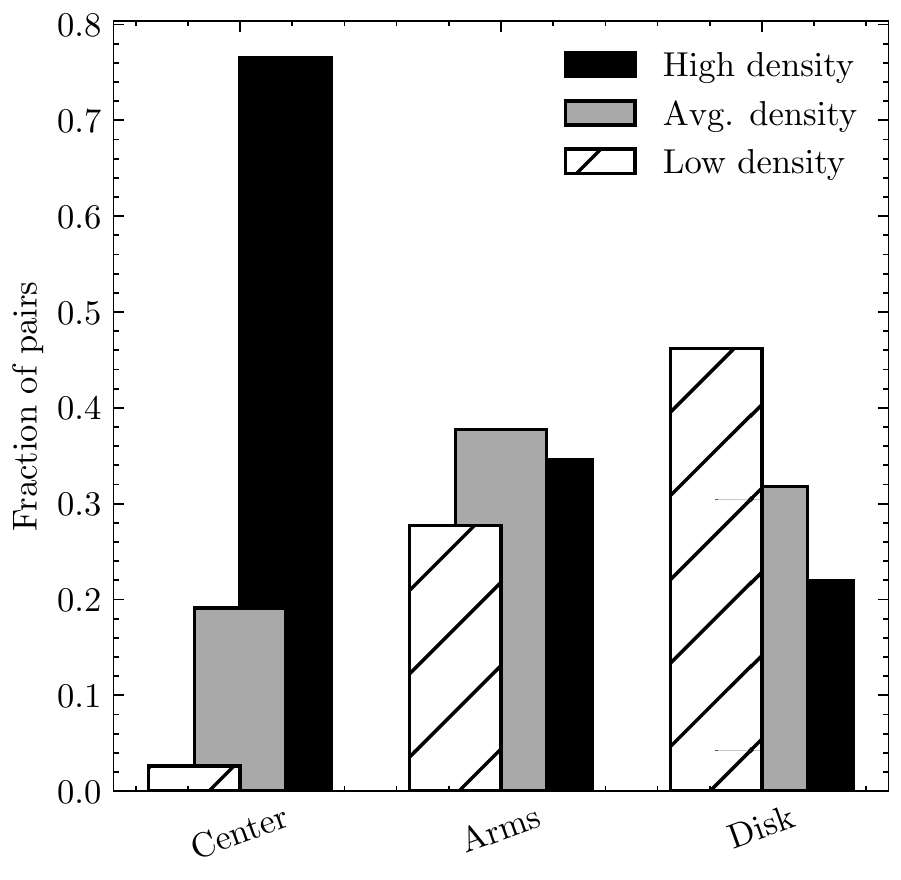}
    \caption{\modifref{Fraction of GMC/\HII{}-region pairs with low,
        medium or high density environments in the galaxies center
        (nucleus and bar), arms, and disk (interarms and outer disk), for a
        \modifreftwo{MOP} of 40\%.}}
    \label{fig:matching:frac:envir}
  \end{figure}}

\newcommand{\FigHistMatchKpc}{%
  \begin{figure}
    \centering %
    \includegraphics[width=\linewidth]{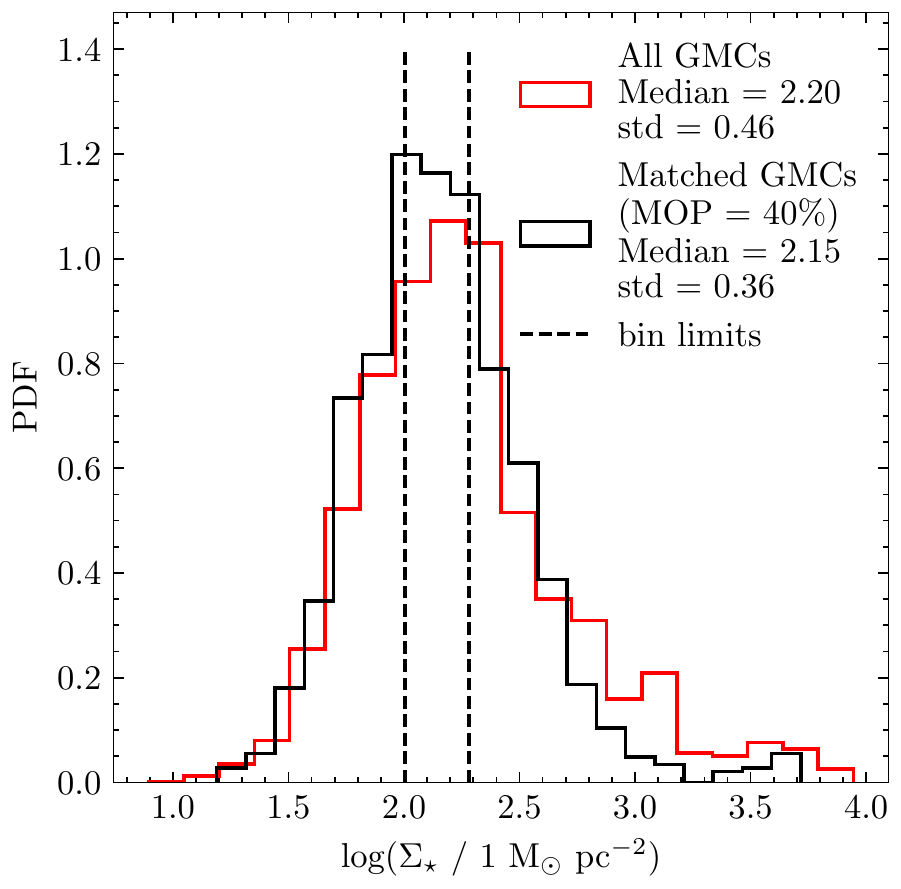}
    \caption{\modifref{Probability density functions of the stellar mass
        surface density within 500\,pc of the matched
        GMC/\HII{}-regions. Global population in red and matched
        population, with a \modifreftwo{MOP} of 40\%, in
        black. The vertical dashed lines show the limits used to define the
        three stellar density environments in
        Section~\ref{sec:dependence:on:kpc}.}}
    \label{fig:matching:frac:kpc}
  \end{figure}}

\newcommand{\FigKpcCorrs}{%
  \setlength{\PanelHeight}{5.2cm}
  \begin{figure*}
    \centering %
    \includegraphics[width=0.87\linewidth]{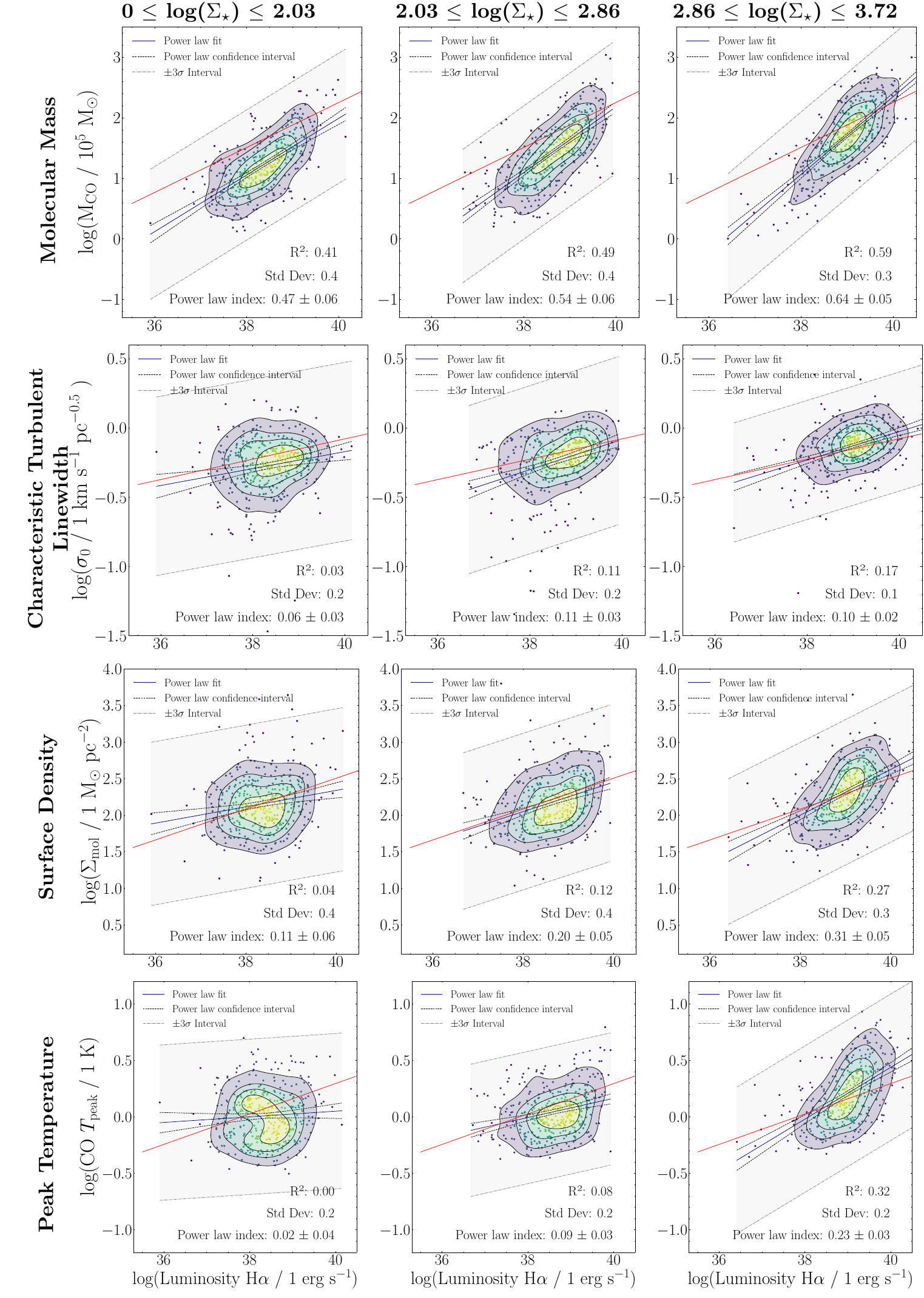}
    \caption{\modifref{Scatter plots of various GMC properties as a
        function of the \Ha{} luminosity for different environmental
        stellar surface densities measured at a scale of 500\,pc, in a
        similar fashion to Fig~\ref{fig:corr:vs:percentage}. Each column
        represents a bin of environmental stellar surface density, as
        defined in Section~\ref{sec:dependence:on:kpc}. From left to right;
        low density, average density and high density
        environments. The red line shows the global correlation from
        Section \ref{sec:global:correlations} with a \modifreftwo{MOP} of
        40\%.}}
    \label{fig:corr:kpc:all}
  \end{figure*}}

\newcommand{\FigSummaryResults}{%
  \begin{figure*}
    \centering %
    \includegraphics[width=0.9\linewidth]{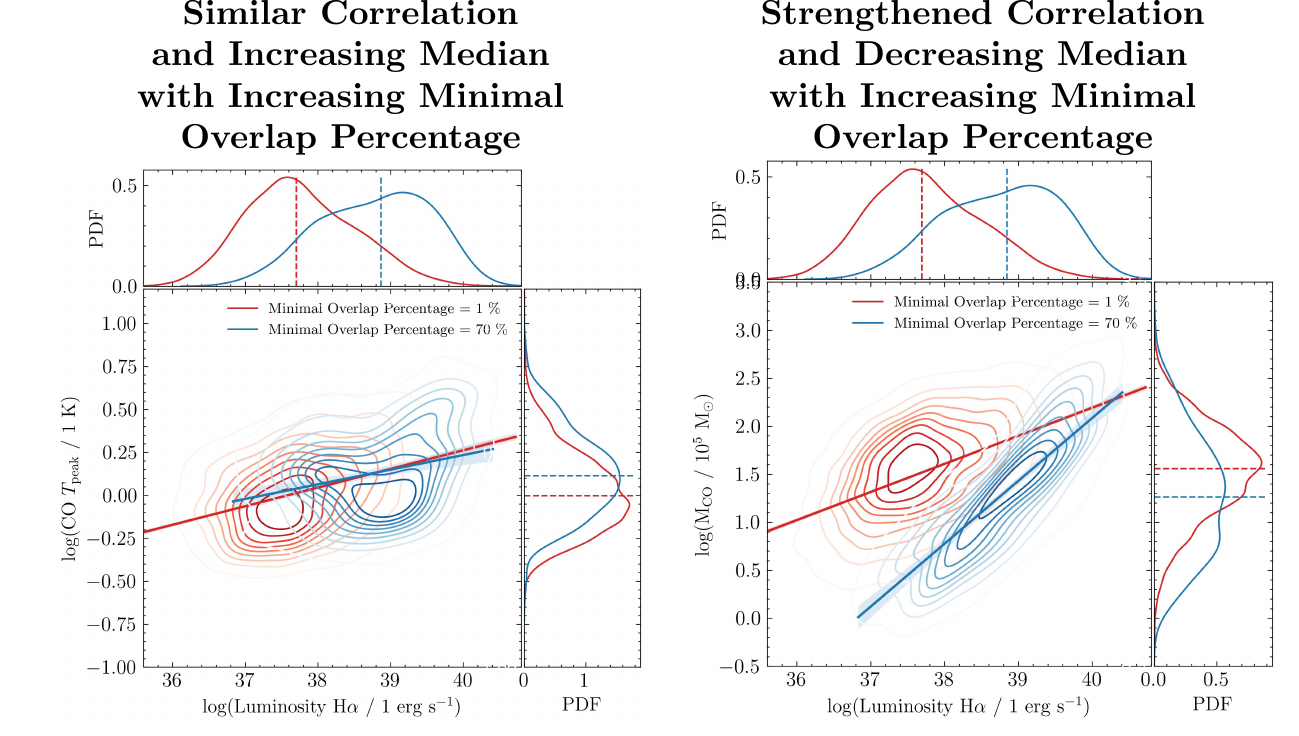}
    \caption{Joint PDFs and their marginalized PDFs of the \modifref{GMCs
        peak temperature (left) and molecular mass (right) with} the
      matched \HII{} region \Ha{} luminosity for two MOP: 1\% in red, and
      70\% in blue.  The joint PDFs are shown as contour plots. The medians
      of the marginalized PDFs and the power laws fitted on the joint PDFs
      are overlaid as straight lines. 
    }
    \label{fig:summary:results}
  \end{figure*}}

\newcommand{\FigLarsonHeyer}{%
  \begin{figure*}
    \centering %
    \includegraphics[width=0.95\linewidth]{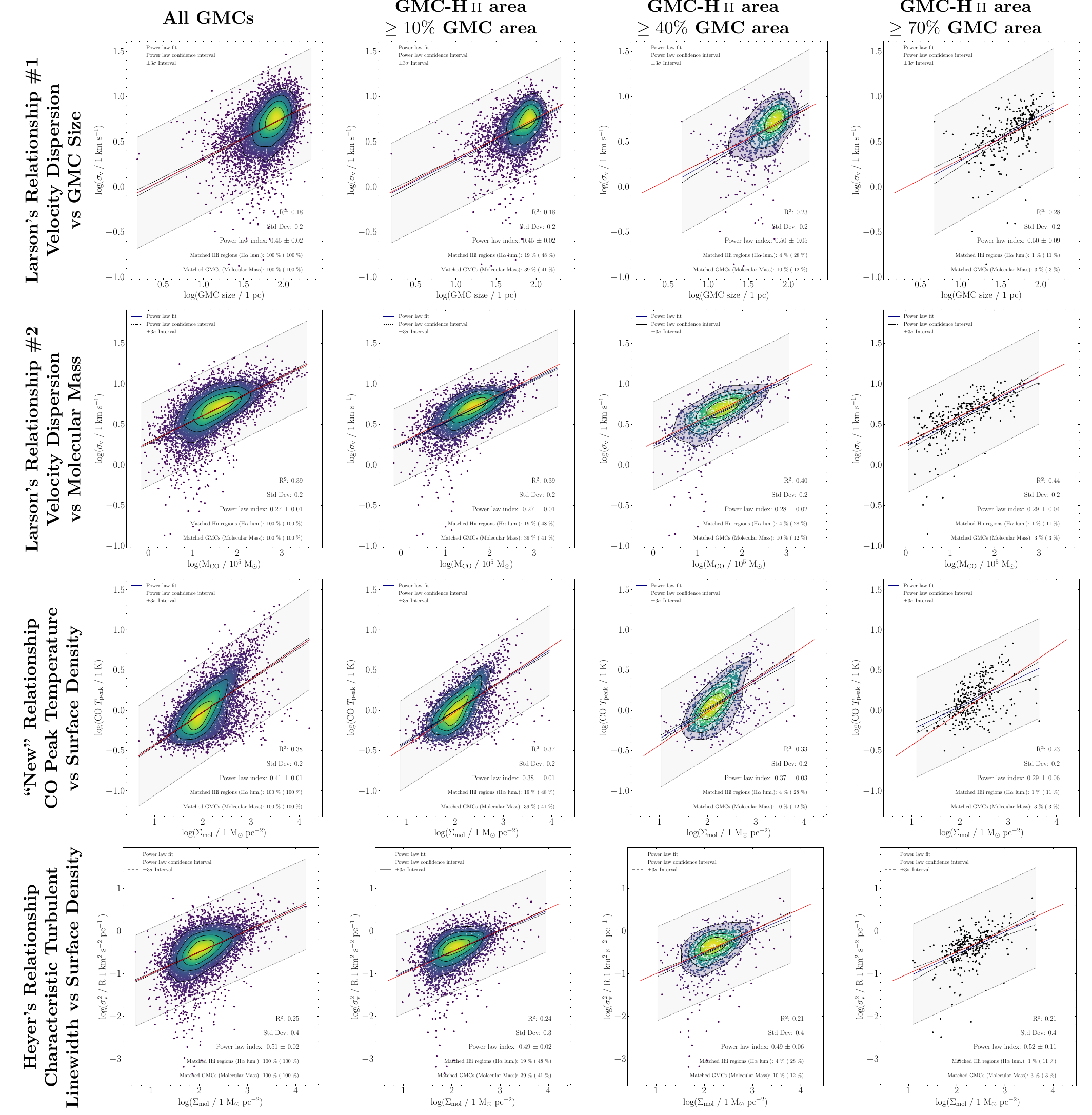}
    \caption{Evolution of Larson's, Heyer's, and peak temperature vs
      surface density correlations for all the GMCs in the sample
      \textbf{first columns,} and three different minimum overlap
      percentage of the \HII/GMC pairs: 10\% for the \textbf{second
        column,} 40\% for the \textbf{third} one, and 70\% for the
      \textbf{fourth} one. The black solid and dotted lines show the power
      law fits and their uncertainty, respectively. The dashed lines show
      the $\pm 3\sigma$ levels from the fitted line. The power law
      parameters are displayed on the bottom right corner of each panel.
      The red solid lines show the power law fit for the parent population
      (all GMCs in the sample).}
    \label{fig:corr:larson}
  \end{figure*}}

\newcommand{\FigTpeakVsMco}{%
  \begin{figure*}
    \centering %
    \includegraphics[width=0.95\linewidth]{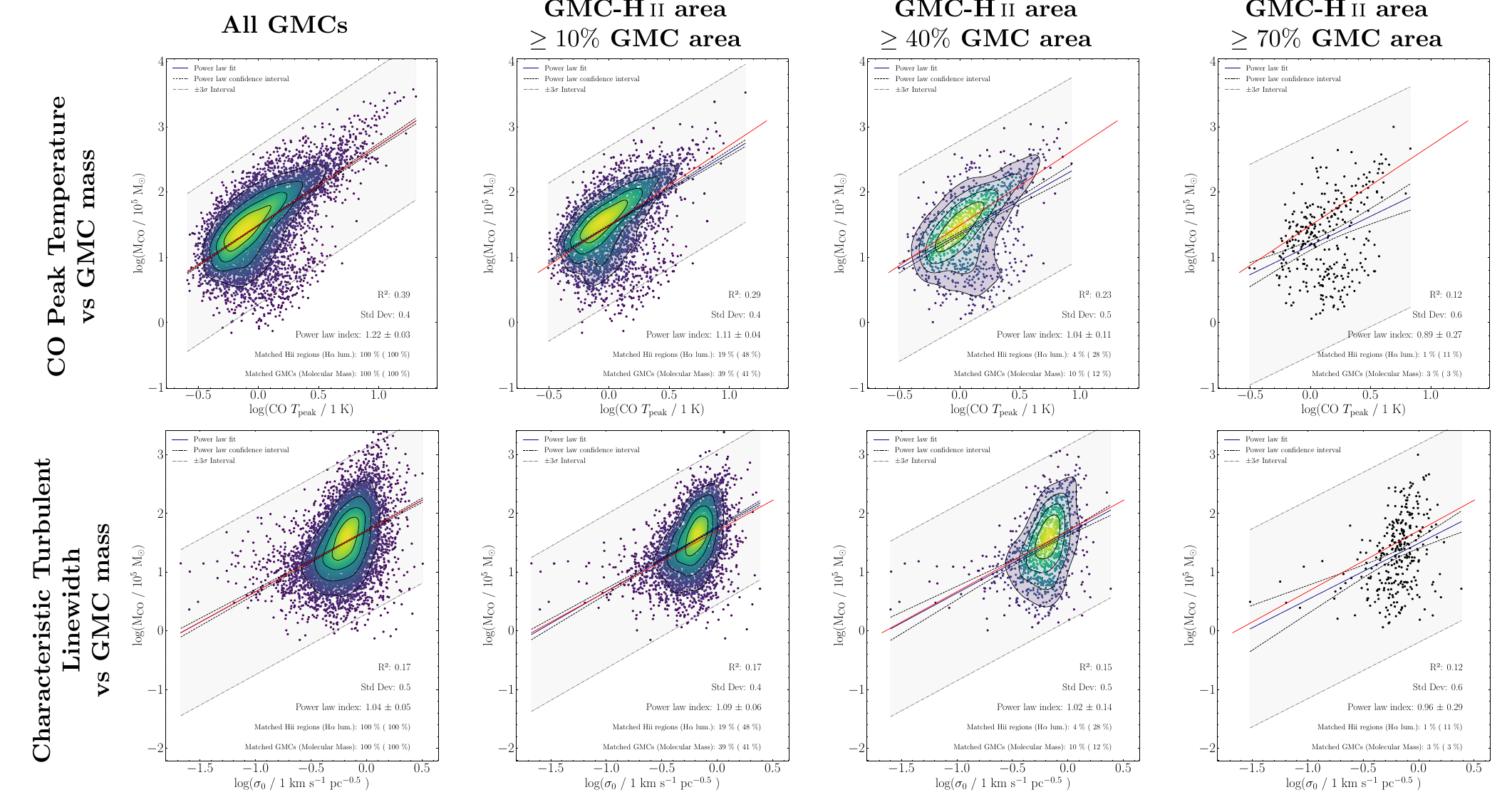}
    \caption{Evolution of joint PDF between the peak temperature
      (\textbf{top}), characteristic turbulent linewidth (\textbf{bottom})
      and the molecular mass for all the GMCs in the sample \textbf{first
        columns,} and three different MOP of the
      \HII/GMC pairs: 10\% for the \textbf{second column,} 40\% for the
      \textbf{third} one, and 70\% for the \textbf{fourth}. The reminder of
      the layout is similar to Fig.~\ref{fig:corr:larson}.}
    \label{fig:tpeak:vs:mco}
  \end{figure*}}

\newcommand{\FigCorrVsGalaxyRsquared}{%
  \begin{figure*}
    \centering %
    \includegraphics[width=0.9\linewidth,page=1]{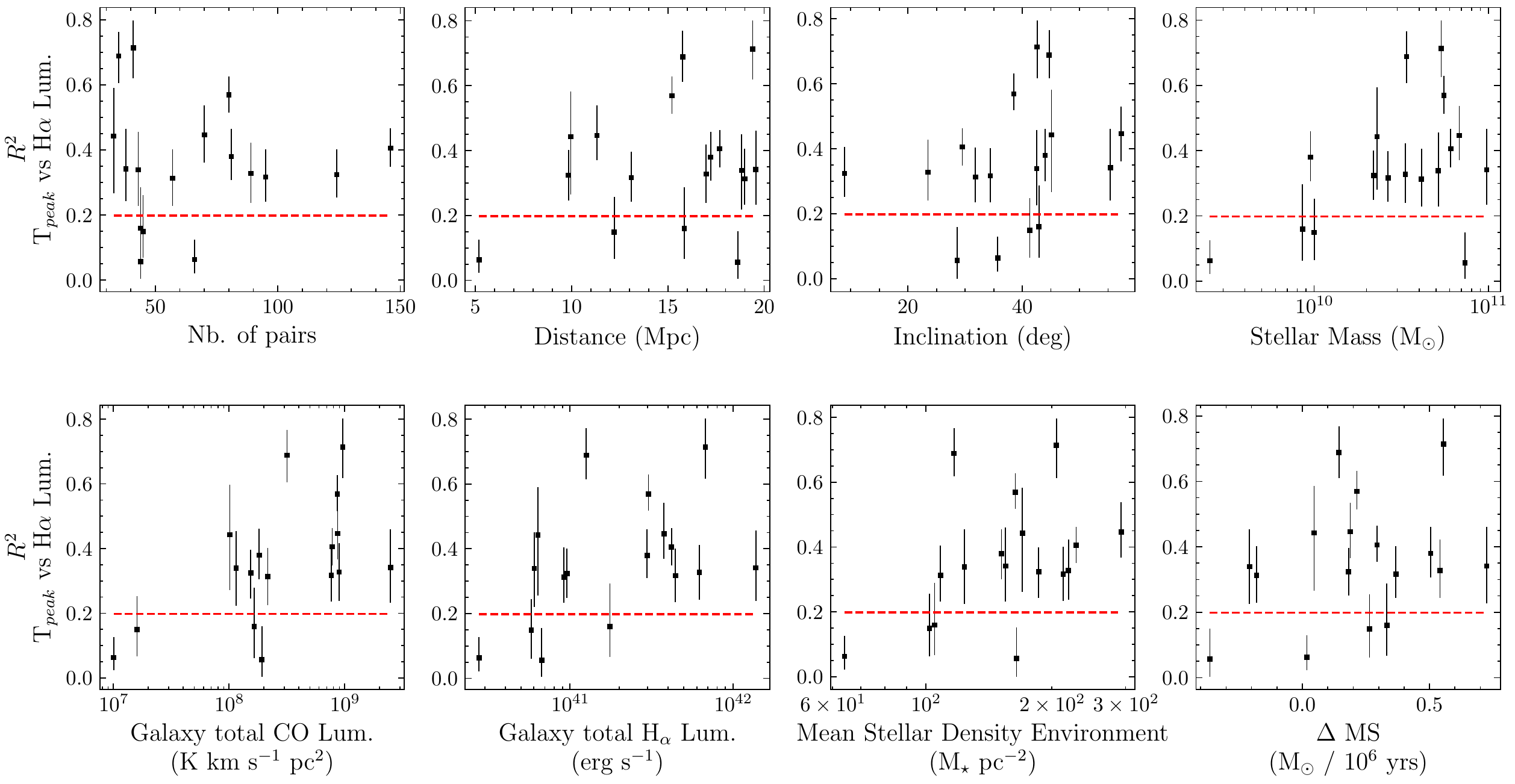}
    \caption{\Rsquared{} coefficients of the temperature at peak with the
      \Ha{} luminosity considering \HII{} region/GMC pairs from each galaxy
      in the sample separately. \textbf{Top from left to right:}
      \Rsquared{} coefficients versus the number of matched \HII{} regions
      and GMCs, distance, inclination and total stellar mass of the host
      galaxy. \textbf{Bottom from left to right:} \Rsquared{} coefficients
      versus the galaxy's CO luminosity, total \Ha{} luminosity, average
      stellar \modifreftwo{surface} density within a 500 pc radius around each of the \HII{}
      region/GMC pairs and the galaxy's offset from the main sequence. The
      red dashed line shows the global \Rsquared{} coefficient from Section
      \ref{sec:global:correlations}.}
    \label{fig:corr:vs:galaxy:rsquared}
  \end{figure*}}

\newcommand{\FigCorrVsGalaxySlope}{%
  \begin{figure*}
    \centering %
    \includegraphics[width=0.9\linewidth,page=1]{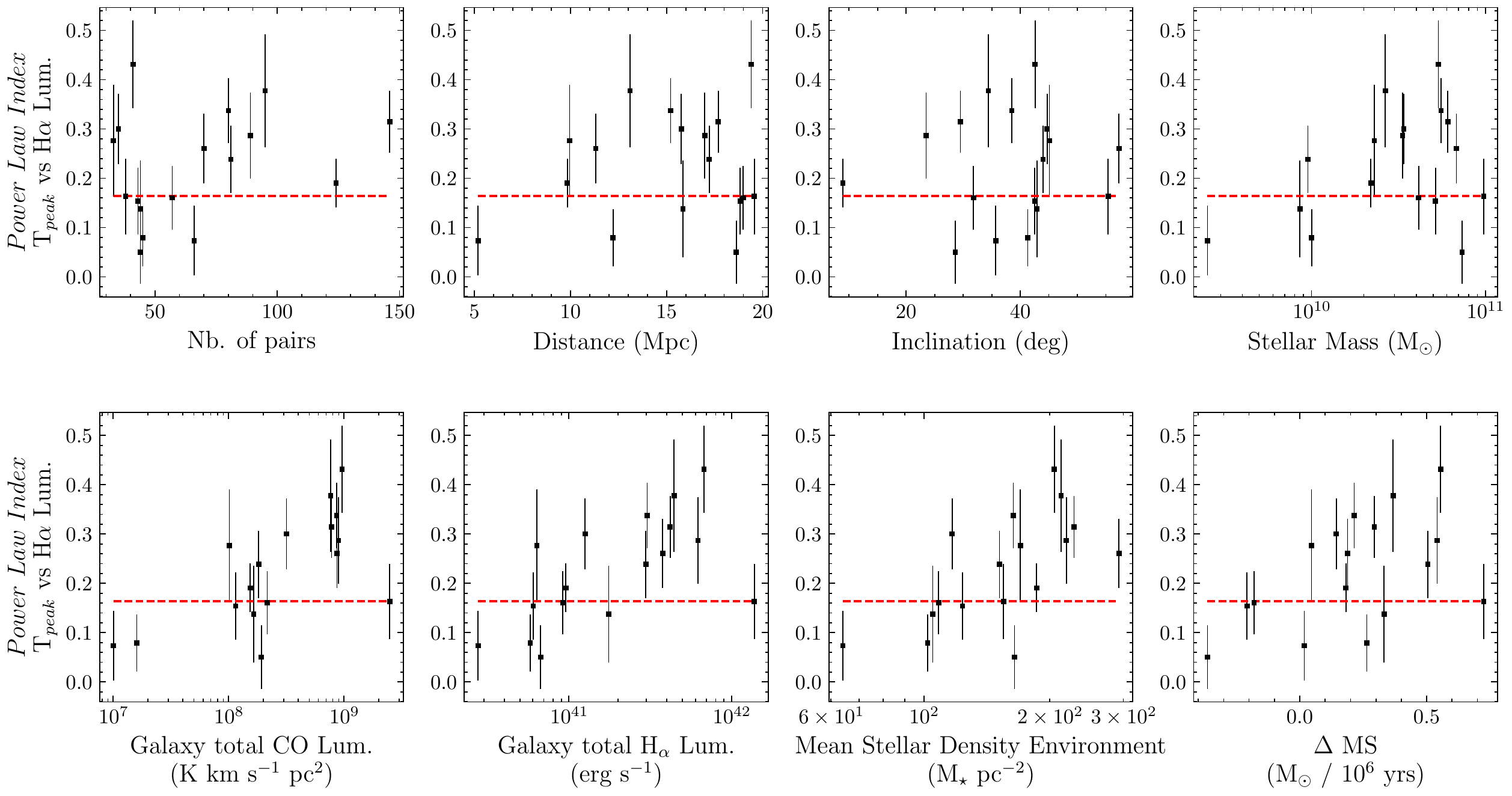}
    \caption{Same as figure \ref{fig:corr:vs:galaxy:rsquared} but the
      \Rsquared{} values are replaced by the power law indices.}
    \label{fig:corr:vs:galaxy:slope}
  \end{figure*}}

\newcommand{\FigGalaxyCorrelations}{%
  \begin{figure*}
    \centering %
    \begin{tabular}{ccc}
    \includegraphics[width=0.3\linewidth,page=1]{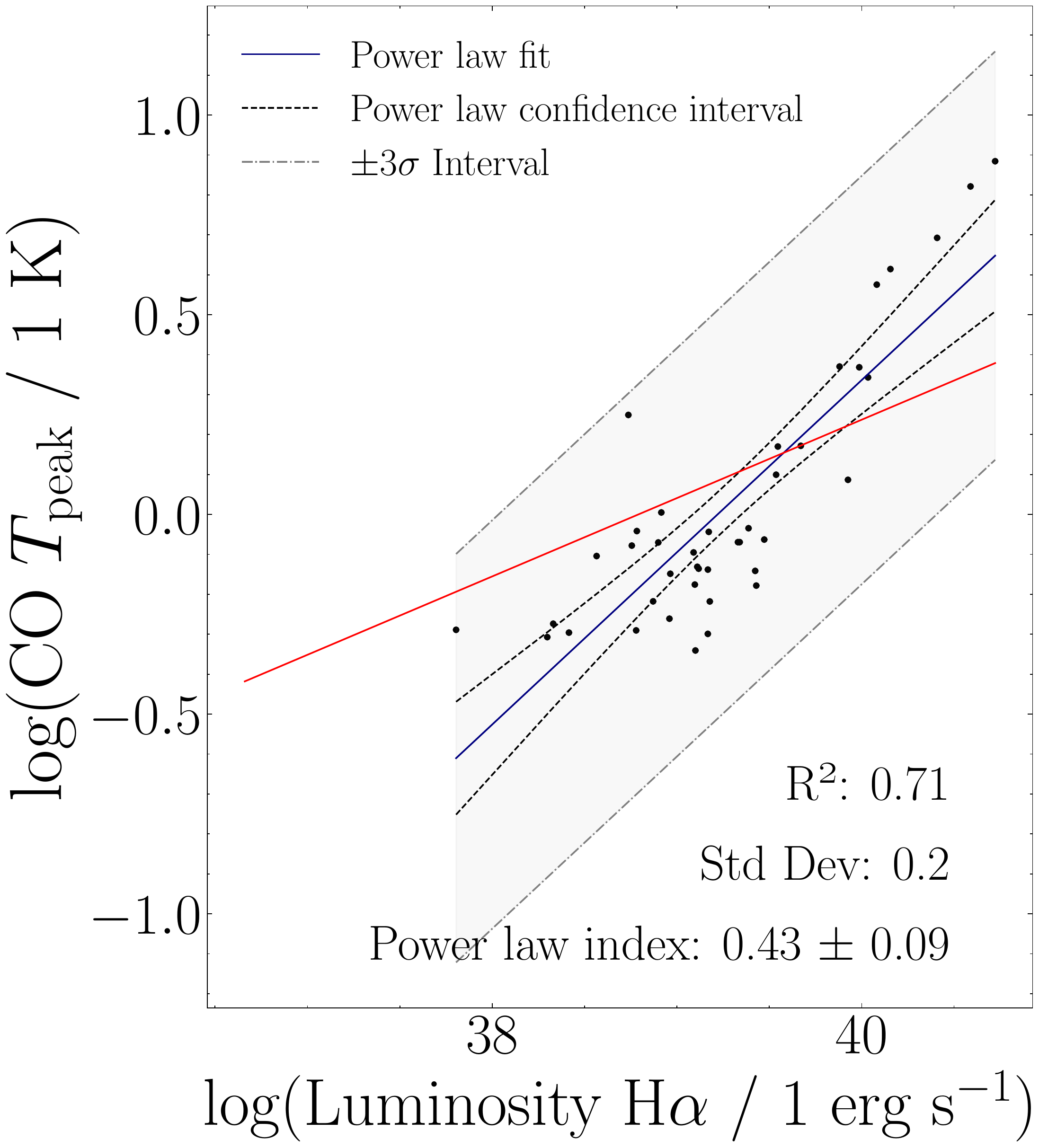} &
    \includegraphics[width=0.312\linewidth,page=1]{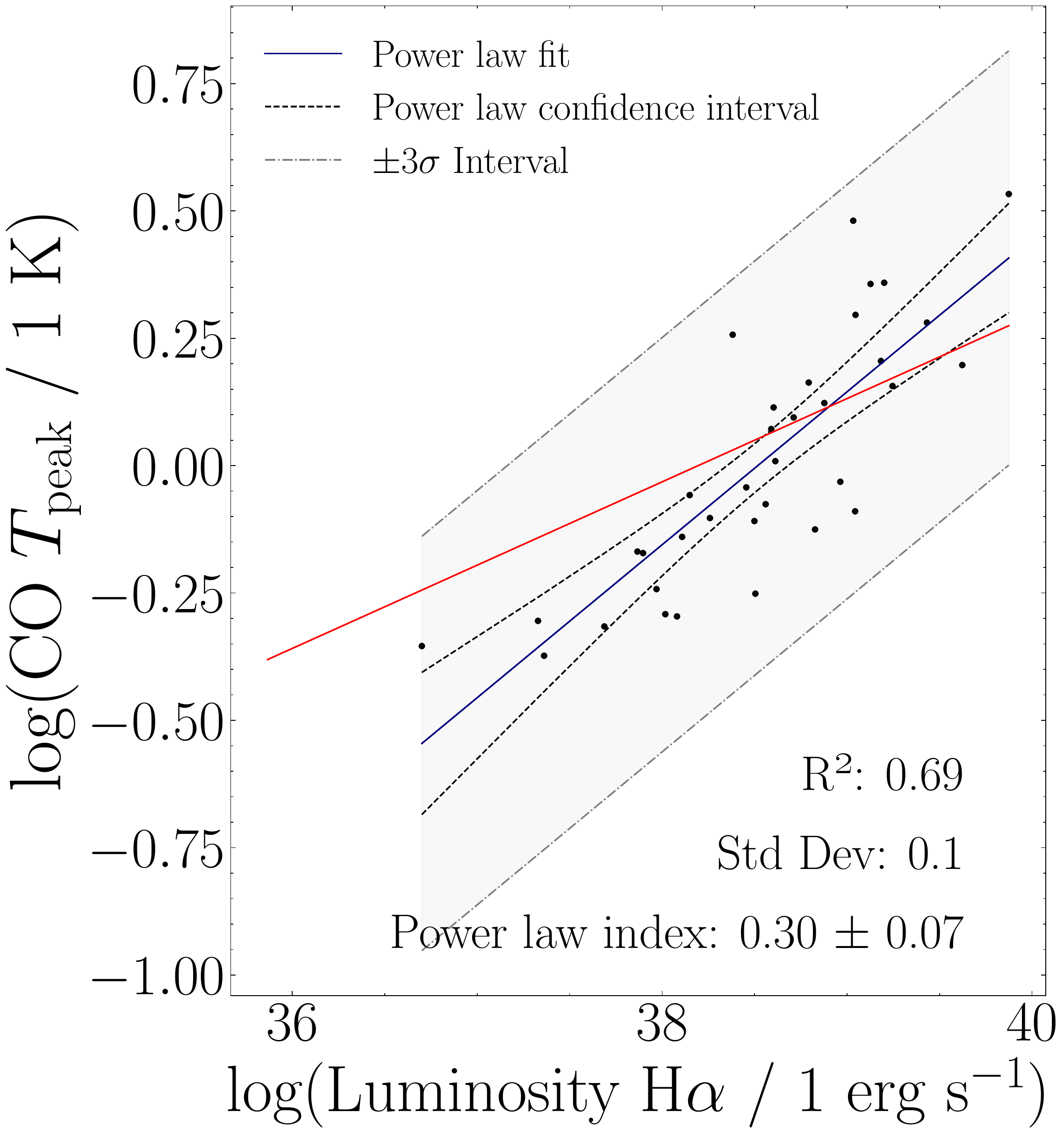} &
    \includegraphics[width=0.3\linewidth,page=1]{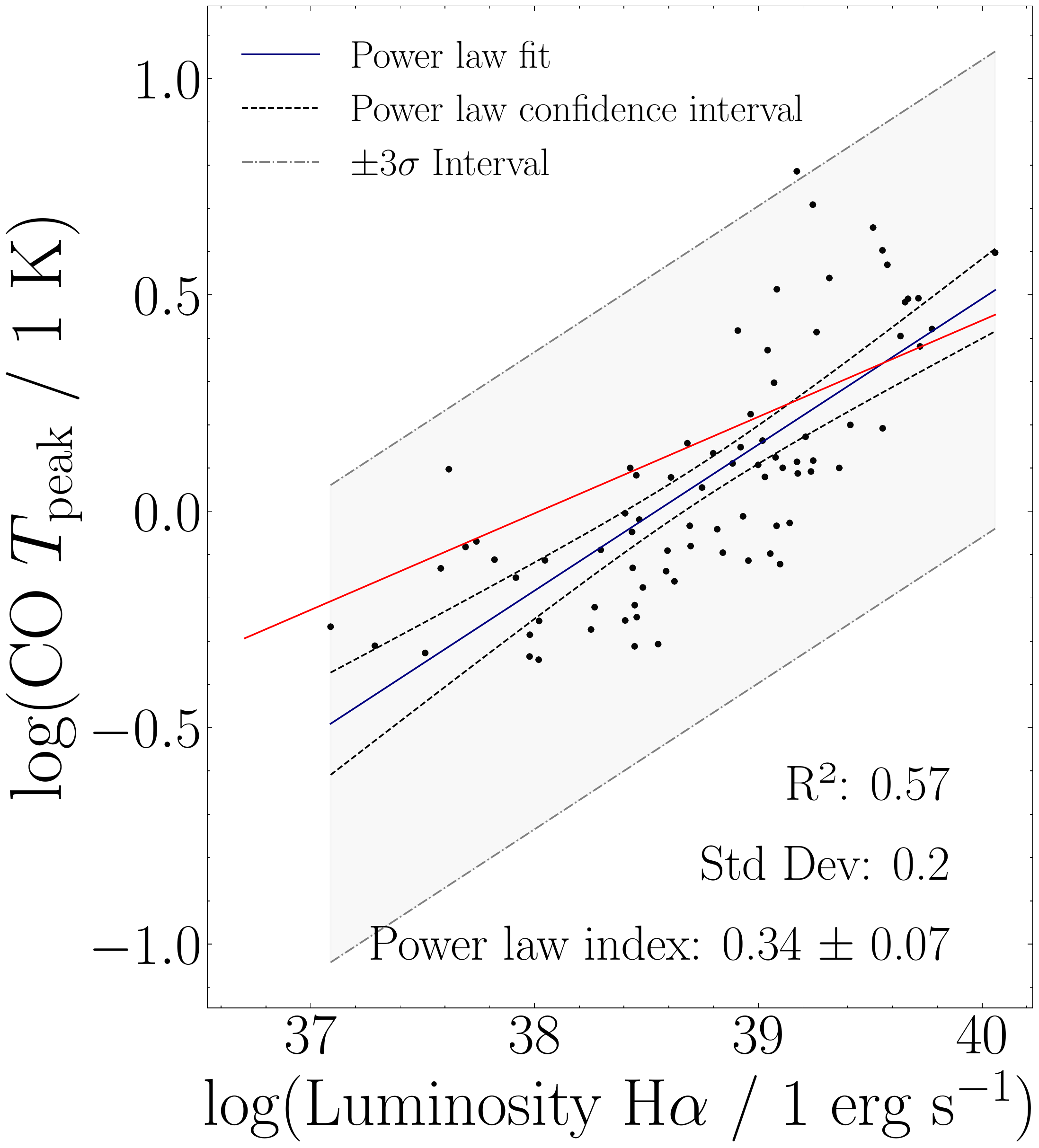} 
    \end{tabular}
    \caption{\modifreftwo{Scatter plots of the CO peak temperature as a
        function of the \Ha{} luminosity for the 3 galaxies in our sample
        (NGC\,1672, NGC\,4535, NGC\,4321) where these two properties are
        the most correlated. In this figure, the MOP is 40\%.}}
    \label{fig:corr:galaxies}
  \end{figure*}}

\newcommand{\FigHaShuffleCorr}{%
  \begin{figure*}
    \centering %
    \includegraphics[width=0.95\linewidth]{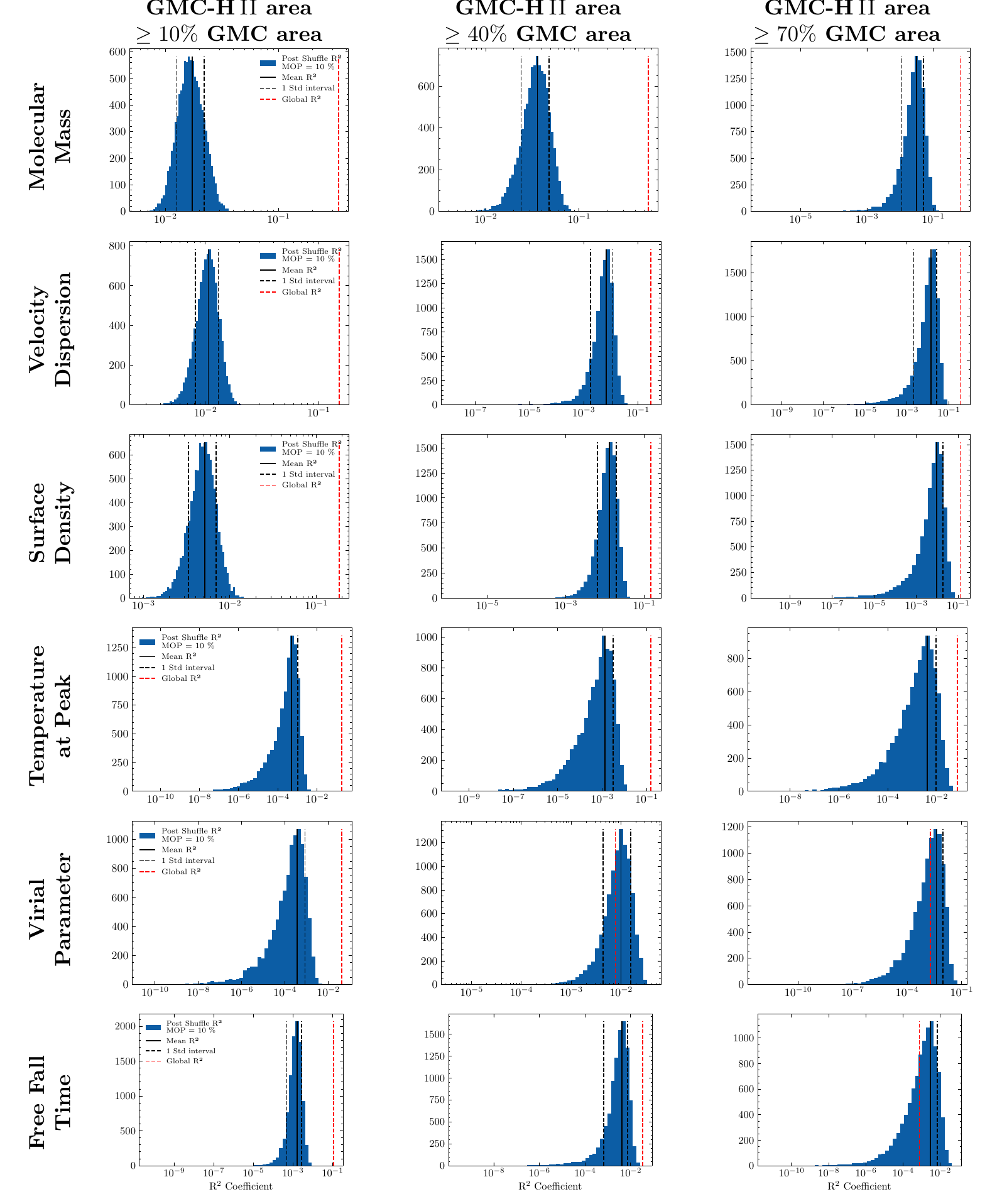}
    \caption{Histograms of the \Rsquared{} coefficient between the
      different GMC properties and the \Ha{} luminosity, after shuffling
      the \Ha{} luminosity values, for three different minimum overlap
      percentage of the \HII/GMC pairs: 10\% for the left column, 40\% for
      the middle one, and 70\% for the right one. The solid lines show the
      mean of the distribution and the dotted lines the 1 standard
      deviation interval around the mean.}
    \label{fig:corr-shuffle-halum}
  \end{figure*}}
  
\newcommand{\FigHaSizeShuffleCorr}{%
  \begin{figure*}
    \centering %
    \includegraphics[width=0.95\linewidth]{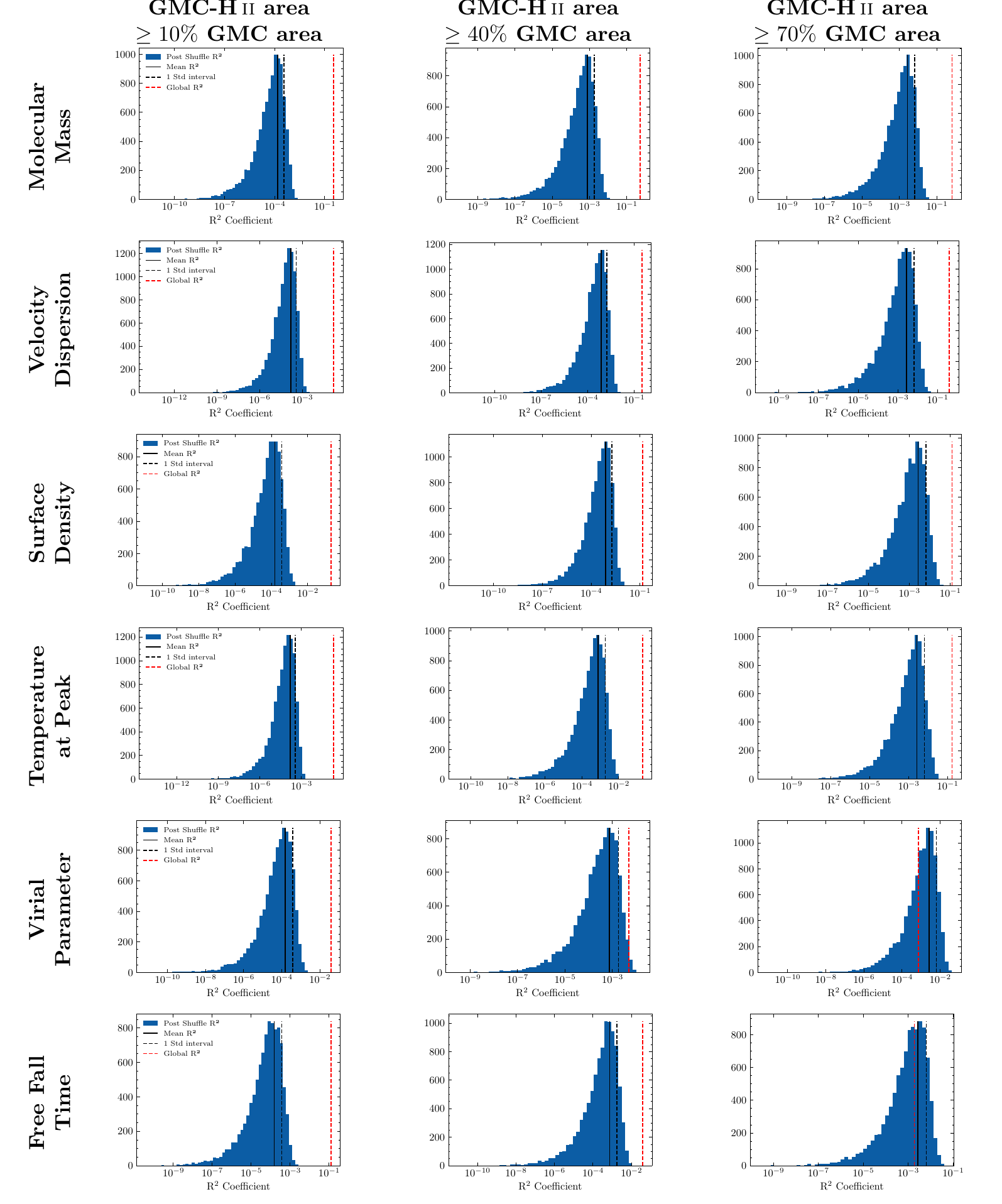}
    \caption{Histograms of the \Rsquared{} coefficient between the
      different GMC properties and the \Ha{} luminosity, after shuffling
      the \Ha{} luminosity values \modifref{while keeping the relationship
        between \Ha{} luminosity and \HII{} region radius values}, for
      three different MOP of the \HII/GMC pairs:
      10\% for the left column, 40\% for the middle one, and 70\% for the
      right one. The solid lines show the mean of the distribution and the
      dotted lines the 1 standard deviation interval around the mean.}
    \label{fig:corr-shuffle-halum-size}
  \end{figure*}}
  
\newcommand{\FigBootstrapTest}{%
  \setlength{\PanelHeight}{5cm}
  \begin{figure*}
    \centering %
    \includegraphics[width=\linewidth]{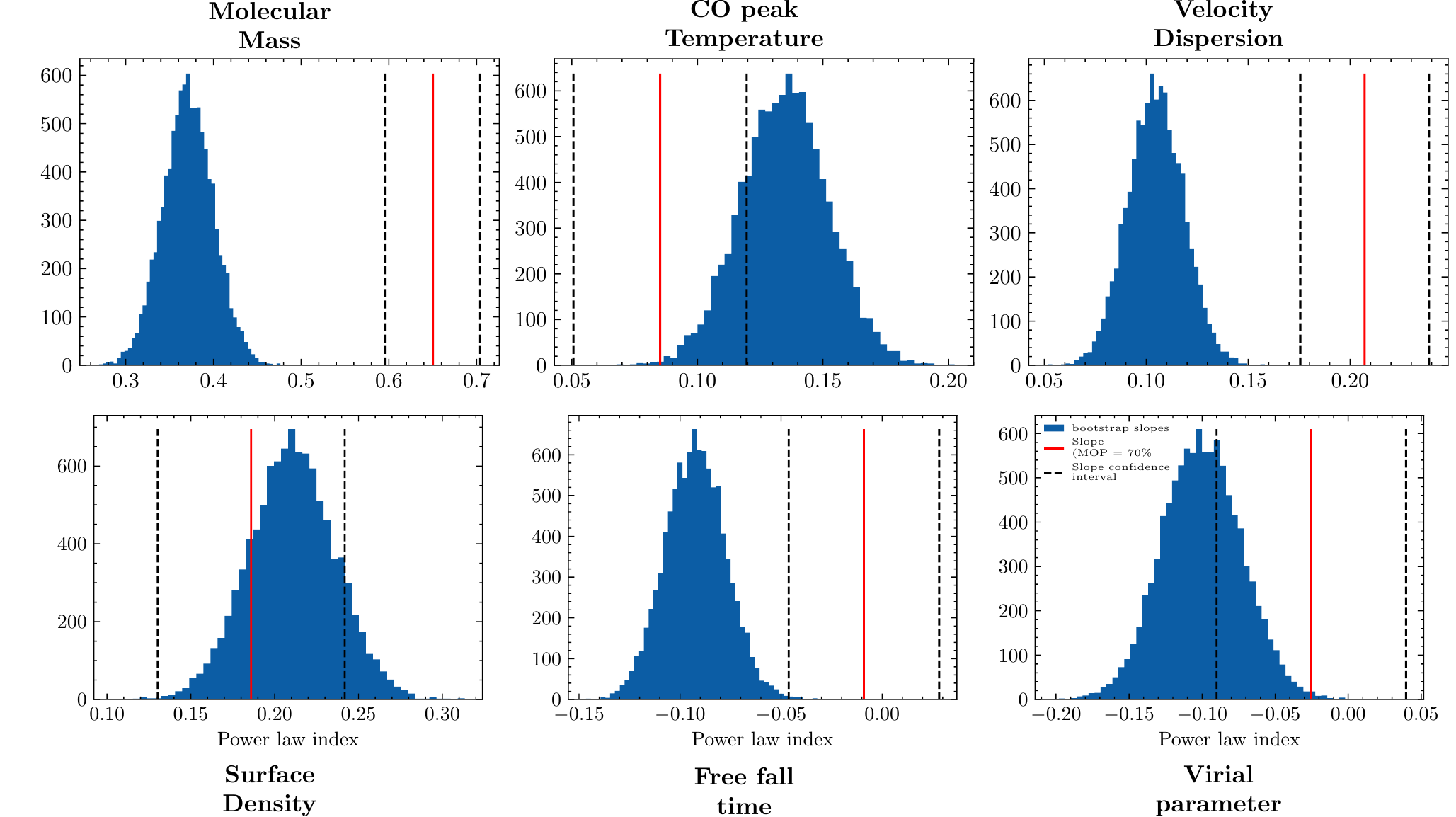}
    \caption{Bootstrap tests of the significance of the power law index
      change with the \modifreftwo{MOP}. The red line represents
      the power law index obtained after matching with a 70\% overlap, and
      the dashed lines its confidence interval. The blue distribution
      represents the power law indices obtained after randomly sampling N
      pairs from the parent distribution (10\% overlap), where N is the
      size of the 70\% minimal overlap distribution. This random sampling
      has been performed 10000 times in order to obtain the
      distribution. This figure shows that the change in power law index
      with the \modifreftwo{MOP} cannot arise from a random
      sampling of the parent population, except for the surface density.}
    \label{fig:bootstraptest}
  \end{figure*}}
  
\pdfmapline{=<nontfm> Times-Roman <8r.enc <Times-Roman.pfb}

\begin{document}

\title{The impact of \HII{} regions on Giant Molecular Cloud properties in
  nearby galaxies sampled by PHANGS ALMA and MUSE} %
\titlerunning{The impact of \HII{} regions on Giant Molecular Cloud
  properties}

\author{%
  Antoine~Zakardjian\inst{\ref{IRAP}} %
  \and Jérôme~Pety\inst{\ref{IRAM},\ref{LERMA}} %
  \and Cinthya~N.~Herrera\inst{\ref{IRAM},\ref{ILL}} %
  \and Annie~Hughes\inst{\ref{IRAP}} %
  \and Elias~Oakes\inst{\ref{UConn},\ref{Chicago}} %
  \and Kathryn~Kreckel\inst{\ref{Heidelberg}} %
  \and Chris~Faesi\inst{\ref{UConn}} %
  \and Simon~C.~O.~Glover\inst{\ref{ITA}} %
  \and Brent~Groves\inst{\ref{ICRAR}} %
  \and Ralf~S.~Klessen\inst{\ref{ITA},\ref{IWR}}%
  \and Sharon Meidt\inst{\ref{UGent}} %
  \and Ashley~Barnes\inst{\ref{aifa}} %
  \and Francesco~Belfiore\inst{\ref{inaf}} %
  \and Ivana~Bešlić\inst{\ref{aifa}} %
  \and Frank~Bigiel\inst{\ref{aifa}} %
  \and Guillermo A. Blanc\inst{\ref{Carne},\ref{DAS}} %
  \and M\'elanie~Chevance\inst{\ref{Heidelberg},\ref{ITA}}%
  \and Daniel~A.~Dale\inst{\ref{uwyo}} %
  \and Jakob~den~Brok\inst{\ref{aifa}} %
  \and Cosima~Eibensteiner\inst{\ref{aifa}} %
  \and Eric~Emsellem\inst{\ref{ESO},\ref{ULyon}} %
  \and Axel Garc\'ia-Rodr\'iguez\inst{\ref{OAN}}
  \and Kathryn Grasha\inst{\ref{ANU},\ref{A3D}} 
  \and Eric~W. Koch\inst{\ref{CFA}} 
  \and Adam~K. Leroy\inst{\ref{OSU}} %
  \and Daizhong~Liu\inst{\ref{MPE}} %
  \and Rebecca~Mc~Elroy\inst{\ref{UOQ}} %
  \and Lukas~Neumann\inst{\ref{aifa}} %
  \and Hsi-An~Pan\inst{\ref{taipei}} %
  \and Miguel~Querejeta\inst{\ref{OAN}} %
  \and Alessandro~Razza\inst{\ref{DAS}} %
  \and Erik~Rosolowsky\inst{\ref{Alberta}} %
  \and Toshiki~Saito\inst{\ref{MPIA}} %
  \and Francesco~Santoro\inst{\ref{MPIA}} %
  \and Eva~Schinnerer\inst{\ref{MPIA}} %
  \and Jiyai~Sun\inst{\ref{McMaster},\ref{CITA}} %
  \and Antonio~Usero\inst{\ref{OAN}} %
  \and Elizabeth~J. Watkins\inst{\ref{Heidelberg}} %
  \and Thomas~Williams\inst{\ref{Oxford}, \ref{MPIA}} %
} %
  
\authorrunning{Zakardjian et al.}

\institute{%
  Institut de Recherche en Astrophysique et Planétologie (IRAP), Université
  Paul Sabatier, Toulouse cedex 4, France
  \label{IRAP} %
  \and IRAM, 300 rue de la Piscine, 38406 Saint Martin d'H\`eres,
  France.\\
  \email{pety@iram.fr} \label{IRAM} %
  \and LERMA, Observatoire de Paris, PSL Research University, CNRS,
  Sorbonne Universit\'es, 75014 Paris, France. \label{LERMA} %
  \and Institut Laue-Langevin, 71 avenue des Martyrs - CS 20156, 38042
  cedex 9 Grenoble, France \label{ILL} %
  \and Department of Physics, University of Connecticut, Storrs, CT, 06269,
  USA \label{UConn}%
  \and Department of Astronomy \& Astrophysics, University of Chicago, 5640
  South Ellis Avenue, Chicago, IL 60637, USA \label{Chicago} \and
  Astronomisches Rechen-Institut, Zentrum f\"{u}r Astronomie der
  Universit\"{a}t Heidelberg, M\"{o}nchhofstra\ss e 12-14, D-69120
  Heidelberg, Germany \label{Heidelberg}%
 \and Universit\"{a}t Heidelberg, Zentrum f\"{u}r Astronomie, Institut f\"{u}r Theoretische Astrophysik, Albert-Ueberle-Str 2, D-69120 Heidelberg, Germany \label{ITA} %
  \and International Centre for Radio Astronomy Research, University of
  Western Australia, 7 Fairway, Crawley, 6009, WA, Australia \label{ICRAR}%
  \and Universit\"{a}t Heidelberg, Interdisziplin\"{a}res Zentrum f\"{u}r
  Wissenschaftliches Rechnen, Im Neuenheimer Feld 205, 69120 Heidelberg,
  Germany \label{IWR}%
  \and Sterrenkundig Observatorium, Universiteit Gent, Krijgslaan 281 S9,
  B-9000 Gent, Belgium \label{UGent}%
  \and Argelander-Institut f\"{u}r Astronomie, Universit\"{a}t Bonn, Auf
  dem H\"{u}gel 71, 53121, Bonn, Germany \label{aifa}%
  \and The Observatories of the Carnegie Institution for Science, 813 Santa
  Barbara Street, Pasadena, CA 91101, USA \label{Carne}%
  \and INAF – Osservatorio Astrofisico di Arcetri, Largo E. Fermi 5, I-50157 Firenze, Italy
	\label{inaf} %
  \and Departamento de Astronomía, Universidad de Chile, Casilla 36-D,
  Santiago, Chile \label{DAS}%
  \and Department of Physics \& Astronomy, University of Wyoming, Laramie,
  WY 82071 USA \label{uwyo}%
  \and European Southern Observatory, Karl-Schwarzschild Stra{\ss}e 2,
  D-85748 Garching bei M\"{u}nchen, Germany \label{ESO}%
  \and Univ Lyon, Univ Lyon 1, ENS de Lyon, CNRS, Centre de Recherche
  Astrophysique de Lyon UMR5574,\\ F-69230 Saint-Genis-Laval,
  France \label{ULyon}%
  \and Research School of Astronomy and Astrophysics, Australian National
  University, Canberra, ACT 2611, Australia \label{ANU}%
  \and ARC Centre of Excellence for All Sky Astrophysics in 3 Dimensions
  (ASTRO 3D), Australia \label{A3D}%
  \and Center for Astrophysics $\mid$ Harvard \& Smithsonian, 60 Garden
  Street, Cambridge, MA 02138, USA \label{CFA}%
  \and Department of Astronomy, The Ohio State University, 140 West 18th
  Avenue, Columbus, Ohio 43210, USA \label{OSU}%
  \and Max-Planck-Institut f\"{u}r extraterrestrische Physik, Giessenbachstra{\ss}e 1, D-85748 Garching, Germany
    \label{MPE} %
  \and School of Mathematics and Physics, University of Queensland, St Lucia 4067, Australia \label{UOQ}%
  \and Department of Physics, Tamkang University, No.151,
  Yingzhuan Road, Tamsui District, New Taipei City 251301,
  Taiwan \label{taipei}%
  \and Observatorio Astron\'{o}mico Nacional (IGN), C/Alfonso XII, 3,
  E-28014 Madrid, Spain \label{OAN} %
  \and Department of Physics, University of Alberta, Edmonton, AB T6G 2E1,
  Canada \label{Alberta} %
  \and Max-Planck-Institut f\"{u}r Astronomie, K\"{o}nigstuhl 17, D-69117,
  Heidelberg, Germany \label{MPIA} %
  \and Department of Physics and Astronomy, McMaster University, 1280 Main
  Street West, Hamilton, ON L8S 4M1, Canada \label{McMaster} %
  \and Canadian Institute for Theoretical Astrophysics (CITA), University
  of Toronto, 60 St George Street, Toronto, ON M5S 3H8, Canada \label{CITA} %
  \and Sub-department of Astrophysics, Department of Physics, University of Oxford, Keble Road, Oxford OX1 3RH, UK \label{Oxford}
} %

\date{Received ; accepted }

\abstract %
{The final stages of molecular cloud evolution involve cloud disruption due
  to feedback by massive stars, with recent literature suggesting the
  importance of early (i.e. pre-supernova) feedback mechanisms.} %
{We aim to determine whether feedback from massive stars in \HII{} regions
  has a measurable impact on the physical properties of molecular clouds at
  a characteristic scale of $\sim$100\,pc, and whether the imprint of
  feedback on the molecular gas depends on the local galactic
  environment.} %
{We identify giant molecular clouds (GMCs) associated with \HII{} regions
  for a sample of \nmuse{} nearby galaxies using catalogs of GMCs and
  \HII{} regions released by the PHANGS-ALMA and PHANGS-MUSE surveys, using
  the overlap of the CO and \Ha{} emission as the key criterion for
  physical association. We compare the distributions of GMC and \HII{}
  region properties for paired and non-paired objects. We investigate
  correlations between GMC and \HII{} region properties among galaxies and
  across different galactic environments to determine whether GMCs that are
  associated with \HII{} regions have significantly distinct physical
  properties to the parent GMC population.} %
{We identify trends between the \Ha{} luminosity of an \HII{} region and
  the CO peak brightness and the molecular mass of GMCs that we tentatively
  attribute to a direct physical connection between the matched objects,
  and which arise independently of underlying environmental variations of
  GMC and \HII{} region properties within galaxies. The study of the full
  sample nevertheless hides a large variability galaxy by galaxy.} %
{At the $\sim$100\,pc scales accessed by the PHANGS-ALMA and PHANGS-MUSE
  data, pre-supernova feedback mechanisms in \HII{} regions have a subtle
  but measurable impact on the properties of the surrounding molecular gas,
  as inferred from CO observations.}%

\keywords{} %
\maketitle{} %

\section{Introduction}

Stellar feedback is a key process in galaxy evolution. Simulated galaxies
without feedback cannot reproduce the observed properties of galaxies: they
cool too rapidly, exhausting their gas supply through the rapid
transformation of gas into stars \citep[e.g.][]{schaye2015}. On cloud
scales, stellar feedback is often invoked to explain the low observed
efficiency of star formation in Giant Molecular Clouds (GMCs), which
convert only $\sim1\%$ of their mass into stars per cloud free-fall time
\citep{Krumholz2014,Fujimoto2016,Utomo2018,Grudic2019,Kim2021b}. On larger
scales, stellar feedback contributes to driving galactic winds and outflows
\citep{Hopkins2012,Bolatto2013}. Stellar feedback processes include
protostellar jets and outflows, stellar winds, direct radiation from stars
that can ionize the gas, re-processed radiation from interstellar dust, and
supernovae \citep{Krumholz2019,Chevance2022b}. While there is consensus
about the overall importance of stellar feedback, much work remains to be
done to quantify the timescales, efficiencies and relative importance of
the various forms of feedback across the diversity of galactic environments
that host star formation.

The impact of stellar feedback on the surrounding molecular gas has been
investigated \modifref{on $\sim$parsec scales} in individual Galactic
star-forming regions
\citep{Pabst2019,Watkins2019,Barnes2020,Grossschedl2021,Luisi2021,Olivier2021}
and in Local Group targets
\citep{Lopez2011,Lopez2014,Mcleod2018,Mcleod2019,Mcleod2020,Mcleod2021}.
\modifref{These studies underline the importance of pre-supernovae
  feedback. Further afield, studies of star formation and feedback in the
  nearby galaxy population have investigated} the properties of \HII{}
regions and GMCs in their galactic context \citep{Barnes2021,rosolowsky21},
the efficiency of star formation within the molecular gas reservoir
\citep{Utomo2018}, the relative spatial configuration and intensity of the
CO and \Ha{} emission \citep{Schinnerer2019,Chevance2020,Pan2022} and the
inferred timescales for various phases of the star formation process.

\modifref{One} outstanding question is whether feedback from star formation
significantly modifies the properties of their natal clouds, not only at
the immediate working surface, but also on the $\sim 50-100\,$pc scales on
which global cloud properties are typically measured. In this paper, we
study the impact of the \HII{} regions on the surrounding molecular gas by
exploring the properties of molecular clouds that are spatially coincident
with \HII{} regions identified in \nmuse\ nearby galaxies. This study
leverages recent data from the PHANGS-ALMA and PHANGS-MUSE
surveys\footnote{\url{http://www.phangs.org}}
\citep{Leroy2021a,Leroy2021b,Emsellem2022}, which have characterized
several $10^4$ GMCs in \nalmagmcs\ galaxies \citep{rosolowsky21} and a
similar number of \HII{} regions in a subset of \nmuse\ galaxies
\citep{Santoro2022}. Our study is complementary to the detailed,
parsec-scale studies of individual star-forming regions that can obtain
estimates of individual feedback terms \citep{Barnes2021,Barnes2022}, since
it provides an overview of how stellar feedback impacts the molecular gas
reservoir across a wide range of galactic environments and interstellar
conditions. \modifref{By analysing the properties of neighbouring clouds
  and \HII{} regions, it likewise offers a complementary view to previous
  pixel-based analyses that have robustly characterised the statistical
  relationship between the molecular gas reservoir and star formation
  activity \citep[most notably to obtain evolutionary timescales of the
  star formation process, e.g.][]{Chevance2020,Chevance2022a,Kim2022} but
  have so far been less focused on the physical properties of the
  star-forming gas.}

\TabGalaxySample{} %

Section~\ref{sec:data:catalogs} summarizes the properties of the
PHANGS-ALMA and PHANGS-MUSE data, and of the GMC and \HII{} region catalogs
derived from those data. Section~\ref{sec:matching:method} presents the
method that we use to match GMCs with \HII{} regions. Section
\ref{sec:matched-vs-global} compares the properties of matched
GMC/\HII{}-regions to the general population of GMCs and \HII{} regions in
our sample galaxies.  In Section \ref{sec:global:correlations}, we
investigate the correlations between GMC properties and \HII{} region
luminosity for the matched GMC/\HII{}-regions.  \modifref{In
  Section~\ref{sec:dependence:on:kpc}, we investigate whether the effects
  of local radiative feedback can be robustly distinguished from
  co-variations of GMC and \HII{} region properties within different}
galactic environments.  Section~\ref{sec:discussion} discusses the results
and Section~\ref{sec:conclusions} summarizes our
findings. Appendices~\ref{sec:appendix:matching:method} to
\ref{sec:appendix:LarsonHeyer} present supplemental figures \modifref{and
  tests of our methodology} that complement the main content of the
article. \modifref{For instance,
  Appendix~\ref{sec:appendix:variability:with:galaxy} examines the
  correlations between GMC properties and \HII{} region luminosity in
  individual galaxies.}

\section{Data and catalogs}
\label{sec:data:catalogs}

In this paper, we compare the GMC and \HII{} region populations of \nmuse\
galaxies in the PHANGS sample with both ALMA \COJtwo{}
\citep{Leroy2021a,Leroy2021b} and MUSE \Ha\ imaging
\citep{Emsellem2022}. Table~\ref{tab:galaxy:sample} summarizes the
properties of our \nmuse\ target galaxies, which are adopted from the
PHANGS Sample Table v1.6 \citep{Leroy2021b}. For the analysis in
Section~\ref{sec:dependence:on:kpc}, we use estimates for the local stellar
surface density presented by \citet{Jiayi2022}. In this section, we
summarize the key properties of the PHANGS-ALMA and PHANGS-MUSE data and of
the GMCs and \HII{}-region catalogs derived from those data.

\subsection{PHANGS-ALMA CO~(2-1) data}

We analyze GMCs that are identified in the PHANGS-ALMA survey of nearby
galaxies \citep{Leroy2021a,Leroy2021b}.  The GMC catalogs that we use are
derived from the combined 12m+7m+TP PHANGS-ALMA \COJtwo{} data cubes, which
have a spectral resolution of $2.5\kms$ and a typical angular resolution of
${\sim} 1''$ to $1.5''$, corresponding to linear resolutions between
$\sim 30$ and $\sim 180$\,pc at the distance of our target galaxies. These
combined cubes are sensitive to emission from all spatial scales. The
PHANGS-ALMA observations were designed to target the region of active star
formation in each galaxy, with a field of view that typically extends to
$R\sim0.3R_{25}$. The mean $1\sigma$ sensitivity of the cubes is
$\sim0.2$\,K per spectral channel. An overview of the PHANGS-ALMA survey
science goals, observing strategy, and data products is presented in
\citet{Leroy2021b}. A detailed description of the PHANGS-ALMA data
processing, including calibration, imaging and combination steps, is
presented in \citet{Leroy2021a}.
The PHANGS-ALMA \COJtwo{} data cubes are available from the PHANGS team
website \footnote{\url{http://www.phangs.org}}, the ALMA archive
\footnote{\url{https://almascience.eso.org/alma-data/lp/PHANGS/}} and the
Canadian Astronomy Data Center (CADC)\footnote{
  \url{https://www.canfar.net/storage/list/phangs/RELEASES/PHANGS-ALMA/}}.

\subsection{PHANGS-ALMA GMC catalogs}
\label{sect:gmccats}

We use the public GMC catalogs released by PHANGS-ALMA on the PHANGS team
website. These catalogs were generated using
\textsc{pycprops}\footnote{\label{footnote:pycprops}\url{https://github.com/phangsteam/pycprops/}},
which is a \textsc{python} implementation of the {\sc cprops} algorithm
originally presented in \citet{rosolowsky06}. A detailed description of the
\textsc{pycprops} methodology, including its application to a subsample of
10 PHANGS-ALMA \COJtwo{} data cubes, is presented in
\citet{rosolowsky21}. The full set of GMC catalogs for \nalmagmcs{}
galaxies in the PHANGS-ALMA survey is presented in a companion paper by
Hughes et al (in prep). We refer the reader to those papers for a complete
description of the PHANGS-ALMA GMC catalogs, but briefly summarize here the
catalog generation process and the derivation of the key quantities that we
use in our analysis.

Catalog generation by \textsc{pycprops} proceeds in two main stages. First,
significant emission in the data cube is identified. The emission is then
segmented into distinct structures (`clouds' for simplicity) by identifying
significant local maxima and uniquely assigning all the emission to each of
these maxima. The criteria used to identify and segment the significant
emission in the PHANGS-ALMA \COJtwo{} data cubes are explained in detail by
\citet{rosolowsky21}. In short, the decomposition for the PHANGS-ALMA GMC
catalogs proceeds by searching for regions larger than the telescope beam
with intensities $>2\sigma$ that are contiguous in position and velocity
with $>4\sigma$ peaks. Compactness is strongly preferred, such that local
maxima are rarely merged into larger structures.

In a second step, the emission associated with each cloud is characterized,
and physical properties of the cloud are determined. \textsc{pycprops} uses
moments of the emission to estimate cloud properties. The moment-based
quantities are corrected for the effects of sensitivity and the finite
resolution of the data before translating them into estimates of physical
quantities. In this paper, we investigate correlations involving the
following GMC properties:
\begin{itemize}
\item {\it The CO peak temperature, $T_{\mathrm{peak}}$}. This is the CO
  brightness temperature, measured at the brightest voxel within the cloud
  boundary.
\item {\it Cloud molecular mass, $M_\mathrm{CO}$.} We use the CO-based GMC
  mass estimate, which is obtained by multiplying the cloud luminosity
  $L_\mathrm{CO}$ by a CO-to-H$_2$ conversion factor,
  $\alpha_{\mathrm{CO}}$. \modifref{The cloud luminosity is the sum of the
    intensity within the cloud boundary, and is related to the mass via
    $M_\mathrm{CO} = \alpha_\mathrm{CO} L_\mathrm{CO}$.  The PHANGS-ALMA
    GMC catalogs implement the $\alpha_\mathrm{CO}$ calibration of
    \citet{sun20}, with an adopted CO($2-1$) to CO($1-0$) line ratio
    $R_{21} = 0.65$.}  The $\alpha_\mathrm{CO}$ adopted for each cloud
  depends only on the local metallicity, which is estimated according to
  the global mass-metallicity scaling relation of \citet{sanchez19} and the
  universal metallicity gradient of \citet{sanchez14}. We refer the reader
  to \citet{sun20} for a more complete discussion.  Due to the restricted
  range of metallicities of our sample galaxies, $L_{\mathrm{CO}}$ and
  $M_\mathrm{CO}$ \modifref{are effectively interchangeable for our
    analysis in this paper.}
\item {\it Cloud surface density, $\Sigma_\mathrm{mol}$.}  We estimate the
  typical molecular gas surface density within the cloud as
  $\Sigma_\mathrm{mol} = M_{\mathrm{CO}} / (2{\pi R^2})$.
\item \modifreftwo{{\it Cloud radius at FWHM, $R$}}. \modifreftwo{The cloud
    size is estimated from the intensity-weighted second moments over the
    two spatial axes of the cube (i.e. the spatial variances $\sigma_x^2$
    and $\sigma_y^2$), and an intensity-weighted covariance term
    $\sigma_{xy}$. These measurements are used to determine the major and
    minor axes of the emission distribution and the cloud's position
    angle. The major and minor axis measurements are corrected for the
    finite sensitivity and angular resolution of the data,
    $\sigma_{\mathrm{maj,corr}}$ and $\sigma_{\mathrm{min,corr}}$. The
    cloud radius is then calculated as
    $R = \eta \sqrt{\sigma_{\mathrm{maj,corr}}
      \sigma_{\mathrm{min,corr}}}$, with $\eta =\sqrt{2\ln 2} = 1.18$.}
\item {\it The characteristic turbulent linewidth at a fiducial scale of
    1\,pc, $\sigma_0$}. To estimate $\sigma_0$, we assume that the
  turbulent structure function within all clouds has an index of 0.5
  \citep[e.g.,][]{Heyer2004}. \modifref{It is related to} the observed
  cloud-scale velocity dispersion $\sigma_v$ according to
  $\sigma_{0} = \sigma_v / \sqrt{R_\mathrm{3D}/\mathrm{1~pc}}.$
  \modifref{Here, $\sigma_v$ is the square root of the intensity-weighted
    variance of the emission along the spectral axis within the cloud
    boundary, and $R_{\mathrm{3D}}$ is an estimate for the
    three-dimensional mean radius of the cloud. $R_{\mathrm{3D}}$ differs
    from the projected FWHM size $R$}. In practice,
  \begin{equation}
    R_{\mathrm{3D}} = \left\{%
      \begin{array}{rl}
        R & ; \quad R\le H/2\\
        \sqrt[3]{\frac{R^2H}{2}} & ; \quad R > H/2~,
      \end{array}\right.
    \label{eq:r3d}
  \end{equation} 
  where $H=100$\,pc is the assumed scale-height of the molecular gas in a
  galactic disk \citep{Malhotra1994}.\\
\item {\it Virial parameter of the whole cloud, $\alpha_{\rm vir}$}. This
  provides an estimate of the relative strength of the gravitational
  binding energy versus the kinetic energy of a molecular cloud. The
  PHANGS-GMC catalogs adopt
  $\alpha_{\rm vir} = 10\times \sigma_v^2 R_{3D}/(G M_{\rm CO})$.\\
\item {\it Free-fall collapse time, $\tau_\emr{ff}$}, which is derived from
  a combination of the above quantities, as
  \begin{equation}
    \tau_\emr{ff} = \sqrt{\frac{3\pi}{32 G \rho}} = \sqrt{\frac{\pi^2 R_\mathrm{3D}^3}{4GM_{\rm CO}}}~.
  \end{equation}
  \modifref{The factor of 4 in the denominator arises from the adopted two
    dimensional Gaussian cloud model, which measures the density assuming
    half the mass is contained within the FWHM cloud size.}
\end{itemize}

\subsection{PHANGS-MUSE Data Description}

We use observations from the Very Large Telescope/Multi Unit Spectroscopic
Explorer (VLT/MUSE; \citealt{Bacon2010}), which is ideally suited for
surveying nearby galaxies given its wide $\sim$1 arcmin field-of-view and
broad wavelength coverage (4800-9300\,\AA) at moderate (R$\sim$2000)
spectral resolution. The PHANGS-MUSE survey (PI: Schinnerer;
\citealt{Emsellem2022}) provides optical integral field unit maps across
the central star-forming disks of \nmuse\ nearby (D$<$19\,Mpc,
1\arcsec$<$100\,pc) spiral galaxies with low to moderate inclination
(<60$^\circ$). The region surveyed in each galaxy is matched to the
PHANGS-ALMA coverage, \modifref{requiring several MUSE pointings per
  galaxy.}

\modifref{The typical seeing during our observations was 0.8\arcsec\ in
  R-band, which corresponds to a typical physical scale of $\sim$70 pc for
  our galaxies. This is well-matched to the PHANGS-ALMA observations, and
  is adequately sampled by the 0.2\arcsec\ pixel size. To create a datacube
  corresponding to the full observed field of view for each galaxy, the
  data for individual pointings were first convolved such that the PSF is
  uniform for each galaxy before combination. We fit all stellar continuum
  emission (spatially binned to a S/N>35) simultaneously with the emission
  lines (where we fit each individual pixel) on the resulting combined
  cube.  The resulting maps of the H$\alpha$ line emission are then used to
  identify the individual \HII\ regions which are used in this work. Full
  details of the data reduction and production of emission line maps for
  PHANGS-MUSE are presented in \cite{Emsellem2022}}.

\subsection{PHANGS-MUSE \HII{} region catalogs}
\label{sect:hiicats}

To locate and characterize the \HII\ regions, we use the nebular catalogs
compiled by \cite{Santoro2022} \modifref{and described fully in
  \cite{groves2023}, which are constructed by applying the}
\texttt{HIIphot} algorithm \citep{Thilker2000} to the PHANGS-MUSE H$\alpha$
line maps.

\modifref{The key physical properties of the identified objects in the
  catalogs are derived using integrated spectra within the footprint of
  each region, assuming Gaussian line profiles and correcting for the
  instrumental dispersion along the spectral axis at location of the \Ha{}
  line \citep[$\sim$49\,km\,s$^{-1}$,][]{Bacon2017}. To select the \HII\
  regions from the nebular catalogs, we select regions that have
  properties} consistent with photo-ionization by massive stars by applying
the [\ion{O}{iii}]/H$\beta$ vs [\ion{N}{ii}]/H$\alpha$
\citep{Kauffmann2003} and [\ion{O}{iii}]/H$\beta$ vs
[\ion{S}{ii}]/H$\alpha$ \citep{Kewley2001} demarcations, excluding any
nebulae where any of these lines has a signal-to-noise ratio of less than
3. We further exclude nebulae that overlap with foreground stars or overlap
with the edge of our fields.

\modifref{In this paper, we work primarily with the total H$\alpha$
  luminosity, $L(\Ha)$ and size of the \HII\ regions. The \HII\ region size
  is defined as the circularized radius equivalent to the pixel area found
  in each \HII\ region footprint. The total H$\alpha$ luminosity is
  obtained from the sum of the H$\alpha$ within the \HII\ region
  footprint. The $L(\Ha)$ measurements that we use are corrected for dust
  extinction: a global extinction correction for each galaxy is applied to
  account for foreground extinction in the Milky Way~\citep{Schlafly2011},
  and an individual extinction correction for each \HII\ region that
  accounts for the dust within the target galaxy is determined using the
  observed Balmer decrement (H$\alpha$/H$\beta$). The latter assumes a
  Milky Way extinction curve \citep{Odonnell1994} with $R_\emr{V}=3.1$. We
  refer the reader to \cite{groves2023} for a more detailed explanation of
  the \HII{} region property definitions and of the corrections that are
  applied for, e.g., extinction and instrumental resolution. As described
  in the catalog paper, derived physical quantities including metallicity
  \citep{Pilyugin2016}, H$\alpha$ line width, and $\Sigma SFR$, are briefly
  considered in our analysis, but not explored in depth.}


\section{\modifref{A method to match GMCs and \HII{} regions}}
\label{sec:matching:method}

\FigMatchingExample{} %
\TabStatOverlapFullSample{} %

Our science goal is to determine whether \HII{} regions influence the
physical properties of GMCs. \modifref{In this section, we describe the
  method that we use to determine whether a GMC and \HII{} region are
  physically associated. Specifically, we work with catalogued measurements
  of the radial velocity of each GMC and \HII{} region, and the pixel masks
  that are generated during the cataloging process to identify the
  boundaries and projected area of each region. The key criterion in our
  matching method is the spatial overlap between the projected areas of the
  GMCs and \HII{} regions.  We use the pixel masks of the GMC and \HII{}
  regions to identify and measure regions of overlap. The \HII{} region
  masks are reprojected using nearest-neighbour interpolation to match the
  astrometry and pixelization scheme of the GMC masks. We superpose the
  binary masks of each galaxy to identify pixels that are common to both
  GMCs and \HII{} region masks. We construct a list of the GMCs that are
  members of regions with common pixels and calculate the percentage of the
  projected area of each GMC that is also present within an \HII{} region
  footprint. We consider an \HII{} region to be paired with a GMC when it
  occupies more than a minimal overlap percentage, MOP, of the GMC's
  projected area. \modifreftwo{We do not, however, consider the percentage
    of the \HII{} region covered by the GMC. This makes our pairing method
    asymmetric. This is physically motivated in the sense that we are
    searching for potential signatures of \HII{} region feedback on GMC
    properties at a scale of $\sim 100\,$pc. We thus prioritise the
    detection of pairs where the \HII{} region covers a significant fraction of
    the GMC area}.  From our initial list of paired regions, we exclude
  candidate pairs where the difference between the radial velocity of the
  GMC and \HII{} region is larger than 10\kms.}

\modifref{We found it necessary to refine our basic approach for the
  relatively common situation where a single GMC has pixels that overlap
  multiple \HII{} regions (or vice versa). Of the total number of GMCs that
  share common pixels with \HII{} regions, approximately half (47\%)
  overlap with more than one \HII{} region. For \HII{} regions, the
  fraction is smaller (19\%). We decided to allow each GMC to have more
  than one associated \HII{} region, but require that \HII{} regions be
  uniquely identified with a GMC, in practice assigning it to the GMC that
  is most covered by the \HII{} region's projected area. We discuss the
  rationale for this approach and compare it to other possible choices,
  e.g., using only exclusive GMCs-\HII{} region pairs, in
  Appendix~\ref{sec:appendix:single-matching}.}

Our matching approach is illustrated in
Fig.~\ref{fig:matching:example:zoom}.  This figure shows the GMCs and
\HII{} regions within a spiral arm region of NGC\,4254.  The left panel
illustrates the results for a MOP of 10\%: here, GMC B is uniquely
matched with \HII{} region 3, while GMC A is matched with both \HII{}
regions 1 and 2. The right panel illustrates a MOP of 70\%: here, only
GMCs A and B remain, and now GMC A is uniquely matched with \HII{} region
2. For most of the analysis in this paper, we set
MOP=40\%. \modifref{Using a higher MOP value would allow us to pinpoint
  the GMCs where \HII{} region feedback is likely to be the most
  pronounced, potentially making it easier to identify any signatures of
  feedback on the cloud properties. However, imposing higher MOP values
  would also reduce the number of identified pairs, making any conclusions
  less general. Our adopted MOP of 40\% is thus a compromise, based on
  both physical and practical considerations.  We present several tests of
  our method, and justify our choice for several user-defined parameters in
  Appendix~\ref{sec:appendix:matching-parameters}.}


\section{Properties of matched GMC-\HII{} regions}
\label{sec:matched-vs-global}

\FigHistGMCandHII{}%

\modifref{We applied our GMC-\HII{} region matching algorithm to the 19
  galaxies that are common to the PHANGS-ALMA and PHANGS-MUSE surveys. In
  NGC\,7496 and IC5332, our matching strategy failed to identity any
  matched GMC and \HII{} region for some or all of the minimum overlap
  percentage thresholds that we used. For the rest of this paper, we
  therefore present results for the remaining 17 galaxies. In this section,
  we compare the properties of matched GMC-\HII{} regions with the typical
  properties of their parent distributions, as well as the impact of the
  adopted MOP on the detection statistics of matched regions and the
  property distribution shapes.}


\subsection{Detection of matched GMC-\HII{} regions}
\label{sec:detstats}

\modifref{Table~\ref{tab:overlap:full:sample} summarizes the overall
  results of our matching strategy, i.e. the number of identified
  GMC-\HII{} region pairs for MOP thresholds of 10, 40 and 70\%, and the
  fraction of CO and \Ha{} luminosity that the paired objects represent in
  each case. Not surprisingly, fewer matched GMC-\HII{} regions are
  identified at higher MOP thresholds. Matched GMC-\HII{} regions barely
  exceed 50\% of their parent populations, even using a low MOP threshold
  of 10\%. This is qualitatively consistent with previous studies that
  found a significant reservoir of quiescent CO-bright gas in galaxies and
  a relatively short timescale for its disruption by stellar feedback
  \citep[e.g.][]{Schinnerer2019, Chevance2022a}. Measured across our full
  sample, for a MOP threshold of 10\%, the CO luminosity associated with
  GMCs in matched GMC-\HII{} regions is $41$\%, which roughly corresponds
  to the number of clouds identified within these regions $39$\%.  While
  relatively fewer in number, the \HII{} regions associated with matched
  GMC-\HII{} regions make a more significant contribution to the total
  \Ha{} luminosity (19.5\% in number vs 48\% in flux for a MOP threshold
  of 10\%). Table~\ref{tab:overlap:per:galaxies} in
  Appendix~\ref{sec:appendix:matched:properties} presents the detection
  statistics and contribution to total CO and \Ha{} luminosity of the
  matched GMC-\HII{} regions for individual galaxies.}

\subsection{Properties of matched GMC-\HII{} regions compared to their
  parent distributions}
\label{sec:hist:overview}

\modifref{Figure~\ref{fig:hist:overview} presents the empirical probability
  density functions (PDF) of the size and luminosity of the \HII{} regions
  and of several physical properties of GMCs (size, luminosity, peak
  temperature, velocity dispersion, surface density, characteristic
  turbulence linewidth, virial parameter, and free-fall
  time). \modifref{The PDFs were computed using a kernel density estimation
    method.} In each panel, the distribution of the parent population is
  shown in red, while the distribution for matched objects using MOP
  criteria of $10$, $40$ and $70$\%\ are shown in blue, green and gold
  respectively.}

\modifref{We conducted Kolmogorov-Smirnov tests
  (\href{https://docs.scipy.org/doc/scipy/reference/generated/scipy.stats.kstest.html}{scipy.stats.kstest})
  and Anderson-Darling tests
  (\href{https://docs.scipy.org/doc/scipy/reference/generated/scipy.stats.anderson_ksamp.html}{scipy.stats.anderson\_ksamp})
  to assess whether the differences in the distributions in
  Fig.~\ref{fig:hist:overview} are statistically significant. If the
  influence of \HII{} regions on GMCs is localized, then we expect that the
  trends in Fig.~\ref{fig:hist:overview} will become more pronounced as we
  increase the adopted MOP. If the trends instead result from the
  covariation of GMC and \HII{} region properties that are themselves due
  to larger scale effects (e.g., the co-location of clouds and \HII{}
  regions in spiral arms), then we would expect the distributions to be
  less sensitive to the adopted minimum overlap percentage.}

\modifref{\modifreftwo{Figure~2d} shows that matched GMCs are on average
  smaller than the parent population (mean radius of 45\,pc vs 69\,pc for a
  MOP threshold of 70\%), whereas
  \modifreftwo{Fig.~\ref{fig:hist:overview}b shows that} matched \HII{}
  regions are on average bigger than the parent population. This difference
  is more pronounced for higher MOP thresholds and is mostly due to the
  asymmetry of our matching criteria. The effect is larger for the \HII{}
  region sizes than for the GMC sizes. This is because our input GMC
  catalogs only contain resolved sources, while the \HII{} region catalogs
  contain a large fraction (80\%) of point sources.}

\modifref{GMCs in the matched populations exhibit a larger fraction of
  clouds with high CO peak brightness than the overall GMC population,
  \modifreftwo{as shown in Fig.~\ref{fig:hist:overview}e}. This cannot be
  due to our matching strategy, since GMCs generally show a marginally
  positive or no trend between their size and CO peak brightness. The
  luminosities of the GMCs and \HII{} regions follow the size trends since
  these properties are correlated to first order. A population of luminous
  GMCs is still evident in the matched populations, however, even though
  there are relatively fewer large GMCs: In
  \modifreftwo{Fig.~\ref{fig:hist:overview}c} the $MOP=70$\% (gold)
  population lies clearly above the parent GMC population (red) at high CO
  luminosity, even though there are more large GMCs in the parent
  population. These trends in size and brightness reinforce each other to
  yield a clear progression in the typical surface density and inferred
  free-fall time \modifreftwo{(Figs.~2g and 2i)} of the matched GMCs, such
  that GMCs in the matched population identified with $MOP=70$\% have
  roughly twice the surface density of the average GMC in the parent
  population, while the inferred typical free-fall time (which varies
  inversely with the surface density) of the matched GMCs is
  correspondingly shorter. }

\modifref{The distributions of cloud properties that involve the GMC
  velocity dispersion are more difficult to distinguish. \modifreftwo{In
    Fig.~\ref{fig:hist:overview}f and j}, the average velocity
  dispersion and virial parameter of the matched GMCs are slightly smaller
  than those of the parent population, which is to be expected from their
  smaller typical size and slightly higher brightness.  If the variation in
  the velocity dispersion was due purely to the cloud's smaller size and
  the Larson size-linewidth relation, then we would expect the
  distributions of the characteristic turbulent linewidth at a fiducial
  scale of 1\,pc to be the same for the matched and parent GMC
  populations. For the higher MOP thresholds, we instead see
  \modifreftwo{in Fig.~\ref{fig:hist:overview}h} a slight shift towards a
  larger average characteristic turbulent linewidth, and a larger fraction
  of clouds with higher characteristic turbulent linewidths, suggesting we
  may statistically detect a contribution to the cloud linewidth in the
  GMCs with matched \HII{} regions above and beyond the effect of cloud
  size. Potentially, this could reflect a slight increase of the turbulence
  within GMCs that closely associated with \HII{} regions. The difference
  is small however (a shift of $<0.1$\kms~pc$^{-0.5}$ in the mean value).}

The results of the tests that we conducted to determine whether the
observed differences in the distributions are statistically are listed in
Table~\ref{tab:KS:test} and Table~\ref{tab:AD:test}. While the
Anderson-Darling test is less sensitive to outliers compared to the
Kolmogorov-Smirnov test, both tests give similar results. For a minimum
overlap percentage of 40\% or above, all the distributions of the matched
GMCs properties are statistically different from their parent
distributions. This is also true for the characteristic turbulent
linewidth, which shows only a small variation in the distribution mean for
different adopted values of the MOP.

\subsection{Intermediate summary}

\begin{itemize}
\item Matched GMC/\HII{}-regions represent a non-negligible fraction of the
  overall GMC population in our sample galaxies. For a 10\% minimum overlap
  percentage, 40\% of the GMCs and one fifth of the \HII{} regions are
  matched, and these matched pairs represent about 50\% of the total CO and
  \Ha{} luminosity that arises from the catalogued objects.
\item All observed trends in the GMC property distributions are significant
  according to standard statistical tests.
\item Matched GMCs have a smaller size and higher typical CO peak
  temperature than a GMC from the overall population.  The relative small
  size of the matched GMCs likely derives from the asymmetry of our adopted
  matching strategy.
\item The observed distributions suggest that the GMCs matched with \HII{}
  regions are denser than a typical GMC.  The larger typical surface
  density is at least partly due to their higher intrinsic brightness at
  the pixel-level, as reflected in their higher typical peak CO brightness
  values.
\item The virial parameter and free-fall time of matched GMCs have lower
  average values than the overall GMC population, which follows from the
  observed trends of their luminosity, size and velocity dispersion.
\item \modifref{The characteristic turbulent linewidth of matched GMCs is
    slightly higher than for a typical GMC when a high ($\geq 40$\%) MOP
    threshold is used.}

\end{itemize}

\modifref{For the rest of the paper, we focus on the mass, the peak
  brightness, the characteristic turbulent linewidth and the molecular gas
  surface density of the GMCs. The three latter ones are intensive cloud
  properties that are often invoked by theories to explain the star
  formation rate and feedback, and do not scale directly with the GMC size
  by construction.}


\section{Correlations between matched GMC properties and the \Ha{}
  luminosity of the \HII{} regions}
\label{sec:global:correlations}

\FigCorrMatrix{} %
\FigCorrVsPercentage{}%

\modifref{In this section, we investigate whether there are significant
  correlations between the properties of the matched GMCs and \HII{}
  regions for our maximum MOP threshold (70\%). We then investigate how
  the adopted MOP threshold influences the observed correlations between
  GMC and \HII{} region properties, focusing on the \Ha{} luminosity of the
  \HII{} region as our primary independent variable, and describe two
  simple randomization tests that we used to determine whether the observed
  correlations could arise by chance. We characterize any observed
  correlations using a simple power-law fit and a goodness-of-fit
  statistic, which we present immediately below.}

\subsection{Characterizing a power law fit}

We characterize correlations by computing an ordinary least square
power-law fit to the observed trends and the associated $\pm 3\sigma$
levels from the fitted line. To assess the quality of the correlation, we
compute the \Rsquared{} coefficient, defined as
\begin{equation}
  \Rsquared = 1 - \frac{\sum_i (Y_i-\langle Y \rangle)^2}{\sum_i (Y_i-F_i)^2},
\end{equation}
where $Y_i$ represents the logarithm of the studied property for GMC $i$,
$\langle Y \rangle$ is the mean of the logarithm of this property over the
studied sample, and $F_i$ represents the values predicted by our fit. The
numerator is proportional to the variance of the \modifref{logarithm of
  the} studied GMC property, while the denominator is proportional to the
variance of the logarithm of the property about the power law fit of \Ha{}
luminosity of the \HII{} region.  A \Rsquared{} value of $1$ indicates that
the correlation is exactly linear in log-log space. A \Rsquared{} value of
$0$ indicates that the predicted properties $Y$ are independent of the
value of the X axis variable. In other words, the higher the \Rsquared{}
value, the better the GMC property is predicted by a power law model from
the \Ha{} luminosity of the \HII{} region.


We also compute confidence intervals on the \Rsquared{} coefficients
through a standard bootstrap approach. We refit a power law using data
samples that were constructed by selecting $N$ pairs from the original data
with replacement where $N$ was the number of original pairs: On average,
36\% of the pairs are different, implying that pairs are typically
duplicated $\sim2$ to 3 times.  We repeat this operation 10\,000 times,
each time computing the \Rsquared{} coefficient between the two quantities
of interest.  We then compute the 95\% confidence interval associated with
the distribution obtained from these 10\,000 trials.

\subsection{Correlations between GMC and \HII{} region properties}

We first explore the correlations between the properties of matched GMC and
\HII{} regions. We compute the \Rsquared{} coefficient between 1) GMC
properties, 2) \HII{} region properties, and 3) properties of GMCs and
\HII{} regions for matched pairs.  Figure~\ref{fig:corr:matrix} presents
the correlation matrices, which are sorted to display the strongest
correlations in the top left corner of each matrix as far as possible. It
shows that the virial parameter and free-fall time of GMCs are essentially
uncorrelated with any of the \HII{} region properties. In contrast, all
other properties of the matched GMCs exhibit some degree of correlation
with the properties of their associated \HII{} regions. The \Ha{}
luminosity and star formation rate are highly correlated with the GMC
luminosity and mass. Moderate correlations with these \HII{} region
properties are observed for the GMC size and velocity dispersion, with
weaker correlations obtained for the GMC surface density, characteristics
turbulent linewidth and CO peak temperature. The \HII{} region velocity
dispersion, metallicity, and stellar surface density exhibit correlations
with the CO luminosity, molecular mass, size and velocity dispersion of the
associated GMCs, but are mostly uncorrelated with the molecular surface
density, characteristic turbulent width and CO peak temperature.

The CO luminosity and mass are tightly correlated, justifying that we use
them interchangeably in the following. The GMC size and velocity dispersion
are also well-correlated with the molecular mass. These correlations
reflect the well-known Larson's scaling relations \citep{Larson1981}. The
characteristic turbulent linewidth shows a strong positive correlation with
the velocity dispersion. As expected from their definitions, the free-fall
time follows the molecular gas surface density.

The \Ha{} luminosity is correlated with all the other \HII{} region
properties that we consider, albeit only weakly with the metallicity and
local stellar surface density. Due to the additional strong covariance
between the \Ha{} luminosity and the size, star formation rate and velocity
dispersion of the \HII{} regions that we observe in
Fig.~\ref{fig:corr:matrix}, we restrict our comparison to studying GMC
properties as a function of the \HII{} region \Ha{} luminosity for the
remainder of the paper.

\subsection{Evolution with the minimum overlap percentage}

Figure~\ref{fig:corr:vs:percentage} presents scatter plots of the GMC
properties as a function of the \Ha{} luminosity for three different MOP
used to identify the GMC/\HII{}-region pairs: 10, 40, and 70\%. To
\modifref{help visualize how} the correlation \modifref{changes} with the
MOP \modifref{that we adopt}, the solid red lines indicate the power-law
fit for a minimum overlap percentage of 10\% in all panels.

The \Ha{} luminosities of the catalogued \HII{} regions in our sample cover
about five orders of magnitude, while the properties of matched GMCs cover
a smaller dynamic range: $\sim1.5$\,dex for peak temperature,
\modifref{$\sim2$\,dex for the characteristic turbulent linewidth}, up to
$\sim3$\,dex for the surface density and molecular mass. \modifref{The
  plots illustrate the correlations indicated by the matrix in
  Fig.~\ref{fig:corr:matrix}: a good correlation between the GMC mass and
  the \Ha{} luminosity of its associated \HII{} region at all MOP
  thresholds, but weaker and more scattered relationships for the intensive
  cloud properties, especially at high MOP values. For a MOP threshold
  of 40\%, the \Rsquared{} values range from 0.12 for the peak brightness}
to 0.55 for the GMC molecular mass, indicating that GMC properties do
indeed show a positive correlation with the \Ha{} luminosity of the \HII{}
region. \modifref{The power-law indices that we obtain range from
  \modifref{0.20} for the GMC surface density to 0.58 for the molecular
  mass. The GMC peak \modifref{brightness} and \modifref{characteristic
    turbulent linewidth} show intermediate power-law indices of
  \modifref{0.12} and \modifref{0.14}, respectively.  There is, however,
  structure in the plots in Fig.~\ref{fig:corr:vs:percentage} that is not
  completely captured by a simple power-law fit, and which is most evident
  for the correlations involving intensive GMCs properties when the
  GMC-\HII{}-region pairs are identified using higher ($\geq40\%$) MOP
  thresholds. Some of this structure in the correlation may be due to
  galactic environment, as we discuss in
  Section~\ref{sec:dependence:on:kpc}}.

\FigCorrFitVsPercentage{}%

In Fig.~\ref{fig:fits:vs:percentage} we plot how the \Rsquared{}
coefficients and the power-law slope of the observed correlations depend on
the adopted MOP. When increasing the minimum overlap percentage from 10
to 70\%, we find that the correlation between GMC molecular mass and the
\Ha{} luminosity of the matched \HII{} regions clearly exhibits a higher
\Rsquared{} coefficient and a steeper power-law slope. The correlation with
the CO peak temperature \modifref{on the other hand marginally} weakens
with increasing MOP. The correlation \modifref{of} the GMC surface
density \modifref{and the characteristic turbulent linewidth}
\modifref{with} \Ha{} luminosity of the matched \HII{} regions show no
dependence (within the confidence interval) on the adopted MOP.

\subsection{Randomization tests to check whether the observed correlations
  are genuine}

We conducted two tests to investigate whether the observed correlations
reflect a potential physical influence of the \HII{} regions on the
GMCs. First, we randomized the \Ha{} luminosities of the \HII{} regions in
the PHANGS-MUSE catalog, holding their positions and sizes constant. The
properties and positions of the GMCs were also left unchanged. We then
computed the \Rsquared{} coefficients for the same set of correlations
described above. We performed this randomization test 10\,000 times in
order to gauge the typical range of \Rsquared{} values that might arise
randomly. We obtained \Rsquared{} values of 0 to 0.04 for all the
properties, i.e., lower than the lowest \Rsquared{} value (0.15) that we
obtained for the actual correlations (see
Fig.~\ref{fig:corr-shuffle-halum}).

To assess whether the observed correlations are an artifact of our matching
strategy, we next randomized both the sizes and the \Ha{} luminosities of
the \HII{} regions, regenerated the list of matched GMC/\HII{} regions, and
constructed the same set of correlations as described above. This yields
typical \Rsquared{} values of 0 to 0.003, much lower than 0.15 (see
Fig.~\ref{fig:corr-shuffle-halum-size}).  We note that completely
randomizing the positions of the \HII{} regions is unfeasible, since the
projected area occupied by \HII{} regions and GMCs represents only 11.2\%
and 10.6\% of the observed field-of-view. Assigning random positions to the
GMCs or \HII{} regions would thus result in no GMC/\HII{}-region pairs
being detected.

\TabStatGalRegion{} %

\subsection{Intermediate summary}

\begin{itemize}
\item A positive correlation between GMC properties (mass, surface density,
  CO peak brightness temperature, and \modifref{characteristic turbulent
    linewidth}) and the \Ha{} luminosity of the matched \HII{} regions
  exists for a MOP of 10\%. The relationship between GMC properties and
  \Ha{} luminosity of the matched \HII{} region can be represented by a
  power law to first order.
\item The dependence of GMC properties on the \Ha{} luminosity of the
  matched \HII{} region varies as a function of the minimum overlap
  percentage in diverse ways. On one hand, the strength of the correlation
  between the \Ha{} luminosity and the molecular mass significantly
  improves as we increase the MOP. The correlations with the GMC surface
  density \modifref{and the characteristic turbulent linewidth} remain
  unchanged, and the correlation with the CO peak brightness temperature
  weakens for higher MOP thresholds.
\item The observed correlations cannot be reproduced in tests that
  randomize the \Ha{} luminosity and/or sizes of the \HII{} regions before
  and after identifying the matched GMC-\HII{} region pairs.
\end{itemize}


\section{\modifref{Correlations within different galactic environments}}
\label{sec:dependence:on:kpc}

\FigHistMatchKpc{} %
\FigBarplotEnvir{} %
\FigKpcCorrs{}%

\modifref{Both the distributions of the GMC properties and their
  correlation with the \HII{} region luminosity vary with the minimum
  overlap percentage that we use to identify a GMC/\HII{} region pair. A
  possible explanation of these results is that GMC properties hosting
  \HII{} regions are modified by local radiative feedback. Another
  possibility is that correlations between GMC and \HII{} region properties
  arise from their co-variation with the local galactic environment.  It is
  thus important to consider whether the matched GMC/\HII{} regions are
  preferentially located in certain galactic environments, and whether the
  galactic environment has an influence on the observed correlations.  In
  this section, we investigate the correlations as a function of the
  kpc-scale environment surrounding the GMC/\HII-region pairs.
  Specifically, we focus on the stellar surface density.  We explored a
  number of other large-scale properties of the galactic environment, such
  as the star formation rate surface density and molecular gas surface
  density, but found that they closely track the stellar surface density
  and give similar results as those presented below.}

\subsection{Method}
\label{sec:kpc:method}

\modifref{To investigate whether there is a dependence on galactic
  environment, we sorted the GMC/\HII{} region pairs into three
  equally-populated bins of the local stellar surface density, which we
  measured within an aperture of radius 500\,pc around each GMC/\HII{}
  region pair using the stellar surface density maps of PHANGS galaxies
  presented by \citet{Jiayi2022}. We used the GMC/\HII{} region pairs
  identified using a \modifreftwo{MOP} of 40 \%, yielding 380 pairs per
  bin. For the sake of simplicity, we refer to the first bin as ``low
  density environments'' (0 $\le$ log($\Sigma_{\star}$ / 1 M$_\odot$
  pc$^{-2}$) $\le$ 2.0), the second bin as ``average density environments''
  (2.0 $\le$ log($\Sigma_{\star}$) $\le$ 2.9) and the last bin as ``high
  density environments'' (2.9 $\le$ log($\Sigma_{\star}$) $\le$ 3.7).}

\modifref{Figure~\ref{fig:matching:frac:kpc} shows the empirical
  probability density function of the stellar surface density surrounding
  the GMC/\HII{} region pairs. The vertical dashed lines represent the bin
  limits. There is an observable difference between the distributions of
  the parent GMC population and of the GMCs that have an interface with an
  \HII{} region (MOP of 40\%).  The number of GMCs that are matched with
  an \HII{} region appears to decrease at the highest stellar surface
  densities. This effect is naturally explained if the timescale of
  significant interaction between GMCs and \HII{} regions is shorter in
  regions of high stellar surface density. }

\subsection{Results}
\label{sec:kpc:results}

\modifref{To illustrate how the stellar surface density bins relate to the
  commonly used dynamical environments, Fig.~\ref{fig:matching:frac:envir}
  presents the relative proportion of low, average and high
  \modifreftwo{stellar surface} density environments that are present in
  the nuclear and bar regions of galaxies (henceforth referred to
  collectively as the center environment), in spiral arms, and in interarm
  regions and galaxy outer disks (henceforth referred to as the disk
  environment).  Galaxy centers are almost exclusively composed of high
  \modifreftwo{stellar surface} density environments. Low, average and high
  \modifreftwo{stellar surface} density environments are present in roughly
  equal proportions within spiral arms. Galaxy disks are charcterized by
  low \modifreftwo{stellar surface} density environments.}

\modifref{Figure~\ref{fig:corr:kpc:all} shows the correlations between
  \Ha{} luminosity and GMC properties for GMC/\HII{}-region pairs for each
  stellar surface density bin.  The correlations between the \Ha{}
  luminosity and the GMC properties are clearly different from one
  environment to the other: the correlations are stronger and steeper in
  the high \modifreftwo{stellar surface} density environments, and less
  pronounced in the low \modifreftwo{stellar surface} density
  environments.}

\modifref{The correlations involving the different GMC properties exhibit
  different trends with stellar surface density. A correlation between GMC
  mass and \HII{} region luminosity persists across all bins, although it
  steepens with increasing stellar surface density. Correlations between
  the \HII{} region luminosity and other intensive GMC properties, on the
  other hand, are weak to non-existent in low \modifreftwo{stellar surface}
  density environments, and only clearly emerge for
  $\log \Sigma_{\star} > 2$}

\subsection{Intermediate summary}

\modifref{
  \begin{itemize}
  \item The correlations between GMC properties and the \Ha{} luminosity
    improve with increasing kpc-scale \modifreftwo{stellar} surface
    density.  The correlations are the strongest in high
    \modifreftwo{stellar} surface density environments, and either weaker
    (for the molecular mass) or non-existent in lower \modifreftwo{stellar}
    surface density environments.
  \item \modifreftwo{There is a strong correlation between the GMC mass and
      the \HII{} region luminosity across the full range of kpc-scale
      stellar surface densities in our sample (Fig.~\ref{fig:corr:kpc:all} top row). In contrast, the GMC surface density or
      peak temperature are uncorrelated with the \HII{} region
      luminosity in low stellar surface density environments, while they
      become clearly correlated in high stellar surface density
      environments.}
  \item \modifreftwo{The overall} correlations between the CO peak
    temperature, molecular gas surface density and the \HII{} region \Ha{}
    luminosity seen in Fig.~\ref{fig:corr:vs:percentage} \modifreftwo{are
      much weaker than the same correlations in the ``dense''
      kpc-environments seen in the right column of
      Fig.~\ref{fig:corr:kpc:all}. This difference may partly} be due to a
    dilution effect.  GMC/\HII{}-region pairs in the low density
    environments accounts for a large fraction of the overall population of
    GMC/\HII{}-region pairs. The fact that they show a weaker correlation
    than average thus dilutes the correlation signature.
  \item \modifreftwo{Comparing Fig.~\ref{fig:corr:vs:percentage}
      and~\ref{fig:corr:kpc:all} shows that the correlations of the
      intensive GMC properties with the \HII{} region luminosity are much
      more sensitive to the kpc-scale stellar density than to the MOP. This
      effect is weaker for the characteristic turbulent linewidth than for
      the peak temperature and surface density of GMCs. In contrast, the
      correlation of the GMC mass with the \HII{} region luminosity is
      about as sensitive to the stellar density as it is to the MOP. This difference of
      behavior suggests that correlations involving the intensive
      properties may stem from co-variations of GMC and \HII{} regions
      properties with galactic-scale environment, while the correlation of
      the GMC mass with the stellar surface density is influenced by the local
      (i.e.\ cloud-scale) effects.}
  \end{itemize}
}


\section{Correlations within individual galaxies}
\label{correlations:with:galaxy}

\modifref{Our combined sample of \HII{} regions and GMCs is drawn from 17
  galaxies. Among their global properties, the inclination and distance of
  the host galaxies could have an impact on our matching procedure, and by
  extension on the correlations that we observe. In this section, we
  investigate whether the correlations described in
  Section~\ref{sec:global:correlations} exist within individual galaxies by
  considering galaxies separately, where presumably observational effects
  such as inclination, distance and resolution do not play a role. For
  completeness, we also explored whether the correlations could be related
  to other galaxy properties, namely the total CO luminosity, total \Ha{}
  luminosity, metallicity, stellar mass and distance from the star-forming
  main sequence.  For this analysis, we use the list of matched GMC/\HII{}
  region pairs identified using a MOP of 40\%. We identify between 29
  (for NGC\,2835) and 227 (for NGC\,1566) pairs in each galaxy in our
  sample, with a mean of 114 pairs per galaxy. This is sufficient to
  consider the correlations between GMC properties and \HII{} region
  luminosity within individual galaxies, but prohibits exploring trends
  with stellar density within them.}

\modifref{We examined the correlations between GMC properties and the
  \HII{} region luminosity using the same procedure as in
  Section~\ref{sec:global:correlations}, measuring the power law slope and
  \Rsquared{} statistic for each. A full list of our per-galaxy results is
  presented in Table~\ref{tab:correlations} of
  Appendix~\ref{sec:appendix:variability:with:galaxy}. For each of the GMC
  properties that we consider, we find significant variability among
  galaxies in terms of the slope and quality of a power law fit. Consistent
  with the global trend in Section~\ref{sec:global:correlations}, the GMC
  mass tends to be well correlated with the \Ha{} luminosity within
  individual galaxies, with moderate (\Rsquared{}>0.3) to strong
  (\Rsquared{}>0.5) correlations obtained for 13 galaxies. Few individual
  galaxies reveal a significant correlation between the \HII{} region
  luminosity and the GMC surface density or characteristic turbulent
  linewidth, again consistent with the global trends. It is striking,
  however, that individual galaxies reveal a much stronger correlation
  between the CO peak brightness and \HII{} region luminosity than can be
  discerned from the global trends, with \Rsquared{} values that are
  comparable to those obtained for the correlations with GMC mass
  (i.e. \Rsquared{}>0.3 for 13 galaxies, and three galaxies with
  \Rsquared{}>0.5). }

\modifref{We explored whether there were any systematic trends between
  these correlation results and properties of the host galaxies by plotting
  the derived power law slopes and \Rsquared{} statistics against galaxy
  distance, inclination, stellar mass, total CO and \Ha{} luminosity, mean
  $\log \Sigma_{\star}$, offset from the star-forming main sequence, and
  number of pairs identified. These results are presented in
  Appendix~\ref{sec:appendix:variability:with:galaxy}. We find no
  dependence of the strength and slope of the observed correlations on the
  inclination or distance to the galaxy, suggesting that the correlations
  are not driven by geometric effects (e.g. viewing angle) or
  resolution. In general, lower mass systems in our sample tend to exhibit
  poorer correlations between the GMC properties and the \HII{} region
  luminosity. Since the typical stellar surface density in galaxy disks
  tends to scale with the galaxy total mass and luminosity, this result is
  consistent with the weaker correlations observed for low
  \modifreftwo{stellar surface} density galactic environments that we
  described in Section~\ref{sec:dependence:on:kpc}.}


\section{Discussion}
\label{sec:discussion}

\FigSummaryResults{} %

In this paper, we searched for evidence for a potential impact of \HII{}
regions on GMC properties \modifref{as measured on spatial scales between
  30 and 180\,pc. We presented a method to match \HII{} regions to a GMC
  based on their relative overlap region.}

We studied 1) the \modifref{distributions of characteristic cloud}
properties (molecular mass, surface density, CO peak temperature,
\modifref{and characteristic turbulent linewidth}) \modifref{of GMCs that
  are associated with \HII{} regions, and how these distributions vary as a
  function of the MOP that we adopt to identify matched objects, and 2)
  correlations between GMC properties and the \Ha{} luminosity of their
  associated \HII{} regions, again as a function of the MOP. To establish
  whether our results are consistent with the observed correlations having
  a local origin (i.e. a direct physical link between the cloud's physical
  state and the \HII{} region), we investigated whether the correlations
  vary with the kpc-scale stellar surface density}, \modifref{and global
  properties of the host galaxy, including observational properties such as
  distance and inclination.}

\modifref{In this section, we briefly summarize the trends that we observe
  and outline three potential physical scenarios that could explain them.}
To conclude, we summarize the lessons learnt during this exercise.

\subsection{Two families of behaviors experienced by the properties of the
  GMC/\HII{}-region pairs with increasing overlap}
\label{subsec:two:different:selections}

Figure~\ref{fig:summary:results} presents the joint and marginalized PDFs
of \modifref{the GMC molecular mass and CO peak temperature} and the
\modifref{\Ha{} luminosity of its associated} \HII{} region for two minimum
overlap percentage \modifref{thresholds}: 1\% in red, and 70\% in
blue. \modifreftwo{A MOP threshold of 1\% yields a good approximation for
  all the GMCs that are contiguous or overlapping with at least one \HII{}
  region.} We used a kernel method to compute all the
PDFs. \modifref{Figure~\ref{fig:summary:results} illustrates that
  increasing the MOP causes two distinct effects on} the correlations
between GMC properties and the \Ha{} luminosity.

First, the CO peak temperature \modifref{(and the other intensive
  properties)} \modifref{demonstrate one behavior when we vary the overlap
  criterion: as we impose a higher MOP increases, the median value of the
  GMC property also significantly increases.}  This behavior reduces the
\modifref{dynamic} range of the x and y axis values, \modifref{reducing}
the \Rsquared{} value of the correlations.

Second, the \modifref{GMC} mass shows a \modifref{different} behavior: its
correlation with the \Ha{} luminosity strengthens \modifref{and
  steepens}. \modifreftwo{Additionally} the median shifts towards lower
\modifref{masses}. \modifreftwo{The latter effect is mostly due to the
  asymmetry of the matching criterion.}

\subsection{Potential Physical Origin}

\modifref{The physical properties of GMCs follow Larson-type scaling
  relations, such that larger GMCs also tend to be more massive and more
  turbulent. Appendix~\ref{sec:appendix:LarsonHeyer} confirms that these
  relationships exist for the GMCs that we identify with \HII{} regions
  using any MOP.} However, these relationships \modifref{fail to} explain
either why the CO peak temperature increases with the MOP or why the
correlation of the GMC mass with the \Ha{} luminosity strengthens when
increasing the minimum overlap percentage. We consider three possible
scenarios to explain these two observed behaviors.
\begin{description}
\item[\textbf{The environmental/coincidence scenario}] The galactic
  environment induces covariation of GMCs and \HII{} region properties
  without a direct causal connection between them.
\item[\textbf{The epigenetic scenario}] Parent GMCs transmit some of the
  properties from their environment to their child \HII{} regions.
\item[\textbf{The radiative feedback scenario}] The child \HII{} regions
  disrupt and alter the intrinsic properties of their parent molecular
  cloud.
\end{description}
The trends identified in our analysis are likely due to a combination of
these scenarios. For instance, there could be a ``background'' co-variation
of matched GMCs and \HII{} regions due to a common environment or to the
heritage of some of the GMC properties to the associated \HII{} regions,
and local effects (e.g., stellar radiative feedback) could become the
predominant driver of the GMC/\HII{}-region correlations only observed for
highly overlapping pairs.

\subsubsection{Environmental / coincidence scenario}

Section~\ref{sec:dependence:on:kpc} first shows that the
\modifreftwo{correlations} between the GMC properties and the \Ha{}
luminosity are sensitive to the \modifref{kpc-scale} stellar surface
density, when considering all the matched GMC/\HII{}-region pairs. For
instance, there is no correlation of the CO peak brightness and surface
density with the \Ha{} luminosity in low density kpc-scale environments,
but these correlations appear and improve systematically as a function of
the kpc-scale stellar density. This means that the kpc environment plays a
role in setting \modifref{the correlations of the GMCs properties with the
  luminosity of the matched \HII{} regions.}

\modifref{However, Sections~\ref{sec:global:correlations} and
  \ref{sec:dependence:on:kpc} also show that the correlation between the
  \Ha{} luminosity and the molecular mass is much more sensitive to the
  \modifreftwo{MOP} than to the kpc environment.}  This argues in favor
of a local origin of \modifref{this behavior.}

\subsubsection{Epigenetic scenario}

Section~\ref{sec:global:correlations} shows a strong positive correlation
between the \HII{} region \Ha\ luminosity and the molecular cloud
mass. Moreover, this correlation increases in steepness and strength when
increasing the MOP. \modifref{ Section~\ref{sec:dependence:on:kpc} also
  showed that this correlation persists even in \modifref{low density
    environments}, where crowding is less of an issue.} This suggests that
this correlation is not driven by coincidental co-variation of the GMC and
\HII{} region properties, but has a local origin. \modifref{An
  evolutionary} link \modifref{offers} a natural local origin: a
\modifref{high-mass} GMC is needed to make a \modifref{large} and thus
luminous \HII{} region. In \modifref{this scenario, GMC properties are
  partly regulated by their kpc-scale environment. This information is
  transmitted to their child \HII{} regions.} This is similar to
epigenetics (i.e.\ the study of how the environment can cause changes that
affect the way genes work).

\subsubsection{\modifref{Stellar} feedback scenario}

The observed spatial de-correlation on small ($<$100pc) scales between
molecular gas and young star forming regions
\citep{Schruba2010,Onodera2010,Grasha2018,Kreckel2018,Kruijssen2018,
  Grasha2019,Kruijssen2019,Schinnerer2019,Chevance2020} indicates that
within the star formation cycle the clearing and dissolution of the natal
molecular birth cloud must occur on relatively short timescales. Recent
results indicate that pre-supernova feedback processes are essential
contributors to this process
\citep{Barnes2021,Mcleod2021,Olivier2021,Barnes2022}, with clearing times
estimated to be only a few Myr \citep{Kim2021a,Chevance2022a,Kim2022}. By
isolating the sub-set of GMCs that overlaps with \HII{} regions, it is
possible that we are catching this process in the act.  If this is the
case, we expect to see some imprint on these feedback processes onto the
parent GMC. Some tentative evidence for this was identified in the
`Headlight cloud' in NGC\,0628 \citep{Herrera2020}, where CO emission
associated with this cloud was suggested to be overluminous due to heating
by the associated \HII{} region.

With our matched GMC and \HII{} region sample, we identify a weak positive
correlation between CO peak brightness and H$\alpha$ luminosity
(Fig.~\ref{fig:corr:vs:percentage}). This correlation is strongest within
\modifref{dense kpc scale environments}, with \Rsquared{} =
\modifref{0.32}. It clearly decreases with increasing overlap percentage
(Fig.~\ref{fig:fits:vs:percentage}), and is essentially absent at minimum
overlap percentages of 70\%. Looking in \modifref{low density kpc scale
  environments}, where the cleanest matches are possible and there should
be less contribution from neighboring or clustered star-forming regions,
the correlation is absent (\Rsquared{} =
\modifref{0.00}). \modifref{Analysis of individual galaxies
  (Section~\ref{correlations:with:galaxy}), on the other hand, tends to
  confirm a trend between CO peak brightness and H$\alpha$ luminosity,
  suggesting the scatter in Figures~\ref{fig:corr:vs:percentage} and
  ~\ref{fig:corr:kpc:all} is at least partly due to galaxy-to-galaxy
  variation.}

The CO peak temperature significantly shifts to higher values when the
MOP increases (see Fig.~\ref{fig:hist:overview} in
Sect.~\ref{sec:matched-vs-global}). \modifref{This shift effect happens
  even in low density environments, and thus probably has a local origin.}
Moreover, the origin of this local effect is completely independent from
the strength of the correlation of the GMC mass with the \Ha{} luminosity
(see \modifref{Appendix~\ref{sec:appendix:LarsonHeyer}}). This independent
effect \modifreftwo{could} be explained by the fact that massive young
stars naturally heat the Photon-Dominated Regions at the interface between
\HII{} regions and GMCs. The larger the MOP, the larger the interface and
thus the more efficient heating. This effect \modifreftwo{would} saturate
the line peak brightness of the low-$J$ CO lines and \modifreftwo{would}
enhance higher-$J$ CO brightness.

We also look for a positive correlation between H$\alpha$ luminosity and
\modifref{characteristic turbulent linewidth}, and find a \modifref{weak}
trend \modifref{for any} MOP. We identify a peak correlation at
\modifref{moderate 40\%} overlap percentage (\Rsquared{} =
0.14). \modifref{Nevertheless, the characteristic turbulent linewidth of
  matched GMCs shifts systematically towards higher values when we adopt a
  higher MOP (Section~\ref{sec:matched-vs-global}), the opposite
  behaviour to $\sigma_{v}$} that decreases in similar conditions. This
could hint to a slight increase of the turbulent motion in GMCs by nearby
\HII{} regions.

\subsection{Studying the impact of \HII{} regions on GMCs: Lessons learnt}

Studying the impact of \HII{} regions on GMCs in nearby galaxies
\modifref{on 30 to 180\,pc scales} is a difficult challenge for several
reasons.
\begin{itemize}
\item The PHANGS \modifref{GMC and \HII{} region catalogs present a large
    number of physical properties for each of these objects.} This is a
  high-dimensional parameter space to be explored.
\item \modifref{Some of the catalog properties are dependent by
    construction (e.g. the mass and size of GMCs).} Even among
  \modifref{physical properties that are in principle independent}, there
  are \modifref{well-established empirical} correlations between GMC
  properties (e.g. Larson-type scaling relations) and between \HII{} region
  properties.
\item The co-variation among properties of GMCs and \HII{} regions
  \modifref{in different galactic environments can be} larger than the
  impact of physical processes \modifref{on cloud-scales and below that
    regulate} the co-evolution of GMC and \HII{} regions.
\end{itemize}
In summary, we are searching for subtle \modifref{signatures} in a
high-dimensional space \modifref{among properties that demonstrate
  pre-existing correlations due to diverse physical origins. This implies
  we need to devise methods that maximize the probability of detecting a
  signature of \HII{} region feedback on their natal GMC, without
  misinterpreting or biasing the results}.

\modifref{The timescale for disrupting GMCs via \HII{} region feedback is
  relatively short compared to the cloud
  lifetime~\citep{Kim2021a,Chevance2022a}. We thus need to search for a
  transient state. This, as well as the considerations outlined above,
  tends to favour using a high MOP to pinpoint the GMCs where \HII{}
  region feedback is most pronounced. The drawback of such a choice is that
  pairs with a large overlap percentage are rare, leading us to combine
  regions from different galaxies. This complicates the interpretation due
  to the different characteristic environments within galaxies of different
  types.}

\modifref{In this paper, we compromised by adopting a fiducial minimum
  overlap percentage of 40\% and studying how the joint PDFs and
  correlations vary as a function of MOP thresholds between 10 and
  70\%. We also used an asymmetric matching criterion to focus on the
  \HII{} regions that are most likely to impact GMC properties on the
  spatial scales that are accessible to PHANGS-ALMA (i.e.  $\sim
  100\,$pc). Increasing the angular resolution of nearby galaxy imaging
  surveys would allow us to better locate the interfaces between matched
  GMCs and \HII{} regions, while maintaining good statistics.  A less
  time-consuming alternative would be to increase the angular resolution of
  the observations towards selected pairs to validate the scenarios
  envisaged here.}

\modifref{The covariation of} GMC properties and \HII{} regions properties
\modifref{is a further complication, since the signature of} feedback can
be mistakenly \modifref{considered as scatter} around pre-existing
correlations, while it actually is a hidden control variable. In other
words, it is important to reveal how this (previously hidden) variable
impacts pre-existing correlations of GMC properties. \modifref{Once again,
  much higher resolution observations of extragalactic high-mass
  star-forming regions would be useful to understand how feedback
  signatures might manifest themselves in lower resolution observations
  over a much larger sample.}


\section{Conclusions}
\label{sec:conclusions}

\modifref{In this paper, we studied the physical properties of GMCs
  associated with \HII{} regions for 17 galaxies in the PHANGS-ALMA
  survey. Our primary goal was to determine whether GMC properties on cloud
  scales (here $30$ to 180\,pc) are modified by star formation feedback. We
  studied the distribution of \modifref{four} cloud properties (GMC mass,
  surface density, CO peak temperature, \modifref{and characteristic
    turbulent linewidth}) and the correlations between these properties
  with the \Ha{} luminosity of the associated \HII{} regions.}  The main
conclusions of our study are:
\begin{enumerate}
\item Matched GMC/\HII{}-regions represent a \modifref{non-negligible
    fraction of the overall population of GMCs and \HII{} regions in
    galaxies.  GMCs have a higher detection rate by number in matched
    regions than \HII{} regions. When considering the detected flux rather
    than the number of detected objects, \HII{} regions in matched pairs
    represent a larger fraction of their host galaxy's \Ha{} luminosity
    than the contribution of GMCs in matched pairs to the galaxy's CO
    luminosity.}
\item \modifreftwo{Matched GMCs tend to be denser than a typical GMC.}
\item \modifref{In all galaxies and environments, the GMC mass (as inferred
    from its CO luminosity) is well-correlated with the \Ha{} luminosity of
    its associated \HII{} region}. \modifref{The GMC} surface density, CO
  peak temperature, and characteristic turbulent linewidth\modifref{
    exhibit weaker correlations with} the \Ha{} luminosity.
\item The galactic environment \modifref{has an impact on the observed}
  correlation between GMC mass and the \Ha{} luminosity. The correlation is
  generally stronger and steeper in the \modifref{dense kpc-scale
    environments (typified by the centers and bars of galaxies)}, and
  weaker in the \modifref{low density kpc-scale enviroments (e.g., outer
    disks and interarms)}.
\item The correlations \modifref{observed using the combined sample of
    matched GMC-\HII{} regions from 17 galaxies obscures some} variability
  galaxy by galaxy.  In particular, \modifref{individual galaxies often
    exhibit} significant correlations between the molecular mass, the CO
  peak temperature and the \Ha{} luminosity. \modifref{The correlation with
    GMC mass persists when the combined sample is considered, but the
    correlation between CO peak temperature and \Ha{} luminosity is highly
    scattered when the matched regions in all galaxies are combined. }
\item \modifref{GMC properties exhibit different behaviours when we adjust
    the MOP that we use to identify matched GMC-\HII{} regions. The
    median value of intensive cloud properties, i.e., the} molecular
  surface density, CO peak temperature, and characteristic turbulent width,
  shifts towards higher values \modifref{when a higher value of MOP is
    adopted.  But the power law index of their correlation with the \Ha{}
    luminosity remains} constant. In contrast, the \modifref{molecular mass
    decreases} with increasing overlap percentage, \modifref{while its}
  correlation with the \Ha{} luminosity strengthens.
\end{enumerate}
\modifref{We propose a scenario where the kpc-scale galactic environment
  regulates the typical properties of GMCs, and massive GMCs are required
  to produce the most luminous \HII{} regions. Focussing on the matched
  GMC-\HII{} regions that we identify using a high overlap percentage, we
  identify variations in the CO peak brightness and (more tentatively)
  characteristic turbulent linewidth that may be a signature of \HII{}
  region feedback on the molecular gas. Generally, however, it is difficult
  to unambiguously identify variations in the physical properties of GMCs
  with the impact of feedback at the spatial scales accessible to
  PHANGS-ALMA.}

\modifref{One way to confirm our results would be \modifreftwo{to} observe
  a sample of extragalactic high mass star-forming regions} at much better
angular resolution in order to fully disentangle the \modifref{effects of
  galactic environment and \HII{} region feedback}. \modifref{Examining}
the relative strengths of the different $J$ CO lines as a function of the
MOP \modifref{would be another useful independent test of the proposed
  link between feedback and CO peak brightness, since we }expect that
higher $J$ CO lines \modifref{would} have \modifref{further} enhanced
brightness. \modifref{For future investigations, we also propose to refine
  the criteria used to identify associated GMC-\HII{} regions to take
  better account of physical quantities that are relevant to cloud
  disruption by stellar feedback. One such approach would be to} compare
the fraction of \Ha{} luminosity in the overlapping region with the
fraction of molecular gas available in the interaction region, i.e. the
number of ionizing photons per molecule of hydrogen, \modifref{rather than
  the simpler projected area criterion used here}.

\begin{acknowledgements}
  This work was carried out as part of the PHANGS collaboration. Based on
  observations collected at the European Southern Observatory under ESO
  programmes 1100.B-0651, 095.C-0473, and 094.C-0623 (PHANGS--MUSE; PI
  Schinnerer), as well as 094.B-0321 (MAGNUM; PI Marconi), 099.B-0242,
  0100.B-0116, 098.B-0551 (MAD; PI Carollo) and 097.B-0640 (TIMER; PI
  Gadotti).

  This work was supported in part by the French Agence Nationale de la
  Recherche through the DAOISM grant ANR-21-CE31-0010 and by the Programme
  National ``Physique et Chimie du Milieu Interstellaire'' (PCMI) of
  CNRS/INSU with INC/INP, co-funded by CEA and CNES.
  
  KK gratefully acknowledges funding from the German Research Foundation
  (DFG) in the form of an Emmy Noether Research Group (grant number
  KR4598/2-1, PI Kreckel).
  
  HAP acknowledges support by the Ministry of Science and Technology of
  Taiwan under grant 110-2112-M-032-020-MY3.
  
  ATB, FB, IB, JdB would like to acknowledge funding from the European
  Research Council (ERC) under the European Union’s Horizon 2020 research
  and innovation programme (grant agreement No.726384/Empire).
  
  CE acknowledges funding from the Deutsche Forschungsgemeinschaft (DFG)
  Sachbeihilfe, grant No. BI1546/3-1.
  
  KG is supported by the Australian Research Council through the Discovery
  Early Career Researcher Award (DECRA) Fellowship DE220100766 funded by
  the Australian Government. KG is supported by the Australian Research
  Council Centre of Excellence for All Sky Astrophysics in 3 Dimensions
  (ASTRO~3D), through project number CE170100013.
  
  SCOG, RSK and EJW acknowledge financial support from DFG via the
  Collaborative Research Center (SFB 881, Project-ID 138713538) 'The Milky
  Way System' (subprojects A1, B1, B2, B8, P2), and they thank for funding
  from the European Research Council in the ERC synergy grant `ECOGAL –
  Understanding our Galactic ecosystem: From the disk of the Milky Way to
  the formation sites of stars and planets' (project ID 855130) as well as
  from the Heidelberg Cluster of Excellence (EXC 2181 - 390900948)
  `STRUCTURES: A unifying approach to emergent phenomena in the physical
  world, mathematics, and complex data'.
  
  MC gratefully acknowledges funding from the DFG through an Emmy Noether
  Research Group (grant numbers CH2137/1-1 and KR4801/1-1), as well as from
  the European Research Council (ERC) under the European Union’s Horizon
  2020 re- search and innovation programme via the ERC Starting Grant
  MUSTANG (grant agreement number 714907).
  
  The work of JS is partially supported by the Natural Sciences and
  Engineering Research Council of Canada (NSERC) through the Canadian
  Institute for Theoretical Astrophysics (CITA) National Fellowship.

 \modifreftwo{ ES and TGW acknowledge funding from the European Research Council (ERC) under the European Union’s Horizon 2020 research and innovation programme (grant agreement No. 694343).}
\end{acknowledgements}

\bibliographystyle{aa} %
\bibliography{main.bib}


\appendix{} %

\section{Supplemental information about the matching method}
\label{sec:appendix:matching:method}

\subsection{Single matching vs multiple matching}
\label{sec:appendix:single-matching}

\modifref{For the spatial overlap matching method used in this paper,
  particular care must be taken when dealing with GMCs that overlap
  multiple \HII{} regions (and vice versa).  One option would be to require
  that GMCs and \HII{} regions form exclusive pairs, i.e.\ ensuring that
  each \HII{} region is uniquely identified to a GMC, which itself is only
  identified with a single \HII{} region. This approach is simple to
  implement, but physically unrealistic: GMCs and \HII{} regions are not
  randomly distributed, but instead preferentially located within
  larger-scale coherent spatial structures, such as spiral arms and
  inter-arm feathers~\citep{schinnerer17}, where star-forming complexes are
  often crowded together. Star formation in a GMC may result in multiple
  \HII{} regions~\citep[see, e.g.,][]{williamsmckee97}. The probability of
  hosting multiple \HII{} regions is more likely for GMCs more massive than
  a few $10^{5}\Msun$, which is true for most of the clouds in the PHANGS
  GMC catalog. Overall, the fraction of associations involving multiple
  GMCs and \HII{} regions is significant. The number of GMCs overlapping
  with multiple \HII{} regions represents 47\% of the total number of
  overlapping GMCs, while the percentage of \HII{} regions that overlap
  with multiple GMCs is 19\% of the total number of overlapping \HII{}
  regions. This approximately holds for any MOP.  We therefore decided to
  allow each GMC to have more than one associated \HII{} region, but
  require that \HII{} regions are uniquely identified with a GMC, in
  practice assigning it to the GMC that overlaps the largest percentage of
  the \HII{} region's projected area.}

\subsection{Asymmetry of the matching criterion}

\FigHistSizeGMCHII{} %

\modifref{We are specifically looking for \HII{} regions that may
  significantly modify the properties of their natal GMC clouds, not only
  at the immediate working surface, but also on the $\sim 50-100\,$pc
  scales on which global cloud properties are typically measured. This
  implies that we are searching for \HII{} regions whose energy content is
  a significant fraction of the GMC one. In particular, we will explicitly
  try to avoid selecting all the small \HII{} regions associated to large
  GMCs because their feedback would have negligible impact and they would
  dilute the searched signatures on the GMC properties.}

\modifref{For completeness, Fig.~\ref{fig:hist:sizeGMCHII} presents the
  probability density functions (PDFs) of the radii of \HII{} regions and
  GMCs across our sample. The size of resolved GMCs or \HII{} regions are
  corrected for the size of ALMA and MUSE point spread function. But this
  plot should not be over-interpreted because the GMC and \HII{}-region
  catalogs treat differently unresolved sources.  The unresolved GMCs, on
  one hand, are filtered out from the PHANGS-ALMA catalog of GMCs, which
  rejects objects when their size cannot be deconvolved from the beam size
  \citep{rosolowsky21}.  On the other hand, the majority ($\sim80$\%) of
  the catalogued \HII{} regions are point sources, for which we only have
  an upper limit of their projected area, corresponding to the point spread
  function of the VLT/MUSE instrument \citep{Santoro2022}. That is why the
  distribution of \HII{} region size exhibits shallow peaks at the
  resolution of the observations, namely $\sim$50\,pc, $\sim$60\,pc and
  $\sim$80\,pc.  Using these definitions, the 10 to 90 percentile range is
  $32-109$\,pc for the GMC radius, and $33-87$\,pc for the \HII{} region
  radius. And there exist \HII{} regions as large as the largest identified
  GMCs.}

\subsection{Overlap parameter choices}
\label{sec:appendix:matching-parameters}

\FigHistVelOffandMOP{}%

\modifref{When matching GMCs and \HII{} regions with the overlap matching
  method coupled with the velocity offset, two parameters are left to the
  user: the \modifreftwo{MOP} and the velocity offset threshold. The left
  panel of Fig.~\ref{fig:hist:overlap} presents the histogram of the number
  of GMC/\HII{} region pairs as a function of their velocity
  offsets. Limiting the velocity offset between a spatially overlapping GMC
  and \HII{} region ensures that the two objects are at least contiguous in
  position-position-velocity space. The uncertainty on the velocity of
  \HII{} regions is \modifref{$\sim 49\kms$,} larger than the uncertainty
  on GMC velocity, and thus the limiting factor on the velocity offset
  threshold. We thus chose a velocity offset threshold of 10\kms. We note
  that limiting the velocity offset to a value lower than 10\kms{}, e.g.,
  5\kms, does not alter our results.  As for the minimal overlap
  percentage, using a 40\% \modifreftwo{MOP} as a fiducial value allows
  us to filter out matched \HII{} regions and GMCs that are potentially
  physically unrelated, while keeping a large enough sample of GMC/\HII{}
  region pairs to compute reliable statistics.  The right panel of
  Fig.~\ref{fig:hist:overlap} presents the histogram of the number of
  GMC/\HII{} region pairs according to the projected area of overlap. The
  distribution of overlap percentages between \HII{} regions and GMCs has a
  mean of 17\%, a median of 10\%, and a standard deviation of 18\%.}

\clearpage{}

\section{Properties of the matched GMC/\HII{} regions in individual
  galaxies}
\label{sec:appendix:matched:properties}

\FigMatchedLocation{} %
\TabStatOverlapPerGalaxies{} %

\modifref{As an example, Fig.~\ref{fig:matching:example:full} compares the
  spatial distribution of matched GMC/\HII{}-regions in NGC\,4254 for a
  MOP of 40\% with the spatial footprints of the full distributions of
  GMCs and \HII{} regions, and the environment masks.}

\modifref{Table~\ref{tab:overlap:per:galaxies} presents the statistics of
  matched GMC-\HII{} regions for each galaxy in our sample. For three
  different MOP thresholds, we list total number of \HII{} regions and
  GMCs in each galaxy, the percentage of \HII{} regions and GMCs that are
  located in matched regions, and their absolute and relative contributions
  to each galaxy's H$\alpha$ and CO
  luminosities. \modifreftwo{Fig.~\ref{fig:corr:galaxies} shows the
    correlation between CO peak temperature and \Ha{} luminosity
    considering \HII{} region/GMC pairs from NGC\,1672, NGC\,4535, and
    NGC\,4321 separately. These 3 galaxies harbor the highest correlation
    coefficient between CO peak temperature and \Ha{} luminosity
    $(\Rsquared \geq 0.5)$.}}

\FigGalaxyCorrelations{} %

\clearpage{}

\section{Supplemental information to assess the robustness of the results}
\label{sec:appendix:robustness}

\TabKSTest{}%
\TabADTest{}%
\FigHaShuffleCorr{}%
\FigHaSizeShuffleCorr{}%
\FigBootstrapTest{}%

\clearpage{}

\section{Supplemental figures about the variability of the properties of
  the matched GMC/\HII{}-regions as a function of the galaxy}
\label{sec:appendix:variability:with:galaxy}

\TabCorrelations{} %
\FigCorrVsGalaxyRsquared{} %
\FigCorrVsGalaxySlope{} %

\modifref{Figures \ref{fig:corr:vs:galaxy:rsquared} and
  \ref{fig:corr:vs:galaxy:slope} present respectively the correlations
  coefficients and power law indices of the correlation between the GMC
  peak brightness and the \HII{} region \Ha{} luminosity, as a function of
  different galaxy-wide properties. The properties in question are: the
  total number of matched GMC/\HII{} regions, the galaxy's distance,
  inclination, stellar mass, total CO luminosity, total \Ha{} luminosity,
  mean stellar density environment and distance from the main sequence. For
  these figures the \modifreftwo{MOP} is 40\%. Reassuringly, the galaxy's
  distance, inclination or number of matched GMC/\HII{} regions doesn't
  have any impact on the found correlations.}

\modifref{To be more quantitative about the effect of inclination and
  metallicity, we computed a linear regression of the correlation
  coefficient \Rsquared{} (of the peak temperature correlation with \Ha{}
  luminosity) against the galaxies inclinations, total CO luminosities and
  metallicities. This will allow us to check whether a galactic property
  has a significant impact on the correlations or not.}

\modifref{For the linear regression between \Rsquared{} and the galaxy
  inclination, the resulting correlation coefficient and slope are
  respectively $0.07 \pm 0.08$ and $0.004 \pm 0.009$. These numbers are
  consistent with 0, so the galaxy inclination has no impact on the
  correlations.  For the linear regression between \Rsquared{} and the
  galaxy metallicity, the resulting correlation coefficient and slope are
  respectively $0.15 \pm 0.18$ and $0.89 \pm 1.15$. These numbers are also
  consistent with 0, so the galaxy metallicity has no impact on the
  correlations.  For comparison, the linear regression between \Rsquared{}
  and the galaxy's total CO luminosity yields a correlation coefficient and
  a slope of respectively $0.32 \pm 0.13$ and $0.16 \pm 0.06$.}

\clearpage{}

\section{Supplemental information about the co-variations of GMC
  properties}
\label{sec:appendix:LarsonHeyer}

\FigLarsonHeyer{} %
\FigTpeakVsMco{} %

The evolution of the correlations of the GMC properties with the \Ha{}
luminosity of the associated \HII{} regions show two different trends. It
is well known that GMC properties co-vary through Larson's and Heyer's
relationships~\citep{Heyer2004, Larson1981}. We ask here which part of the
uncovered trends are controlled by these co-variations of the GMC
properties.

Figure~\ref{fig:corr:larson} shows the joint PDFs of the quantities that
are known to co-vary in our sample~\citep{rosolowsky21}.
\begin{description}
\item[\textbf{Larson's first relationship}] links the GMC velocity
  dispersion to its radius through
  \begin{equation}
    \sigma_v \propto R^{0.45\pm0.02}.
  \end{equation}
\item[\textbf{Larson's second relationship}] links the GMC velocity
  dispersion to its mass through
  \begin{equation}
    \sigma_v \propto M_\emr{CO}^{0.27\pm0.01}.
  \end{equation}
\item[\textbf{Heyer's relationship}] links the characteristic turbulent
  linewidth to the surface density through
  \begin{equation}
    \sigma_0 \propto \Sigma_\emr{gas}^{0.49\pm0.02}.
  \end{equation}
\end{description}
Moreover, Fig.~\ref{fig:corr:larson} shows that all these power-law
relations are stable within the credibility intervals when increasing the
MOP. Unlike the correlations between different GMC properties and the
\HII{} region luminosity, Larson's and Heyer's relationships as well as the
correlation between the CO peak temperature and the molecular surface
density are insensitive to the minimum overlap percentage, especially in
terms of power law index.

The similar evolution with the MOP of the correlations of the molecular
surface density, CO peak temperature, and characteristic turbulent
linewidth with the \Ha{} luminosity leads us to ask whether there is an
additional co-variation between the CO peak temperature and the molecular
surface density. Indeed, Fig.~\ref{fig:corr:larson} suggests that there is
a relationship, even though imperfect, through
\begin{equation}
  T_\emr{peak} \propto \Sigma_\emr{gas}^{0.38\pm0.01}.
\end{equation}
Figure~\ref{fig:tpeak:vs:mco} shows the evolution of the joint PDF of the
CO peak temperature and the GMC mass with the minimum overlap
percentage. We find a positive correlation
\begin{equation}
  T_\emr{peak} \propto M_\emr{CO}^{1.07\pm0.04},
\end{equation}
with a \Rsquared{} value of 0.27. This positive correlation implies that
the two behaviors uncovered between the GMC properties and the \Ha{}
luminosities are independent. Indeed, an anti-correlation between the CO
peak temperature and the GMC mass is needed to explain the shift of the
marginalized PDF of these two quantities in opposite directions. In other
words, the presence of an \HII{} region more or less close to its parent
GMC modifies some of the intrinsic scaling relationships between GMC
properties.

\end{document}

%% file: definitions.tex
\usepackage{graphicx}
\usepackage{natbib}
\usepackage{xcolor}
\usepackage{multirow}
\usepackage{ulem}
\usepackage{hyperref}

\usepackage{CJKutf8}
\usepackage{comment}
\usepackage{makecell}
\usepackage{subcaption}

\newcommand{\emm}[1]{\ensuremath{#1}}
\newcommand{\emr}[1]{\emm{\mathrm{#1}}}
\newcommand{\unit}[1]{\emr{\,#1}}
\newcommand{\e}[1]{\emm{\times 10^{#1}}}

\newcommand{\COJtwo}{CO(2$-$1)}
\newcommand{\kms}{\unit{km\,s^{-1}}}
\newcommand{\Msun}{\unit{M_{\odot}}}

\newcommand{\Ha}{\emr{H\alpha}}
\newcommand{\HII}{\ion{H}{II}}
\newcommand{\Rsquared}{\emm{R^2}}

\newlength{\PanelHeight}

\newcommand{\modifref}[1]{{#1}}
\newcommand{\modifreftwo}[1]{{#1}}
